\PassOptionsToPackage{dvipsnames}{xcolor}
\documentclass[acmsmall,screen,authorversion,nonacm]{acmart}

\usepackage[utf8]{inputenc}
\usepackage[dvipsnames]{xcolor}
\usepackage{amsmath, amsthm}
\usepackage{thmtools}
\usepackage{mathtools}
\usepackage{graphicx}
\usepackage{wrapfig}
\usepackage{makecell}
\usepackage{parskip}
\usepackage{array}
\usepackage{tabularx}
\usepackage{tikz}
\usepackage{lineno}
\usepackage{csquotes}
\usepackage{thm-restate}
\usepackage{stackengine}
\usepackage{diagbox}
\usepackage{verbatim}
\usepackage{multirow}
\usepackage{stmaryrd}
\usepackage[inline]{enumitem}
\usepackage{listings}
\usepackage{subcaption}
\usepackage{hyperref}
\usepackage{pgfplots}
\usepackage[framemethod=tikz]{mdframed}
\usepackage[ruled,linesnumbered]{algorithm2e}
\usepackage[capitalise]{cleveref}
\makeatletter
\@ifundefined{newcounteralias}{}{%
  \let\RM@orig@newcounteralias\newcounteralias
  \renewcommand*\newcounteralias[2]{%
    \@ifundefined{c@#1}{\RM@orig@newcounteralias{#1}{#2}}{}}%
}
\makeatother
\usepackage{todonotes}
\usepackage{booktabs}
\usepackage{pgfplotstable}
\usepackage{colortbl}
\usepackage{mathpartir}
\usepackage[group-separator={,}, group-minimum-digits=4]{siunitx}
\usetikzlibrary{arrows.meta}
\usetikzlibrary{shapes,snakes}
\usetikzlibrary { decorations.pathmorphing, decorations.pathreplacing, decorations.shapes, }
\pgfdeclarelayer{background layer}
\pgfdeclarelayer{foreground layer}
\pgfsetlayers{background layer,main,foreground layer}

\usetikzlibrary{arrows,positioning,automata}
\graphicspath{ {./} }

\tikzset{
>=stealth',
punkt/.style={
       rectangle,
       rounded corners,
       draw=black, very thick,
       text width=6.5em,
       minimum height=2em,
       text centered},
pil/.style={
       ->,
       thick,
       shorten <=2pt,
       shorten >=2pt,},
main/.style={
    draw, rectangle,inner sep=4pt,
    }
}

\renewcommand*\descriptionlabel[1]{\hspace\labelsep
  \normalfont{#1}}

\makeatletter
\newcommand{\customlabel}[2]{%
   \protected@write \@auxout {}{\string \newlabel {#1}{{#2}{\thepage}{#2}{#1}{}} }%
   \hypertarget{#1}{#2}
}
\makeatother

\newcommand{\ToolName}{\textsc{RoboKAT}}

\definecolor{mGreen}{rgb}{0,0.6,0}
\definecolor{mGray}{rgb}{0.5,0.5,0.5}
\definecolor{mPurple}{rgb}{0.58,0,0.82}
\definecolor{backgroundColour}{rgb}{0.95,0.95,0.92}
\definecolor{mybluecolor}{rgb}{0.2549019607843137, 0.2117647058823529, 0.9823529411764706}

\newlist{compactitem}{itemize}{3} %
\setlist[compactitem]{label={\Large \textbullet},nosep,leftmargin=*}
\crefname{compactitemi}{Item}{Items}

\newlist{compactenum}{enumerate}{3} %
\setlist[compactenum]{label=(\arabic*), nosep,leftmargin=*}
\crefname{compactenumi}{Item}{Items}
\crefname{tcrule}{Rule}{Rules}
\crefname{wfcond}{Condition}{Conditions}

\newcommand{\Nats}{\mathbb{N}}
\newcommand{\IntSet}[1]{[#1]}

\newcommand{\PSPACE}{\mathbf{PSPACE}}
\newcommand{\NPSPACE}{\mathbf{NPSPACE}}

\newcommand{\tuple}[1]{\langle #1 \rangle}

\newcommand{\KAT}{\mathsf{KAT}}

\newcommand{\expr}{\mathsf{exp}}

\newcommand{\Prog}{\mathsf{Pseq}}

\newcommand{\ConcProg}{\mathsf{P}}

\newcommand{\ExpressionDom}{\mathsf{Exp}}
\newcommand{\InstructionDom}{\mathsf{Inst}}
\newcommand{\NDet}{\mathsf{NDet}}
\newcommand{\LabelDom}{\mathsf{Lab}}
\newcommand{\ValueDom}{\mathsf{Val}}
\newcommand{\LocationDom}{\mathsf{Loc}}
\newcommand{\RegisterDom}{\mathsf{Reg}}
\newcommand{\reg}{d}  %
\newcommand{\kexp}{\varphi}  %
\newcommand{\rsym}{b}  %
\newcommand{\ThreadDom}{\mathsf{Tid}}
\newcommand{\EventDom}{\mathsf{E}}
\newcommand{\ReadDom}{\mathsf{R}}
\newcommand{\WriteDom}{\mathsf{W}}

\newcommand{\RMWDom}{\mathsf{RMW}}
\newcommand{\FenceDom}{\mathsf{F}}
\newcommand{\AnnDom}{\mathsf{Ann}}
\newcommand{\failann}[1]{\overline{#1}}
\newcommand{\FailDom}{\mathsf{FCAS}}

\newcommand{\Events}{E}

\newcommand{\pc}{\mathsf{pc}}
\newcommand{\PC}{\mathsf{PC}}

\newcommand{\mrel}{\mathsf{rel}}
\newcommand{\macq}{\mathsf{acq}}
\newcommand{\mrlx}{\mathsf{rlx}}
\newcommand{\msc}{\mathsf{sc}}
\newcommand{\macqrel}{\mathsf{acq\text{-}rel}}

\newcommand{\op}{\mathsf{opr}}
\newcommand{\tid}{\mathsf{tid}}
\newcommand{\event}{e}
\newcommand{\id}{\mathsf{id}}
\newcommand{\llab}{\mathsf{lab}}
\newcommand{\src}{\mathsf{src}}
\newcommand{\tgt}{\mathsf{tgt}}
\newcommand{\ann}{\mathsf{ann}}
\newcommand{\lloc}{\mathsf{loc}}

\newcommand{\val}{\mathsf{val}}
\newcommand{\rd}{\mathtt{r}}

\newcommand{\wt}{\mathtt{w}}
\newcommand{\fc}{\mathtt{f}}
\newcommand{\rmw}{\mathtt{rmw}}

\definecolor{colorPO}{rgb}{0.75,0.5,0.25}
\colorlet{colorSB}{black!75}
\definecolor{colorRF}{rgb}{0,0,1}
\definecolor{colorMB}{rgb}{0.8,0,0}
\colorlet{colorCO}{colorMB}
\colorlet{colorMO}{colorMB}
\colorlet{colorOB}{black}
\definecolor{colorFR}{rgb}{0.45,0.12,0.05}
\colorlet{colorECO}{black}
\colorlet{colorCOM}{black}
\colorlet{colorSW}{black}
\definecolor{colorHB}{rgb}{0,0.4,0}
\definecolor{colorPPO}{rgb}{0.93,0,0.55}
\colorlet{colorRSEQ}{black}
\colorlet{colorSC}{black}
\colorlet{colorGW}{black}
\colorlet{colorPSC}{black}
\colorlet{colorREL}{black}
\colorlet{colorSO}{black}
\colorlet{colorWB}{olive}
\colorlet{colorDOB}{violet}
\colorlet{colorRMW}{brown}
\colorlet{colorVARa}{Fuchsia}
\colorlet{colorVARb}{RedOrange}
\colorlet{colorVARc}{WildStrawberry}
\colorlet{colorDP}{Fuchsia}

\newcommand{\identity}[1]{[#1]}
\newcommand{\seqbef}{\textcolor{colorSB}{\mathsf{sb}}}
\newcommand{\po}{\textcolor{colorPO}{\mathsf{po}}}

\newcommand{\poloc}{\textcolor{colorPO}{\mathsf{poloc}}}
\newcommand{\rf}{\textcolor{colorRF}{\mathsf{rf}}}

\newcommand{\rfe}{\textcolor{colorRF}{\mathsf{rfe}}}

\newcommand{\mb}{\textcolor{colorMB}{\mathsf{mb}}}
\newcommand{\mo}{\textcolor{colorMO}{\mathsf{mo}}}

\newcommand{\fr}{\textcolor{colorFR}{\mathsf{fr}}}

\newcommand{\eco}{\mathsf{eco}}

\newcommand{\hb}{\textcolor{colorHB}{\mathsf{hb}}}

\newcommand{\sw}{\mathsf{sw}}

\newcommand\locx[1]{{{#1}_{x}}}
\newcommand{\acyclic}{\mathsf{acy}}
\newcommand{\irreflexive}{\mathsf{irr}}
\newcommand{\rempty}{\mathsf{emp}}
\newcommand{\Ax}[1]{\textup{(#1)}} %

\newcommand{\ttest}{\tau}
     
\newcommand{\RelBase}{\mathsf{Rel}} 

\newcommand{\mm}{\mathsf{MM}}
\newcommand{\scmm}{\mathsf{SC}}
\newcommand{\wramm}{\mathsf{WRA}}
\newcommand{\ramm}{\mathsf{RA}}
\newcommand{\rlxmm}{\mathsf{Relaxed}}
\newcommand{\arlxmm}{\mathsf{Relaxed\text{-}Acyclic}}

\newcommand{\sramm}{\mathsf{SRA}}

\newcommand{\rcmm}{\mathsf{RC20}}
\newcommand{\tsomm}{\mathsf{TSO}}
\newcommand{\psomm}{\mathsf{PSO}}
\newcommand{\powermm}{\mathsf{POWER}}

\newcommand{\rcelevenmm}{\mathsf{RC11}}
\newcommand{\wrcelevenmm}{\mathsf{WRC11}}
\newcommand{\lramm}{\mathsf{LRA}}
\newcommand{\ocamlmm}{\mathsf{OCaml}}
\newcommand{\wasmmm}{\mathsf{Wasm}}
\newcommand{\armmm}{\mathsf{Armv8}}
\newcommand{\lkmm}{\mathsf{LKMM}}
\newcommand{\riscmm}{\mathsf{RISC\text{-}V}}
\newcommand{\immmm}{\mathsf{IMM}}
\newcommand{\rmomm}{\mathsf{RMO}}
\newcommand{\psimm}{\mathsf{PSI}}

\newcommand{\pw}{\mathsf{pw}}
\newcommand{\cpw}{\mathsf{cpw}}

\newcommand{\Bag}{\mathcal{X}}

\newcommand{\LTS}{\mathcal{T}}

\newcommand{\States}{\mathcal{Q}}
\newcommand{\StBase}{Q}
\newcommand{\StName}[2]{\if\relax\detokenize{#1}\relax\StBase_{\mathit{#2}}\else\StBase^{#1}_{\mathit{#2}}\fi}
\newcommand{\ActSet}[1][]{\StName{#1}{act}}  %
\newcommand{\EdgSet}[1][]{\StName{#1}{edg}}
\newcommand{\InSet}[1][]{\StName{#1}{in}}    %
\newcommand{\OutSet}[1][]{\StName{#1}{out}}  %
\newcommand{\RepMap}[1][]{\LamName{#1}{rep}}  %
\newcommand{\FstSet}[1][]{\StName{#1}{fst}}  %
\newcommand{\LstSet}[1][]{\StName{#1}{lst}}  %
\newcommand{\SrcSet}[1][]{\StName{#1}{src}}  %
\newcommand{\PndSet}[1][]{\StName{#1}{pnd}}  %
\newcommand{\SumPath}[1][]{\StName{#1}{pth}} %
\newcommand{\SumCyc}[1][]{\StName{#1}{cyc}}  %
\newcommand{\LamName}[2]{\if\relax\detokenize{#1}\relax\lambda_{\mathit{#2}}\else\lambda^{#1}_{\mathit{#2}}\fi}
\newcommand{\ThrMap}[1][]{\LamName{#1}{thr}}  %
\newcommand{\LocMap}[1][]{\LamName{#1}{loc}}  %
\newcommand{\LabMap}[1][]{\LamName{#1}{lab}}  %
\newcommand{\TrMap}[1][]{\LamName{#1}{tr}}    %
\newcommand{\Alphabet}{\Sigma}
\newcommand{\Trans}{\Delta}
\newcommand{\InitState}{\mathsf{init}}

\newcommand{\LTo}[1]{\xrightarrow{\raisebox{-0.5ex}[0ex][0ex]{\footnotesize $#1$} } }
\newcommand{\LTransTo}[1]{\xhookrightarrow{\raisebox{-0.5ex}[0ex][0ex]{\footnotesize $#1$} } }

\newcommand{\Paragraph}[1]{\smallskip\noindent{\bf #1}}
\newcommand{\SubParagraph}[1]{\smallskip\noindent{\em #1}}

\newcommand{\ProjectGraph}[3]{{#1}{\downharpoonright}_{#2}^{{#3}} }

\newcommand{\Lang}[1]{\mathcal{L}(#1)}

\newcommand{\ExecutionGraphsOf}[2]{\llbracket{#1}\rrbracket_{#2}}

\newcommand{\GraphsMM}[1]{\llbracket{#1}\rrbracket}

\newcommand{\CFEIntro}{\textsc{Intro}}
\newcommand{\CFEForget}{\textsc{Forget}}
\newcommand{\CFEIntrosb}{\textsc{Intro}_{\seqbef}}
\newcommand{\CFEIntromb}{\textsc{Intro}_{\mb}}
\newcommand{\CFEIntrorf}{\textsc{Intro}_{\rf}}
\newcommand{\CFEIntroR}[1]{\textsc{Intro}_{#1}}

\newcommand{\dom}[1]{\textsc{dom}(#1)}
\newcommand{\img}[1]{\textsc{img}(#1)}

\newcommand{\Ops}{\mathrm{Ops}}

\newcommand{\TestDom}{\mathsf{Tests}}

\tikzset{
   every path/.style={>=stealth},
   po/.style={->,color=colorPO, thick},
   seqbef/.style={->,color=colorSB, thick},
   ppo/.style={->,color=colorPPO, thick},
   sw/.style={->,color=colorSW, thick},
   rf/.style={->,color=colorRF,dashed, thick},
   mrf/.style={->,color=colorRF,dashed, thick},
   urf/.style={->,color=colorRF, thick},
   fr/.style={->,color=colorFR,dashed, thick},
   hb/.style={->,color=colorHB,thick, thick},
   mo/.style={->,color=colorMO,dotted, very thick},
   dob/.style={->,color=colorDOB, very thick},
   ob/.style={->,color=colorOB, very thick},
   rmw/.style={->,color=colorRMW,thick, thick},
   rseq/.style={->,color=colorRSEQ,thick,dotted, thick},
   com/.style={->,color=colorCOM,thick, thick},
   mb/.style={->,color=colorMB,dotted, very thick},
}

\tikzset{
    ncbar angle/.initial=90,
    ncbar/.style={
        to path=(\tikztostart)
        -- ($(\tikztostart)!#1!\pgfkeysvalueof{/tikz/ncbar angle}:(\tikztotarget)$)
        -- ($(\tikztotarget)!($(\tikztostart)!#1!\pgfkeysvalueof{/tikz/ncbar angle}:(\tikztotarget)$)!\pgfkeysvalueof{/tikz/ncbar angle}:(\tikztostart)$)
        -- (\tikztotarget)
    },
    ncbar/.default=0.5cm,
}

\tikzset{round left paren/.style={ncbar=0.5cm,out=110,in=-110}}
\tikzset{round right paren/.style={ncbar=0.5cm,out=70,in=-70}}

\tikzset{
  katpic/.style    = {scale=1.0, every node/.style={transform shape}},
  katnode/.style   = {draw, circle, inner sep=0pt, minimum size=5.2mm, font=\small},
  katnodesmall/.style   = {draw, circle, inner sep=0pt, minimum size=2.6mm, font=\small},
  katacc/.style    = {katnode, double, double distance=0.8pt},
  kattrans/.style  = {->, >={Stealth[length=3.5pt]}, line width=0.7pt},
  katstart/.style  = {kattrans, shorten >=1pt},
  every loop/.style = {kattrans, looseness=7},
  katlab/.style    = {font=\small, inner sep=1.2pt},
    evnode/.style = {draw, rounded corners=2pt, inner sep=2pt, font=\footnotesize},
  evsb/.style   = {kattrans, colorSB},
  evrf/.style   = {kattrans, colorRF},
  evmb/.style   = {kattrans, colorMB},
}
\newcommand{\Ksb}{{\color{colorSB}\seqbef}}
\newcommand{\Krf}{{\color{colorRF}\rf}}
\newcommand{\Krfinv}{{\color{colorRF}\rf^{-1}}}
\newcommand{\Kmb}{{\color{colorMB}\mb}}

\colorlet{colorLocX}{RoyalBlue}
\colorlet{colorLocY}{BurntOrange}
\colorlet{colorLocXDark}{colorLocX!85!black}
\colorlet{colorLocYDark}{colorLocY!75!black}
\newcommand{\lx}[1]{{\color{colorLocXDark}#1}}
\newcommand{\ly}[1]{{\color{colorLocYDark}#1}}
\colorlet{colorLocId}{ForestGreen}
\colorlet{colorLocLt}{Plum}
\newcommand{\locid}{{\color{colorLocId!80!black}p}}
\newcommand{\loclt}{{\color{colorLocLt!80!black}q}}

\newcommand{\GLang}[1]{\mathcal{G}(#1)}

\newcommand{\pwSC}{\pw_{\scmm}}

\colorlet{colorCa}{BurntOrange}
\colorlet{colorCb}{NavyBlue}
\colorlet{colorCc}{BrickRed}
\newcommand{\cA}{{\color{colorCa!55!black}1}}
\newcommand{\cB}{{\color{colorCb!75!black}2}}
\newcommand{\cC}{{\color{colorCc!75!black}3}}

\newcommand{\ArtifactUrl}{https://github.com/kobusgiovanna/robustness-metatheorem-formalization}
\title{A Unifying Metatheorem on Weak-Memory Robustness}

\author{Giovanna Kobus Conrado}
\orcid{0000-0001-9474-6505}
\affiliation{%
  \institution{Aarhus University}
  \city{Aarhus}
  \country{Denmark}
}
\email{gkc@cs.au.dk}

\author{Stephan Spengler}
\orcid{0009-0009-5722-8843}
\affiliation{%
  \institution{Uppsala University}
  \city{Uppsala}
  \country{Sweden}
}
\email{stephan.spengler@it.uu.se}

\author{Mohamed Faouzi Atig}
\orcid{0000-0001-8229-3481}
\affiliation{%
  \institution{Uppsala University}
  \city{Uppsala}
  \country{Sweden}
}
\email{mohamed\_faouzi.atig@it.uu.se}

\author{Andreas Pavlogiannis}
\orcid{0000-0002-8943-0722}
\affiliation{%
  \institution{Aarhus University}
  \city{Aarhus}
  \country{Denmark}
}
\email{pavlogiannis@cs.au.dk}

\begin{document}

\begin{abstract}
Memory models specify the behaviors that concurrent programs may exhibit under compiler and hardware optimizations, which often deviate from Sequential Consistency ($\mathsf{SC}$).
\emph{Robustness} is the property of a program $P$ not exhibiting any weak behaviors when executing under a weak memory model $\mathsf{MM}$, compared to $\mathsf{SC}$.
As such, the algorithmic problem of deciding whether $P$ is robust against $\mathsf{MM}$ has been studied extensively, and has been repeatedly shown to be in $\PSPACE$ for many memory models, such as TSO, PSO, and Release/Acquire.
However, each model has required its own, dedicated proof, raising the natural question of whether robustness admits a unifying treatment across memory models.

In this work, we answer this question affirmatively for a broad class of memory models definable in Kleene Algebra with Tests (KAT).
We prove that robustness against any prefix-closed, KAT-definable memory model is in $\PSPACE$.
This metatheorem recovers nearly all prior $\PSPACE$ robustness results, and establishes $\PSPACE$ membership for several new models, including the full RC11 and the OCaml memory model.
We have further formalized the correctness of our approach in Lean.
On the other hand, we prove that robustness against the KAT-definable $\rlxmm$ model is undecidable, and thus
prefix-closedness cannot be dropped, in general.
Finally, we implement our algorithm in a prototype tool $\ToolName$, and demonstrate its flexibility by checking robustness against nine memory models, several of which are not handled by prior work.
\end{abstract}


\maketitle

\section{Introduction}\label{SEC:INTRO}

Sequential Consistency ($\scmm$) is the standard view of communication in concurrency, whereby a thread's modification of shared memory is immediately visible to the whole system~\cite{Lamport1979}.
This allows program executions to admit a simple, sequential representation, that reflects the mental model of most practitioners in concurrent programming.
In practice, however, interthread communication is considerably more involved, owing to compiler optimizations as well as architectural features such as buffers, caches, and speculative execution.
These features can cause threads to hold divergent views of the shared memory, and produce highly complex and counterintuitive \emph{weak} behaviors.
\emph{Memory models}~\cite{Alglave2012} are formal specifications of the behaviors that a program may exhibit in such a system, and have become standard in programming (e.g., for C/C++~\cite{Boehm2008}, Java~\cite{Manson2005}, Rust~\cite{Rustonomicon}, OCaml~\cite{Dolan2018}, the Linux Kernel~\cite{Alglave2018}) and hardware design (e.g., SPARC/x86-TSO~\cite{Sewell2010}, Arm~\cite{Pulte2017}).

One desirable property of a program $\ConcProg$ executing under a memory model $\mm$ is \emph{robustness}, which specifies that $\ConcProg$ does not exhibit any weak behaviors under $\mm$~\cite{Shasha1988,Bouajjani2011}.
A robust program, thus, runs with the efficiency of the optimizations allowed by $\mm$, while following the simple mental and reasoning model of $\scmm$.
Robust programs further admit $\scmm$-based verification techniques, which are far more efficient and lightweight, as typically verification under weak memory is of much higher complexity~\cite{Atig2010,Lahav2022}, or even undecidable~\cite{Atig2010,Abdulla19,Conrado2026}.

As such, algorithmically deciding whether a program is robust against a memory model has been studied extensively across many models.
Case after case, the problem was proven to be decidable in $\PSPACE$ for $\tsomm$~\cite{Bouajjani2011,Bouajjani2013}, $\psomm$~\cite{Derevenetc2015}, $\rmomm$~\cite{Derevenetc2015}, $\powermm$~\cite{Derevenetc14}, $\ramm$~\cite{Lahav2019}, and a C11-style model~\cite{Margalit2021}.
Analogous results for transactional models further imply a $\PSPACE$ bound for the RMW-free fragments of SRA and WRA~\cite{Beillahi19}.
At this point, it is natural to ask whether each case indeed requires individual treatment, or they all fall under the umbrella of a single, root argument.
\emph{Is there a unifying theory for weak-memory robustness?}

\Paragraph{Our contributions.}
This paper answers this question affirmatively for a broad class of memory models.
Our starting point is \textsc{Kater}~\cite{Kokologiannakis2023}, which observes that most popular memory models are expressible in Kozen's Kleene Algebra with Tests (KAT)~\cite{Kozen97}.
We show that KAT-definability suffices to decide the robustness question uniformly and, complexity-wise, optimally, under a natural prefix-closedness condition.
We summarize here our main contributions, while we refer to the coming sections for formal definitions and proofs.

We consider program executions represented as standard execution graphs~\cite{Batty2011,Lahav2017} $G=\tuple{\Events,\seqbef,\rf,\mb}$, where
$\Events$ is a set of events comprising the execution,
$\seqbef$ represents the execution order of each thread,
$\rf$ is a reads-from relation pairing reading events with the writes they observe, and
$\mb$ represents the order in which writes modify the shared memory.
Our primary result is a unifying metatheorem for weak-memory robustness, which we have also formalized in Lean\footnote{
Available at \url{\ArtifactUrl}.
The formalization proves our reduction of robustness to the emptiness of an automaton, and that the number of states of this automaton is at most single-exponential in the input size.
The remaining argument, which proves the problem in $\PSPACE$, is not formalized, as Mathlib does not support space complexity.
Nevertheless, this argument is standard textbook material.
}.

\begin{restatable}{theorem}{thmpspaceprefix}\label{THE:PSPACE_PREFIX}
Given any concurrent program $\ConcProg$ and any prefix-closed memory model $\mm$ defined in $\KAT$, deciding whether $\ConcProg$ is robust against $\mm$ is in $\PSPACE$.
\end{restatable}

Prefix-closedness is the natural property that the execution of any program can be constructed incrementally as the program runs, and is satisfied by most common memory models.
\cref{THE:PSPACE_PREFIX} unifies the $\PSPACE$ robustness of nearly all prior work\footnote{The exceptions are POWER and RMO, in which executions have richer signatures containing dependencies.
}, and extends the results for $\sramm$ and $\wramm$ from their RMW-free fragments to the full models.
Moreover, it proves that robustness is in $\PSPACE$ for several \emph{new}  memory models, e.g.,
the full $\rcelevenmm$~\cite{Lahav2017}, 
the $\ocamlmm$ memory model~\cite{Dolan2018},
$\lramm$~\cite{Singh24}, and 
$\psimm$~\cite{Raad18} (in its shared-memory realization with unary transactions)
\footnote{
For C11-style models (e.g. $\rcelevenmm$), we omit non-atomics, as these can be handled separately in $\PSPACE$ --- see \cref{SEC:APP_MODELS} for details on this, as well as the $\KAT$ definitions of several models mentioned throughout the paper.
}.
This closes all memory models that were left open in~\cite{Lahav2019} (with $\psimm$ restricted to unary transactions).
We further remark that robustness is at least as hard as state reachability under $\scmm$, against any memory model that can simulate $\scmm$ (which is often easy, e.g., by utilizing CAS operations or memory fences~\cite{Alglave2010}).
In such cases, robustness becomes $\PSPACE$-hard, and thus \cref{THE:PSPACE_PREFIX} is tight.

We prove \cref{THE:PSPACE_PREFIX} by showing that non-robust programs always contain a regular set of weak executions that can be represented efficiently.
This follows the high-level approach of Courcelle's celebrated work on MSO for graphs~\cite{Courcelle2012}; however, Courcelle's theorem gives a much larger, non-elementary complexity bound.
To arrive at $\PSPACE$, we first prove that $\scmm$-consistent executions have bounded pathwidth.
This, together with the prefix-closedness of $\mm$, implies that non-robust programs always exhibit weak executions of bounded pathwidth.
Thus, the search for a witness reduces to searching for a bounded-pathwidth execution of $\ConcProg$ that is $\mm$-consistent and $\scmm$-inconsistent.
To perform this search, we define an algebra of graph operations, so that a finite sequence of such operations represents the path decomposition of an execution.
We then develop dedicated automata for verifying that such an execution constitutes a robustness violation witness, i.e., that it is 
\begin{enumerate*}[label=(\roman*)]
    \item valid for $\ConcProg$,
    \item $\mm$-consistent, and
    \item $\scmm$-inconsistent.
\end{enumerate*}
Crucially, we ensure that each automaton transition can be verified in polynomial space, leading to our $\PSPACE$ bound.

Our automata-theoretic treatment further implies that robustness is always efficiently \emph{monitorable} in a dynamic setting, e.g., in the spirit of the tool \textsc{Sober} for TSO~\cite{Burckhardt2008}, or the recent 
\textsc{Rsan} for a C11-style model~\cite{Margalit2025}.
Thus, although such tools rest on clever insights and memory-model specific optimizations to be efficient, their existence also follows from a unifying algorithmic principle.

Although the prefix-closedness condition of \cref{THE:PSPACE_PREFIX} is satisfied by most common memory models, it is not universal.
E.g., the, $\KAT$-definable, $\rlxmm$ model is not prefix-closed.
It is, thus, natural to ask whether this condition is necessary.
We prove that it cannot be dropped, in general.

\begin{restatable}{theorem}{therelaxed}\label{THE:RELAXED}
Robustness against $\rlxmm$ is undecidable, already over programs with two threads and four memory locations.
\end{restatable}

Finally, we implement the algorithm behind \cref{THE:PSPACE_PREFIX} in a prototype tool $\ToolName$, which is parametric on the memory model.
We use $\ToolName$ to check robustness against nine memory models, including $\rcelevenmm$, $\ocamlmm$, $\sramm$  and $\psimm$, for which, to our knowledge, no sound and complete robustness checker existed.
Although its generality comes at a natural performance cost compared to model-specific tools, $\ToolName$ gives the correct verdict on each of 180 litmus test/model pairs within seconds, and finds all known violations in the \textsc{Rocker} benchmarks~\cite{Lahav2019}.

\section{Preliminaries}\label{SEC:PRELIM}

In this section, we fix notation and recall automata, introduce execution graphs, programs, and memory models, and define robustness, the main subject of this work.

\subsection{Background}\label{SUBSEC:PRELIM_BACK}

\Paragraph{General notation.}
Given an integer $i$, we let $\IntSet{i}=\{1,\dots, i\}$.
For a binary relation $R$, we write $R^?$, $R^+$, $R^*$, and $R^{-1}$ for its reflexive, transitive, and reflexive-transitive closures and its inverse, respectively.
We denote by $R_1;R_2$ the composition of $R_1$ and $R_2$.
We write $\irreflexive(R)$, $\acyclic(R)$, and $\rempty(R)$ to indicate that $R$ is irreflexive, acyclic, and empty, respectively. 
We occasionally call $\tuple{\event_1, \event_2}\in R$ an edge, written $\event_1\LTo{R}\event_2$.
Given a set $S$, we write $\identity{S}$ for the identity relation on $S$.
A binary relation $R$ is a \emph{successor relation} on a finite set $X$ if it is the transitive reduction of some total order on $X$.

\Paragraph{Labeled transition systems.}
A \emph{labeled transition system} (LTS) is a tuple $\LTS=\tuple{\States,\Alphabet,\Trans,\InitState}$, where
\begin{enumerate*}[label=(\roman*)]
\item $\States$ is a finite set of states,
\item $\Alphabet$ is a finite alphabet,
\item $\Trans\subseteq \States\times (\Alphabet\cup\{\epsilon\}) \times \States$ is a transition relation, and
\item $\InitState\in \States$ is the initial state.
\end{enumerate*}
We write $p\LTransTo{\alpha}q$ for $\tuple{p,\alpha,q}\in \Trans$, and call the transition \emph{silent} if $\alpha=\epsilon$.
For a transition $\delta=\tuple{p,\alpha,q}$, we let $\src(\delta)=p$, $\tgt(\delta)=q$, and $\llab(\delta)=\alpha$.
A \emph{run} of $\LTS$ is a sequence $\InitState\LTransTo{\alpha_1}p_1\LTransTo{\alpha_2}\dots\LTransTo{\alpha_n}p_n$ of transitions starting at $\InitState$.
The \emph{language} $\Lang{\LTS}\subseteq \Alphabet^*$ contains every word $\alpha_1\alpha_2\dots\alpha_n$ that induces a run of $\LTS$.
The \emph{size} of $\LTS$ is $|\LTS|\triangleq|\States|+|\Trans|$.

\Paragraph{Word automata.}
A \emph{word automaton} is a tuple $\mathcal{A} = \tuple{\States, \Alphabet, \Trans, \InitState, \States_F}$, where $\tuple{\States, \Alphabet, \Trans, \InitState}$ is an LTS and $\States_F \subseteq \States$ is a set of \emph{accepting} states.
The \emph{language} $\Lang{\mathcal{A}}$ of $\mathcal{A}$ contains every word $\alpha_1\alpha_2\dots\alpha_n$ that induces a run in $\mathcal{A}$ ending in $\States_F$.
$\mathcal{A}$ is \emph{deterministic} if $\Trans$ is a function with no silent transitions.

\subsection{Execution Graphs}\label{SUBSEC:PRELIM_EXEC}

\Paragraph{Program domains.}
We consider concurrent programs consisting of a finite set of threads $\ThreadDom$.
Each thread executes events, and communicates with others through a finite set $\LocationDom$ of shared \emph{locations} that store \emph{values} from a finite set $\ValueDom$.
Events further carry an \emph{annotation} from a finite set $\AnnDom$.

\Paragraph{Labels and events.}
A \emph{label} is a \emph{read} $\rd_a(t,x,v)$, a \emph{write} $\wt_a(t,x,v)$, a \emph{read-modify-write (RMW)} $\rmw_a(t,x,v_{\rd},v_{\wt})$, or a \emph{fence} $\fc_a(t)$, where $t\in\ThreadDom$ is a thread, $x\in\LocationDom$ is a location, $v,v_{\rd},v_{\wt}\in\ValueDom$ are values, and $a\in\AnnDom$ is an annotation (e.g., an access mode, such as $\mrlx$ in $\rcelevenmm$).
We let $\LabelDom$ be the set of labels.
An \emph{event} $\event=\tuple{\id,\llab}$ consists of an identifier $\id\in\Nats$ and a label $\llab\in\LabelDom$, and $\EventDom$ is the set of events.
We write $\llab(\event)$, $\tid(\event)$, $\op(\event)\in\{\rd,\wt,\rmw,\fc\}$, and $\ann(\event)$ for the label, thread, operation, and annotation of $\event$,
and $\lloc(\event)$, $\val_{\rd}(\event)$, and $\val_{\wt}(\event)$ for its location and the values it reads and writes, when defined (e.g., a fence has no location and no value).
We use the same notation for labels.
We identify events with their labels when identifiers are irrelevant (e.g., $\wt_a(t,x,v)$), and may further omit components that are irrelevant or clear from the context (e.g., $\wt(x)$ or $\fc_{\macq}$).
We let $\ReadDom$, $\WriteDom$, $\RMWDom$, and $\FenceDom$ be the sets of read, write, RMW, and fence events, respectively, and use the same names for the corresponding sets of labels.
For $X\subseteq\EventDom$ or $X\subseteq\LabelDom$, $t\in\ThreadDom$, and $x\in\LocationDom$, we let $X^t$ and $\locx{X}$ be the elements of $X$ of thread $t$ and of location $x$, respectively.

\Paragraph{Execution graphs.}
Under weak memory, program executions are often represented as execution graphs~\cite{Lahav2017}, also called candidate executions~\cite{Batty2011}.
An \emph{execution graph} is a tuple $G=\tuple{\Events,\seqbef,\rf,\mb}$, where $\Events\subseteq\EventDom$ is a finite set of events,
and $\seqbef$, $\rf$, and $\mb$ are binary relations over $\Events$ that are \emph{well-formed}, in the following sense.

\begin{compactitem}
\item The \emph{sequenced-before} relation $\seqbef$ records the order in which events are issued by each thread, by relating each event to its immediate successor. 
Formally, $\seqbef$ is well-formed if it is the union of successor relations on $\Events^t$, one per $t\in\ThreadDom$.
\item The \emph{reads-from} relation $\rf$ relates each read/RMW event to the write/RMW event it reads from.
Formally, $\rf$ is well-formed if $\rf\subseteq\bigcup_{x\in\LocationDom}(\locx{\WriteDom}\cup\locx{\RMWDom})\times(\locx{\ReadDom}\cup\locx{\RMWDom})$, $\rf^{-1}$ is a function with domain $\Events\cap(\ReadDom\cup\RMWDom)$, and $\val_{\wt}(\event_1)=\val_{\rd}(\event_2)$ for each $\event_1\LTo{\rf}\event_2$.
\item The \emph{modifies-before} relation $\mb$ orders the writes/RMWs on each location.
Formally, $\mb$ is well-formed if it is the union of successor relations on $\locx{\Events}\cap(\WriteDom\cup\RMWDom)$, one per $x\in\LocationDom$.
\end{compactitem}
We write $G.\Events$, $G.\seqbef$, $G.\rf$, and $G.\mb$ for the components of $G$, and omit $G$ when it is clear from the context.
Related literature often defines graphs slightly differently, with $\po\triangleq\seqbef^{+}$ instead of $\seqbef$ and $\mo\triangleq\mb^{+}$ instead of $\mb$, but for our purposes, it is more natural to not consider these relations transitively closed.
We call a pair $\tuple{\Events,\seqbef}$, i.e., an execution graph without $\rf$ and $\mb$, an \emph{$\seqbef$-graph}.

\subsection{Programs}\label{SUBSEC:PRELIM_PROG}

We consider programs given as source code, in a toy language~\cite{Lahav2019} with registers, assignments, nondeterministic assignments, conditional branching, and memory operations.
Besides the domains $\ValueDom$, $\LocationDom$, $\ThreadDom$ and $\AnnDom$, each thread $t$ has a finite set $\RegisterDom_t$ of \emph{registers}; these sets are pairwise disjoint, and we let $\RegisterDom\triangleq\bigcup_{t\in\ThreadDom}\RegisterDom_t$.

\Paragraph{Syntax.}
Expressions $\ExpressionDom$ and instructions $\InstructionDom$ are defined by the following grammar.
\[
\begin{array}{r@{\ }l}
\ExpressionDom  \ni \expr & ::= \reg \mid v \mid \expr + \expr \mid \expr = \expr \mid \expr \neq \expr \mid \dots \\
\InstructionDom \ni \mathit{inst} & ::= \reg := \expr \mid \reg := \NDet(\ValueDom') \mid \mathtt{if}\ \expr\ \mathtt{goto}\ n \\
  & \mathrel{\phantom{::=}} \mid x :=_a \expr \mid \reg :=_a x \mid \reg :=_a \mathsf{CAS}(x, \expr, \expr) \mid \mathsf{fence}_a 
\end{array}
\]
where $v\in\ValueDom$, $\ValueDom'\subseteq\ValueDom$, and $\reg\in\RegisterDom$.
We interpret $+$, $=$, $\neq$, and the other binary operators as functions $\ValueDom \times \ValueDom \to \ValueDom$, and we assume that $\ValueDom$ contains two distinct values $0$ and $1$, that comparisons such as $=$ and $\neq$ return $1$ if they hold and $0$ otherwise, and that $\mathtt{if}$ treats $0$ as false and every other value as true.
The first three instructions are local, with $\NDet(\ValueDom')$ representing a nondeterministic choice of a value in $\ValueDom'$.
The next three instructions access the shared memory, respectively performing a write, read, and compare-and-swap operation on location $x$.
The annotation $a\in\AnnDom$ may have some impact on the synchronization that these accesses induce, depending on the memory model.
Finally, the last instruction is a fence, also carrying an annotation $a\in\AnnDom$.
The code of a thread $t$ is a finite instruction sequence $S_t\colon \IntSet{n_t} \to \InstructionDom$, where $n_t$ is the number of instructions of $t$.
The thread runs $S_t(1), S_t(2), \dots$ in order, except that a $\mathtt{goto}$ may jump to another line, and it terminates once it leaves $\IntSet{n_t}$.

\Paragraph{Semantics.}
Each thread $S_t$ induces an LTS $\Prog_t=\tuple{\States_t, \LabelDom^t, \Trans_t, \InitState_t}$ in the standard way; we give a high-level description here, and refer to \cref{SUBSEC:APP_PRELIM_PROG} for the formal definition.
\begin{compactitem}
\item The states $\States_t$ are pairs $\tuple{\pc, \phi}$ of a \emph{program counter} $\pc \in \IntSet{n_t+1}$, pointing to the next instruction, where $n_t+1$ stands for termination and a jump outside $\IntSet{n_t}$ goes to $n_t+1$,
 and a \emph{store} $\phi\colon \RegisterDom_t \to \ValueDom\cup\{\bot\}$, which holds the contents of the registers.
 Note that $|\States_t|=(n_t + 1) \cdot (|\ValueDom|+1)^{|\RegisterDom_t|}$.
We lift $\phi$ to expressions as usual, with $\phi(\expr)=\bot$ whenever $\expr$ mentions a register holding $\bot$, i.e., an uninitialized register.
\item The alphabet $\LabelDom^t$ is the set of labels of events of $t$.
\item The transition relation $\Trans_t$ executes the instruction $S_t(\pc)$, provided that the expressions it evaluates are not $\bot$.
Local instructions lead to silent transitions that simply update $\pc$ and $\phi$.
Every other instruction emits the label of the event it executes, e.g., $x :=_a \expr$ emits $\wt_a(t,x,\phi(\expr))$.
A read $\reg :=_a x$ may emit $\rd_a(t,x,v)$ for any $v\in\ValueDom$, and stores $v$ in $\reg$; the actual value it observes is fixed later.
Likewise, $\reg :=_a \mathsf{CAS}(x, \expr_R, \expr_W)$ emits an RMW label if it reads $\phi(\expr_R)$, and a read label otherwise.
To distinguish it from the read of a read instruction, a failed CAS annotates its read label with a dedicated sibling of its annotation.
\item The initial state $\InitState_t$ is $\tuple{1, \phi_0}$, where $\phi_0$ maps every register to $\bot$.
\end{compactitem}

\Paragraph{Concurrent programs.}
We interchangeably use $S_t$ and $\Prog_t$ to refer to the \emph{sequential program} of thread $t$; their distinction will be clear from the context. 
Likewise, the \emph{concurrent program} $\ConcProg$ is represented as a collection of sequential programs $\tuple{S_t}_{t\in \ThreadDom}$ or their respective LTSs $\tuple{\Prog_t}_{t\in \ThreadDom}$.
Given an $\seqbef$-graph $G=\tuple{\Events,\seqbef}$ and a thread $t\in\ThreadDom$, let $\ProjectGraph{G}{\seqbef}{t}\triangleq \llab(\event_1)\llab(\event_2)\dots \llab(\event_n)$, where $\event_1,\event_2,\dots,\event_n$ enumerates $\Events^{t}$ in $\seqbef$-order.
The \emph{graph language} of $\ConcProg$ is
\[
  \GLang{\ConcProg}\triangleq\{G\colon \text{$G$ is an $\seqbef$-graph and } \ProjectGraph{G}{\seqbef}{t}\in \Lang{\Prog_{t}} \text{ for every } t\in\ThreadDom\},
\]
the $\seqbef$-graphs in which each thread executes according to its own sequential program.

\subsection{Memory Models}\label{SUBSEC:PRELIM_MM}

A memory model specifies which behaviors a concurrent system may exhibit.
In our setting, behaviors are execution graphs, and a model allows those that avoid certain forbidden cycles.
Following~\cite{Kokologiannakis2023}, we express these patterns over Kleene Algebra with Tests ($\KAT$).

\Paragraph{Tests.}
A \emph{test} is a set of labels $\ttest\subseteq\LabelDom$, and we let $\TestDom\triangleq 2^{\LabelDom}$ be the set of tests.
We let $G.\ttest\triangleq\{\event\in G.\Events\colon \llab(\event)\in\ttest\}$ be the events of $G$ satisfying $\ttest$, and we write $\identity{\ttest}$ for $\identity{G.\ttest}$.
In practice, tests are combinations of equalities on the components of a label, such as $\WriteDom\cap\locx{\LabelDom}$, the writes on $x$, or $\{\llab\in\LabelDom\colon\ann(\llab)=\mrlx\}$, the labels annotated with $\mrlx$.
We assume throughout that every test occurring in a memory model is of this form,
so $\llab\in\ttest$ can be decided in time polynomial in $|\ThreadDom| + |\LocationDom| + |\ValueDom| + |\AnnDom|$.

\Paragraph{$\KAT$ expressions.}
Our relation symbols are $\RelBase\triangleq\{\seqbef,\rf,\mb,\seqbef^{-1},\rf^{-1},\mb^{-1}\}$; $G.\rsym$ denotes the corresponding relation of $G$.
So, the relations of an execution graph (together with their inverses) are the building blocks of memory models.
$\KAT$ expressions are given by the grammar
\[
  \kexp ::= \identity{\ttest} \mid \rsym \mid \kexp \cup \kexp \mid \kexp;\kexp \mid \kexp^{+} \mid \kexp^{-1},
\]
where $\ttest\in\TestDom$ and $\rsym\in\RelBase$.
An expression $\kexp$ denotes a binary relation $G.\kexp$ over $G.\Events$, defined as expected: $G.\identity{\ttest}\triangleq\identity{G.\ttest}$, $G.(\kexp_1\cup \kexp_2)\triangleq G.\kexp_1\cup G.\kexp_2$, $G.(\kexp_1;\kexp_2)\triangleq G.\kexp_1;G.\kexp_2$, $G.\kexp^{+}\triangleq(G.\kexp)^{+}$, and $G.\kexp^{-1}\triangleq(G.\kexp)^{-1}$.
We also write $\kexp^{?}\triangleq \kexp\cup\identity{\LabelDom}$ and $\kexp^{*}\triangleq(\kexp^{+})^{?}$.
The \emph{size} $|\kexp|$ of $\kexp$ is its number of occurrences of tests, base relation symbols and operators.
Compared to~\cite{Kokologiannakis2023}, we add the inverse operator, and keep tests as sets of labels rather than as Boolean combinations of predicates.

\Paragraph{Derived relations.}
Common relations are abbreviations of KAT expressions, such as
$\po\triangleq\seqbef^{+}$, $\mo\triangleq\mb^{+}$, $\hb\triangleq(\po\cup\rf)^{+}$, and $\rfe\triangleq\bigcup_{t\in\ThreadDom}\identity{\LabelDom^{t}};\rf;\identity{\LabelDom\setminus\LabelDom^{t}}$.
We call such expressions \emph{derived} relations, and the size of a derived relation is the size of the KAT expression defining it; e.g., $\rfe$ has size $O(|\ThreadDom|)$.
The from-read relation is usually defined as $\fr\triangleq(\rf^{-1};\mo)\setminus\id$, where $\id$ is identity.
It relates a read to the writes that are $\mo$-after the write it reads from, except itself (e.g., in case it is an RMW).
Sensible models that use $\fr$ also require every RMW to read from its $\mb$-predecessor, in which case $\fr$ is $\KAT$-definable as $\fr\triangleq\identity{\ReadDom};\rf^{-1};\mo\cup\identity{\RMWDom};\rf^{-1};\mo;\mo$ --- throughout the paper, we use the latter definition when defining various memory models.

\Paragraph{Memory models.}
A \emph{memory model} $\mm$ is a $\KAT$ expression.
An execution graph $G$ is \emph{consistent in $\mm$}, denoted $G\models\mm$, if $G.\mm$ is irreflexive, and \emph{inconsistent in $\mm$} otherwise.
We let $\GraphsMM{\mm}=\{G\colon G$ is an execution graph and $G\models\mm\}$
be the set of $\mm$-consistent execution graphs.
Some memory models are commonly expressed using several emptiness, acyclicity and irreflexivity conditions, but all these are expressible in our setup:
$\rempty(\kexp)$ is equivalent to $\irreflexive(\kexp;\kexp^{-1})$,
$\acyclic(\kexp)$ is equivalent to $\irreflexive(\kexp^+)$, and 
$\irreflexive(\kexp_1)\land \irreflexive(\kexp_2)$ is equivalent to $\irreflexive(\kexp_1\cup \kexp_2)$.
For example, the standard Sequential Consistency~\cite{Lamport1979} and the fragment of C11 with release/acquire atomics~\cite{Lahav2019} are
\[
  \scmm\triangleq(\po\cup\rf\cup\mo\cup\fr)^{+}
  \qquad\text{and}\qquad
  \ramm\triangleq\hb\cup\mo;\hb\cup\mo;\hb;\rf^{-1}\cup\mo;\mo;\rf^{-1}\ .
\]
In particular for $\ramm$, the four relations connected under $\cup$ are known, in order, as 
\emph{irreflexivity of $\hb$},
\emph{write coherence},
\emph{read coherence}, and
\emph{atomicity}.

\Paragraph{Program semantics.}
Given a memory model $\mm$, the \emph{semantics} of $\ConcProg$ under $\mm$ is $\ExecutionGraphsOf{\ConcProg}{\mm}\triangleq\{G\in\GraphsMM{\mm}\colon \tuple{G.\Events, G.\seqbef}\in\GLang{\ConcProg}\}$, i.e., the execution graphs of $\ConcProg$ that are consistent in $\mm$.

\Paragraph{Prefix-closedness.}
Let $G=\tuple{E,\seqbef,\rf,\mb}$ be an execution graph.
An event $\event\in E$ is \emph{maximal} in $G$ if it has no outgoing $\seqbef$-edge and no outgoing $\rf$-edge.
For a maximal event $\event\in E$, we write $G\setminus\{\event\}$ for the execution graph induced by $E\setminus\{\event\}$, i.e., with $\seqbef$ and $\rf$ restricted to $E\setminus\{\event\}$ and $\mb$ replaced by the transitive reduction of $\mb^+\cap\bigl((E\setminus\{\event\})\times(E\setminus\{\event\})\bigr)$.
A memory model $\mm$ is \emph{prefix-closed}~\cite{Kokologiannakis19} if every non-empty $G\models\mm$ has a maximal event $\event$ such that $G\setminus\{\event\}\models\mm$.
As a maximal event is the last event of its thread, and the language of an LTS is closed under prefixes, $G\setminus\{\event\}$ is then an execution of every program of which $G$ is an execution, i.e., $G\in\ExecutionGraphsOf{\ConcProg}{\mm}$ implies $G\setminus\{\event\}\in\ExecutionGraphsOf{\ConcProg}{\mm}$.
Intuitively, prefix-closed memory models allow executions to be constructed incrementally.
Most common memory models are prefix-closed.

\Paragraph{Robustness.}
We can now define robustness, the main focus of this paper.
As is standard, $\scmm$ is the strongest memory model we consider, i.e., for any memory model $\mm$, we have 
$\GraphsMM{\scmm}\subseteq \GraphsMM{\mm}$.
Given additionally a program $\ConcProg$, robustness asks whether $\ConcProg$ exhibits any weak behaviors under $\mm$.
Formally, $\ConcProg$ is \emph{robust} against $\mm$ iff $\ExecutionGraphsOf{\ConcProg}{\scmm}=\ExecutionGraphsOf{\ConcProg}{\mm}$.
A \emph{robustness violation witness} (or simply \emph{witness}) is an execution graph $G\in \ExecutionGraphsOf{\ConcProg}{\mm}\setminus \ExecutionGraphsOf{\ConcProg}{\scmm}$.
\cref{FIG:ROB_EXAMPLE} shows a non-robust program and a witness.

\tikzset{
  evnode/.style = {inner sep=2pt, font=\footnotesize},
  evx/.style    = {evnode, fill=colorLocX!15},
  evy/.style    = {evnode, fill=colorLocY!15},
  evsb/.style   = {kattrans, colorSB},
  evrf/.style   = {kattrans, colorRF, dashed},
  evmb/.style   = {kattrans, colorMB, dotted},
  evfr/.style   = {kattrans, colorFR, dashdotted}, 
}

\begin{figure}
\centering
\begin{tabular}{@{}p{.285\textwidth}@{\hspace{.08\textwidth}}p{.45\textwidth}@{}}
\centering
$\vcenter{\hbox{%
\begin{tikzpicture}[katpic, katlab/.append style={font=\footnotesize}]
  \foreach \y/\i/\t in {0.72/0/{\Prog_{t_0}}, 0/1/\Prog_{t_1}, -0.72/2/\Prog_{t_2}} {
    \node[font=\footnotesize] at (-0.85,\y) {$\t$};
    \node[katnodesmall] (n\i a) at (0,\y)   {};
    \node[katnodesmall] (n\i b) at (1.2,\y) {};
    \node[katnodesmall] (n\i c) at (2.4,\y) {};
    \draw[katstart] (-0.5,\y) -- (n\i a);
  }
  \draw[kattrans] (n0a) -- node[katlab, above=0pt]{$\lx{\wt(x,0)}$} (n0b);
  \draw[kattrans] (n0b) -- node[katlab, above=0pt]{$\ly{\wt(y,0)}$} (n0c);
  \draw[kattrans] (n1a) -- node[katlab, above=0pt]{$\lx{\wt(x,1)}$} (n1b);
  \draw[kattrans] (n1b) -- node[katlab, above=0pt]{$\ly{\rd(y,0)}$} (n1c);
  \draw[kattrans] (n2a) -- node[katlab, above=0pt]{$\ly{\wt(y,1)}$} (n2b);
  \draw[kattrans] (n2b) -- node[katlab, above=0pt]{$\lx{\rd(x,0)}$} (n2c);
\end{tikzpicture}%
}}$ &
\centering
$\vcenter{\hbox{%
\begin{tikzpicture}[katpic, katlab/.append style={font=\footnotesize}]
  \node[font=\footnotesize] at (0,0.42)   {$t_0$};
  \node[font=\footnotesize] at (2.0,0.42) {$t_1$};
  \node[font=\footnotesize] at (4.0,0.42) {$t_2$};
  \node[evx] (ix) at (0,0)      {$\event_1\colon\wt(x,0)$};
  \node[evy] (iy) at (0,-1.3)   {$\event_2\colon\wt(y,0)$};
  \node[evx] (a)  at (2.0,0)    {$\event_3\colon\wt(x,1)$};
  \node[evy] (b)  at (2.0,-1.3) {$\event_4\colon\rd(y,0)$};
  \node[evy] (c)  at (4.0,0)    {$\event_5\colon\wt(y,1)$};
  \node[evx] (d)  at (4.0,-1.3) {$\event_6\colon\rd(x,0)$};
  \draw[evsb] (ix)  to node[katlab, left=0pt, pos=.2]{$\seqbef$} (iy);
  \draw[evsb] (a)  to node[katlab, left=0pt, pos=.14]{$\seqbef$} (b);
  \draw[evsb] (c)   to node[katlab, right=0pt, pos=.2]{$\seqbef$} (d);
  \draw[evmb] (ix) -- node[katlab, above=1pt]{$\mb$} (a);
  \draw[evrf] (iy) -- node[katlab, below=1pt]{$\rf$} (b);
  \draw[evmb] (iy) to node[katlab, above left=0pt, pos=.1]{$\mb$} (c);
  \draw[evrf]  (ix) to node[katlab, below left=0pt, pos=.08]{$\rf$} (d);
  \draw[evfr] (b) to node[katlab, below right=-1pt, pos=.97]{$\fr$} (c);
  \draw[evfr] (d) to node[katlab, above right=-1pt, pos=.03]{$\fr$} (a);
\end{tikzpicture}%
}}$
\tabularnewline[-9.4pt] %
\centering \subcaption{A program $\ConcProg$.}\label{SUBFIG:ROB_PROG} &
\centering \subcaption{A witness $G$ of the non-$\ramm$-robustness of $\ConcProg$.}\label{SUBFIG:ROB_GRAPH}
\tabularnewline
\end{tabular}
\caption{
A non-$\ramm$-robust program illustrating store buffering (\subref{SUBFIG:ROB_PROG}),
and a witness exhibiting the $\scmm$-cycle $\event_3 \LTo{\seqbef} \event_4 \LTo{\fr} \event_5 \LTo{\seqbef} \event_6 \LTo{\fr} \event_3$ (\subref{SUBFIG:ROB_GRAPH}).
\label{FIG:ROB_EXAMPLE}
}
\end{figure}
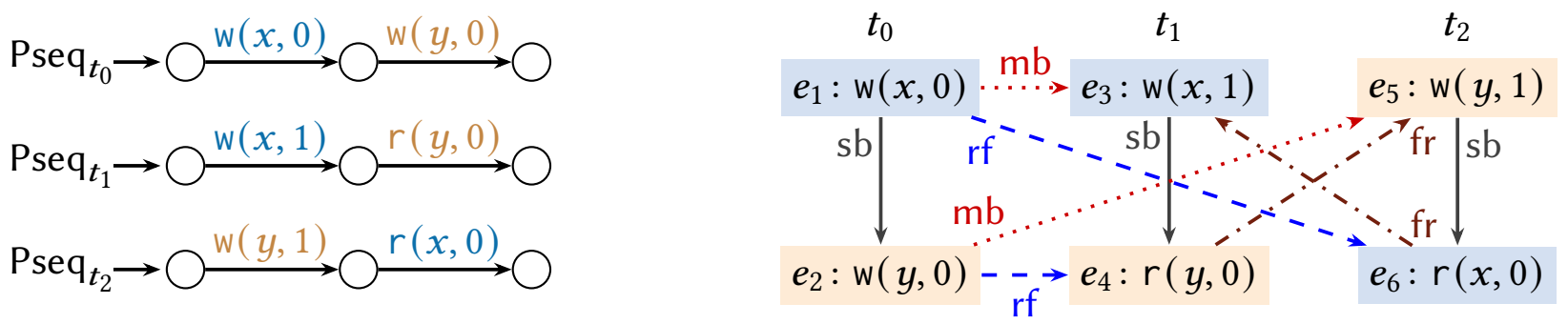

\subsection{$\KAT$ Automata}\label{SUBSEC:PRELIM_KAT}

Our algorithm for robustness does not deal with $\KAT$ expressions directly.
Instead, we compile memory models into nondeterministic automata using standard procedures~\cite{Kozen01}, similarly to \cite{Kokologiannakis2023}.

\Paragraph{$\KAT$-NFAs.}
A \emph{$\KAT$-NFA} is a word automaton $\mathcal{K}=\tuple{\States,\TestDom\cup\RelBase,\Trans,\InitState,\States_F}$ over the alphabet of tests and base relation symbols, such that $\Trans$ has no silent transitions.
So, $\Lang{\mathcal{K}}$ is the set of words $l_1 l_2\cdots l_k$ over $\TestDom\cup\RelBase$ for which there is a run $\InitState\LTransTo{l_1}q_1\LTransTo{l_2}\cdots\LTransTo{l_k}q_k$ of $\mathcal{K}$ with $q_k\in\States_F$.
More generally, we say that $\mathcal{K}$ has a \emph{run} on $w=l_1 l_2\cdots l_k$ from $q$ to $q'$ if there are transitions $q\LTransTo{l_1}q_1\LTransTo{l_2}\cdots\LTransTo{l_k}q'$ in $\Trans$.

\Paragraph{Realizations.}
Let $G$ be an execution graph and $w=l_1 l_2\cdots l_k$ a word over $\TestDom\cup\RelBase$.
A \emph{realization} of $w$ in $G$ is a sequence of events $\event_0,\event_1,\dots,\event_k\in G.\Events$
such that, for each $1\leq i\leq k$, $\tuple{\event_{i-1},\event_i}\in G.l_i$ if $l_i\in\RelBase$, and $\event_{i-1}=\event_i\in G.l_i$ if $l_i\in\TestDom$.
The realization goes from $\event_0$ to $\event_k$, and it is \emph{cyclic} if $\event_0=\event_k$.
So, a realization traverses one edge of $G$ at a time, and a test checks the label of the current event.

\Paragraph{Compilation.}
Every $\KAT$ expression can be compiled into a $\KAT$-NFA whose words realize exactly the edges of the expression, as captured formally by the following lemma.

\begin{restatable}[\cite{Kozen01}]{lemma}{lemkatcompile}\label{LEM:KAT_COMPILE}
For every $\KAT$ expression $\kexp$, there is a $\KAT$-NFA $\mathcal{K}_{\kexp}$ with $O(|\kexp|)$ states and transitions, constructible in time $O(|\kexp|)$, such that, for every execution graph $G$ and $\event,\event'\in G.\Events$, we have $\tuple{\event,\event'}\in G.\kexp$ iff some $w\in\Lang{\mathcal{K}_{\kexp}}$ has a realization in $G$ from $\event$ to $\event'$.
\end{restatable}

\begin{figure}
\centering
\begin{tabular*}{\textwidth}{@{}p{89.3pt}@{\hspace{16pt}}p{115.1pt}@{\extracolsep{\fill}}p{146pt}@{}}%
\centering
$\vcenter{\hbox{%
\begin{tikzpicture}[katpic, x=0.8cm, y=0.8cm, katnode/.append style={font=\scriptsize, minimum size=4.2mm}, katlab/.append style={font=\scriptsize}]
  \node[katnode] (q0) at (0,2.2)    {$\InitState$};
  \node[katacc]  (q)  at (0,0.2)    {$q_1$};
  \node[katnode] (r)  at (1.45,1.25) {$q_3$};
  \node[katnode] (p)  at (1.45,0.2) {$q_2$};
  \node[katnode] (u)  at (2.5,2.2) {$q_4$};
  \node[katnode] (v)  at (2.5,1.25) {$q_5$};
  \draw[katstart] (-0.65,2.2) -- (q0);
  \draw[kattrans, bend right=15] (q0) to node[katlab, left=0pt, pos=.58, align=right]{$\Ksb,$\\[-2pt]$\Krf,$\\[-2pt]$\Kmb$} (q);
  \draw[kattrans, bend right=15] (q) to node[katlab, right=0pt, pos=.35]{$\identity{\LabelDom}$} (q0);
  \draw[kattrans] (q0) -- node[katlab, above right=-2pt, pos=.5]{$\identity{\ReadDom}$} (r);
  \draw[kattrans] (q0) -- node[katlab, above=0pt, sloped, pos=.5]{$\identity{\RMWDom}$} (u);
  \draw[kattrans] (r) -- node[katlab, right=1pt, pos=.3]{$\Krfinv$} (p);
  \draw[kattrans] (p) -- node[katlab, below=1pt, pos=.3]{$\Kmb$} (q);
  \draw[kattrans] (u) -- node[katlab, right=1pt]{$\Krfinv$} (v);
  \draw[kattrans, bend left=20] (v) to node[katlab, below right=-1pt]{$\Kmb$} (p);
\end{tikzpicture}%
}}$ &
\centering
$\vcenter{\hbox{%
\begin{tikzpicture}[katpic, x=0.8cm, y=0.8cm, katnode/.append style={font=\scriptsize, minimum size=4.2mm}, katlab/.append style={font=\scriptsize}]
  \node[katacc]  (h)  at (1.2,2.45)     {$q_1$};
  \node[katnode] (q0) at (0,2.45)     {$\InitState$};
  \node[katnode] (m)  at (0,1.6)     {$q_2$};
  \node[katacc]  (h2) at (-0.9,0.8)  {$q_3$};
  \node[katnode] (m2) at (0.9,0.8)   {$q_4$};
  \node[katacc]  (qf) at (0,0.0)   {$q_5$};
  \draw[katstart] (-0.65,2.45) -- (q0);
  \draw[kattrans] (q0) -- node[katlab, above=1pt]{$\Ksb,\Krf$} (h);
  \draw (h) edge[loop right] node[katlab, right]{$\Ksb,\Krf$} (h);
  \draw[kattrans] (q0) -- node[katlab, right=1pt]{$\Kmb$} (m);
  \draw (m) edge[loop below] node[katlab, below=0pt]{$\Kmb$} (m);
  \draw[kattrans] (m)  -- node[katlab, above left=0pt,  pos=.7]{$\Ksb,\Krf$} (h2);
  \draw (h2) edge[loop left] node[katlab, left]{$\Ksb,\Krf$} (h2);
  \draw[kattrans] (m)  -- node[katlab, above right=0pt, pos=.7]{$\Kmb$} (m2);
  \draw (m2) edge[loop right] node[katlab, right]{$\Kmb$} (m2);
  \draw[kattrans] (h2) -- node[katlab, below left=-1pt,  pos=.4]{$\Krfinv$} (qf);
  \draw[kattrans] (m2) -- node[katlab, below right=-1pt, pos=.4]{$\Krfinv$} (qf);
\end{tikzpicture}%
}}$ &
\centering
$\vcenter{\hbox{%
\begin{tikzpicture}[katpic, katlab/.append style={font=\scriptsize},
    evx/.append style={inner sep=1.2pt}, evy/.append style={inner sep=1.2pt}]
  \foreach \i/\e/\st in {0/{\event_3}/evx, 1/{\event_4}/evy, 2/{\event_2}/evy,
                         3/{\event_5}/evy, 4/{\event_6}/evx, 5/{\event_1}/evx,
                         6/{\event_3}/evx}
    \node[\st, anchor=base] (e\i) at (0.8*\i,0) {$\e$};
  \newcommand{\tstep}[1]{\big\downarrow\rlap{$\scriptscriptstyle #1$}}
  \foreach \i/\qs in {0/{\InitState}, 1/{q_1\\\tstep{\identity{\LabelDom}}\\\InitState\\\tstep{\identity{\ReadDom}}\\q_3}, 2/{q_2},
                      3/{q_1\\\tstep{\identity{\LabelDom}}\\\InitState}, 4/{q_1\\\tstep{\identity{\LabelDom}}\\\InitState\\\tstep{\identity{\ReadDom}}\\q_3},
                      5/{q_2}, 6/{q_1}}
    \node[katlab, anchor=north] at (0.8*\i,-0.25) {$\renewcommand{\arraystretch}{1.1}\begin{array}{@{}c@{}}\qs\end{array}$};
  \foreach \i/\j/\c/\l in {0/1/colorSB/{\seqbef}, 1/2/colorRF/{\rf^{-1}},
                           2/3/colorMB/{\mb},     3/4/colorSB/{\seqbef},
                           4/5/colorRF/{\rf^{-1}}, 5/6/colorMB/{\mb}} {
    \draw[kattrans, \c] (e\i) -- (e\j);
    \node[katlab, \c, anchor=base] at (0.8*\i+0.4,0.22) {$\l$};
  }
\end{tikzpicture}%
}}$
\tabularnewline[-9.4pt] %
\centering \subcaption{$\scmm$.}\label{SUBFIG:KAT_SC_NF} &
\centering \subcaption{$\ramm$.}\label{SUBFIG:KAT_RA_NF} &
\centering \subcaption{A cycle of \cref{SUBFIG:ROB_GRAPH}.}\label{SUBFIG:ROB_RUN}
\tabularnewline
\end{tabular*}
\caption{(\subref{SUBFIG:KAT_SC_NF}) and (\subref{SUBFIG:KAT_RA_NF}) show $\KAT$-NFAs for $\scmm$ and $\ramm$.
(\subref{SUBFIG:ROB_RUN}) shows the cycle of \cref{SUBFIG:ROB_GRAPH} that drives (\subref{SUBFIG:KAT_SC_NF}) from $\InitState$ to the accepting state $q_1$, so $G\not\models\scmm$; each column lists the states at one event, joined by the tests the event satisfies.}
\label{FIG:KAT_MMS}
\end{figure}
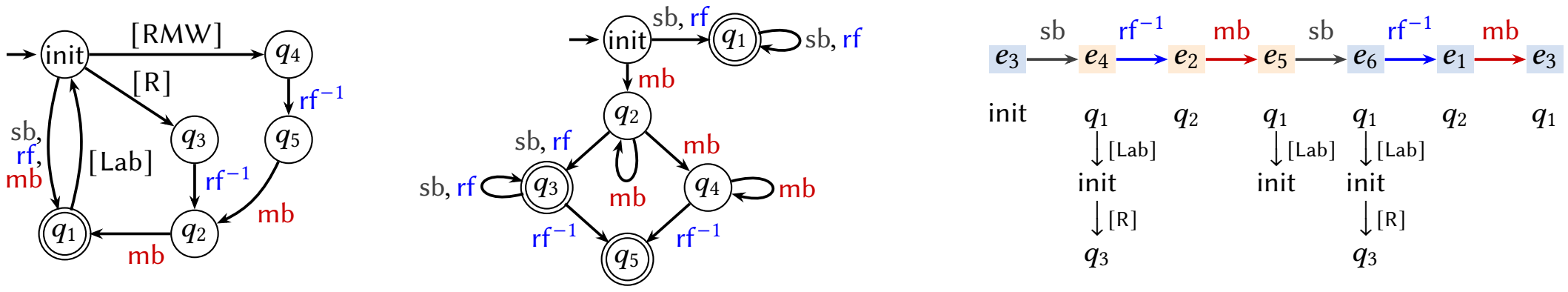

We say that a $\KAT$-NFA $\mathcal{K}$ is a \emph{$\KAT$-NFA for} a memory model $\mm$ if, for every execution graph $G$, $G\models\mm$ iff no word of $\Lang{\mathcal{K}}$ has a cyclic realization in $G$.
By \cref{LEM:KAT_COMPILE}, $\mathcal{K}_{\mm}$ is a $\KAT$-NFA for $\mm$, but $\mm$ may have others.
\cref{FIG:KAT_MMS} illustrates $\KAT$-NFAs for $\scmm$ and $\ramm$ on an example.

\section{The Pathwidth of Execution Graphs}\label{SEC:PW}

In this section, we present the graph-theoretic notion of pathwidth, and prove the first result towards our 
\cref{THE:ROB_PSPACE}:~against prefix-closed memory models, a non-robust program always exhibits a witness whose pathwidth is bounded in terms of $|\ThreadDom|$ and $|\LocationDom|$.

\subsection{Path Decompositions and Pathwidth}\label{SUBSEC:PW_PRELIMS}

Pathwidth is a graph-theoretic notion that measures how closely a graph resembles a path~\cite{pw}.

\Paragraph{Path decompositions and pathwidth.}
Let $G=\tuple{\Events,\seqbef,\rf,\mb}$ be an execution graph, and call $\event\LTo{\rsym}\event'$ with $\rsym\in\{\seqbef,\rf,\mb\}$ the edges of $G$.
A \emph{path decomposition} of $G$ is a sequence $\Bag_1, \Bag_2, \ldots, \Bag_n$ of subsets of $\Events$, called \emph{bags}, such that
\begin{enumerate*}[label=(\roman*)]
\item every event appears in some bag, i.e., $\bigcup_{i \in \IntSet{n}} \Bag_i = \Events$,
\item every edge appears in some bag, i.e., for every edge $\event\LTo{\rsym}\event'$ of $G$ there exists some $i\in \IntSet{n}$ with $\event,\event'\in \Bag_i$, and
\item every event appears in a contiguous subsequence of bags, i.e., for all $i,j,k \in \IntSet{n}$ with $i \leq j \leq k$, we have $\Bag_i \cap \Bag_k \subseteq \Bag_j$.
\end{enumerate*}
The \emph{width} of a path decomposition is $\max_{i\in \IntSet{n}} |\Bag_i|-1$.
The \emph{pathwidth} of $G$, written $\pw(G)$, is the smallest width among the path decompositions of $G$.

\Paragraph{Very nice path decompositions.}
It is often convenient to work with decompositions that change by one event at a time, and consider one edge at a time.
A path decomposition $\Bag_1, \Bag_2, \ldots, \Bag_n$ of $G$ is \emph{very nice} if $\Bag_1 = \Bag_n = \emptyset$ and each bag $\Bag_{i+1}$, for $i\in\IntSet{n-1}$, can be labeled with one of the following steps, so that the corresponding condition holds.
\begin{compactitem}
\item \emph{Introduce $\event$}, for $\event\in\Events$: $\Bag_{i+1} = \Bag_i \cup \{\event\}$, and $\event$ appears in no bag $\Bag_j$ with $j\leq i$.
\item \emph{Forget $\event$}, for $\event\in\Events$: $\Bag_{i+1} = \Bag_i \setminus \{\event\}$, and $\event\in\Bag_i$.
\item \emph{Introduce $\event\LTo{\rsym}\event'$}, for an edge $\event\LTo{\rsym}\event'$ of $G$: $\Bag_{i+1} = \Bag_i$, and $\event,\event'\in\Bag_i$.
\end{compactitem}
Moreover, every edge is introduced exactly once, i.e., each edge labels exactly one bag.

\begin{lemma}[\cite{paramalgos}]\label{LEM:NICE_PW}
    Every execution graph $G=\tuple{\Events,\seqbef,\rf,\mb}$ has a very nice path decomposition of width $\pw(G)$ and at most $2|\Events|+|\seqbef|+|\rf|+|\mb|+1$ bags.
\end{lemma}

\subsection{Pathwidth and Memory Models}\label{SUBSEC:PW_MM}

In general, execution graphs of weak memory models  have unbounded pathwidth, even without their $\mb$-edges~\cite{Conrado26}.
The pathwidth of $\scmm$-consistent graphs, however, is bounded, and this often suffices to bound the pathwidth of robustness-violation witnesses.

\tikzset{
  evrf/.style   = {kattrans, colorRF, dashed},
  evmb/.style   = {kattrans, colorMB, dotted},
  evx/.style    = {evnode, fill=colorLocX!15},
  evy/.style    = {evnode, fill=colorLocY!15},
  evhl/.style   = {draw, line width=0.9pt},
  bag/.style    = {draw, rounded corners=2pt, inner sep=4pt,
                   font=\scriptsize, align=center, anchor=north},
  baghl/.style  = {bag, line width=1.2pt},
}

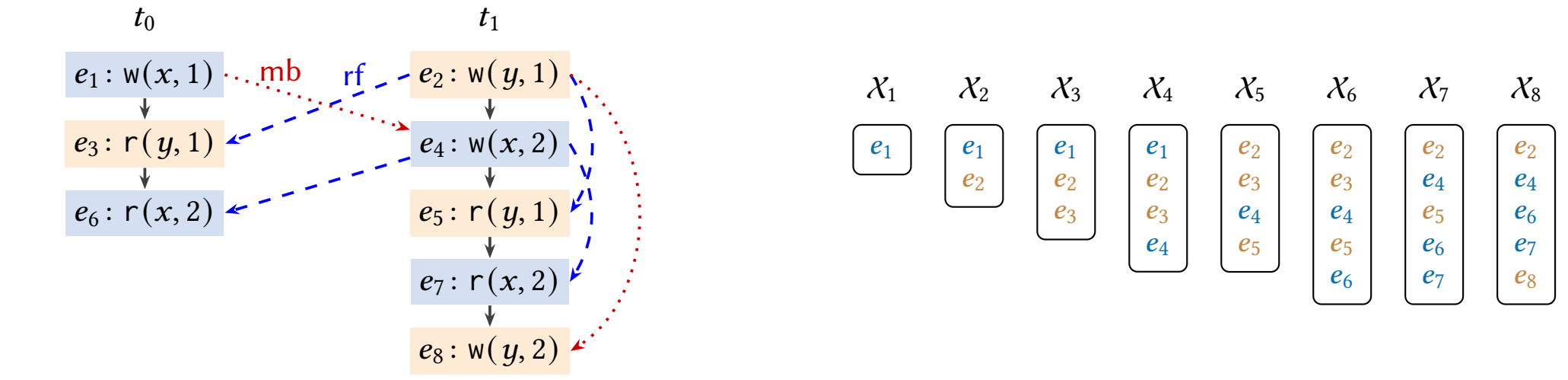
\begin{figure}
\centering
\begin{tabular*}{\textwidth}{@{}p{.45\textwidth}@{\extracolsep{\fill}}p{.5\textwidth}@{}}
\centering
$\vcenter{\hbox{%
\begin{tikzpicture}[katpic, katlab/.append style={font=\footnotesize}]
  \node[font=\footnotesize] at (0,0.48)   {$t_0$};
  \node[font=\footnotesize] at (3.0,0.48) {$t_1$};
  \node[evx]       (e1) at (0,0)      {$\event_1\colon\wt(x,1)$};
  \node[evy] (e3) at (0,-0.6)  {$\event_3\colon\rd(y,1)$};
  \node[evx] (e6) at (0,-1.2)  {$\event_6\colon\rd(x,2)$};
  \node[evy] (e2) at (3.0,0)      {$\event_2\colon\wt(y,1)$};
  \node[evx] (e4) at (3.0,-0.6)  {$\event_4\colon\wt(x,2)$};
  \node[evy] (e5) at (3.0,-1.2)  {$\event_5\colon\rd(y,1)$};
  \node[evx]       (e7) at (3.0,-1.8)  {$\event_7\colon\rd(x,2)$};
  \node[evy]       (e8) at (3.0,-2.4)  {$\event_8\colon\wt(y,2)$};
  \draw[evsb] (e1) -- (e3);
  \draw[evsb] (e3) -- (e6);
  \draw[evsb] (e2) -- (e4);
  \draw[evsb] (e4) -- (e5);
  \draw[evsb] (e5) -- (e7);
  \draw[evsb] (e7) -- (e8);
  \draw[evrf] (e2.180) -- node[katlab, above=1pt, pos=.3]{$\rf$} (e3.0);
  \draw[evrf] (e4.190) -- (e6.0);
  \draw[evrf] (e2.0) to[bend left=32] (e5.0);
  \draw[evrf] (e4.0) to[bend left=32] (e7.0);
  \draw[evmb] (e1.0) -- node[katlab, above=1pt, pos=.3]{$\mb$} (e4.170);
  \draw[evmb] (e2.0) to[bend left=58] (e8.0);
\end{tikzpicture}%
}}$ &
\centering
$\vcenter{\hbox{%
\begin{tikzpicture}[katpic]
  \foreach \i in {1,...,8}
    \node[font=\scriptsize] at (0.8*\i-0.8,0.3) {$\Bag_\i$};
  \node[bag]   at (0.00,0) {$\lx{\event_1}$};
  \node[bag]   at (0.8,0) {$\lx{\event_1}$\\$\ly{\event_2}$};
  \node[bag]   at (1.6,0) {$\lx{\event_1}$\\$\ly{\event_2}$\\$\ly{\event_3}$};
  \node[bag]   at (2.4,0) {$\lx{\event_1}$\\$\ly{\event_2}$\\$\ly{\event_3}$\\$\lx{\event_4}$};
  \node[bag]   at (3.2,0) {$\ly{\event_2}$\\$\ly{\event_3}$\\$\lx{\event_4}$\\$\ly{\event_5}$};
  \node[bag] at (4.0,0) {$\ly{\event_2}$\\$\ly{\event_3}$\\$\lx{\event_4}$\\$\ly{\event_5}$\\$\lx{\event_6}$};
  \node[bag]   at (4.8,0) {$\ly{\event_2}$\\$\lx{\event_4}$\\$\ly{\event_5}$\\$\lx{\event_6}$\\$\lx{\event_7}$};
  \node[bag]   at (5.6,0) {$\ly{\event_2}$\\$\lx{\event_4}$\\$\lx{\event_6}$\\$\lx{\event_7}$\\$\ly{\event_8}$};
\end{tikzpicture}%
}}$
\tabularnewline[-9.4pt] %
\centering \subcaption{An $\scmm$-consistent graph $G$, with $\sigma=\event_1,\dots,\event_8$.}\label{SUBFIG:PW_GRAPH} &
\centering \subcaption{The path decomposition of $G$, following \cref{LEM:PW_SC_REDUCED}.}\label{SUBFIG:PW_BAGS}
\tabularnewline
\end{tabular*}
\caption{A graph $G$ over $\ThreadDom=\{t_0,t_1\}$ and $\LocationDom=\{x,y\}$ (\subref{SUBFIG:PW_GRAPH}), and a path decomposition of $G$ of width 4 (\subref{SUBFIG:PW_BAGS}).}
\label{FIG:PW}
\end{figure}

\Paragraph{The pathwidth of $\scmm$.}
Let $G$ be some $\scmm$-consistent execution graph.
Recall that $G\models\scmm$ iff $(\po\cup\rf\cup\mo\cup\fr)^+$ is irreflexive in $G$.
Let $\sigma=\event_1,\event_2,\dots,\event_n$ be a linear extension of $(\po\cup\rf\cup\mo\cup\fr)^+$, and write $\event\prec\event'$ when $\event$ precedes $\event'$ in $\sigma$.
We construct a path decomposition of $G$ where the $i$-th bag consists of $\event_i$ together with
\begin{enumerate*}[label=(\roman*)]
\item the $\prec$-largest event of $\Events^{t}\cap\{\event_1,\dots,\event_{i-1}\}$, for each $t\in\ThreadDom$, and
\item the $\prec$-largest event of $\locx{\Events}\cap(\WriteDom\cup\RMWDom)\cap\{\event_1,\dots,\event_{i-1}\}$, for each $x\in\LocationDom$,
\end{enumerate*}
whenever these exist.
In words, while scanning $\sigma$, the bags remember the last event of each thread and the last write/RMW on each location.
This covers every $\seqbef$- and $\mb$-edge, which joins an event to the previous event of its thread or location, respectively.
It also covers every $\rf$-edge, since, under $\scmm$, every read/RMW observes the latest write/RMW on its location.
See \cref{FIG:PW}.

\begin{restatable}{lemma}{lempwscreduced}\label{LEM:PW_SC_REDUCED}
For every execution graph $G$, if $G \models \scmm$ then $\pw(G) \leq |\ThreadDom| + |\LocationDom|$.
\end{restatable}

Hereinafter, we write $\pwSC\triangleq|\ThreadDom|+|\LocationDom|$ for the bound of \cref{LEM:PW_SC_REDUCED}.

\Paragraph{Pathwidth-bounded witnesses.}
Given $\ell\in\Nats$, a robustness-violation witness $G$ is \emph{$\ell$-$\pw$-bounded} if $\pw(G)\leq\ell$.
A memory model $\mm$ \emph{admits $\ell$-$\pw$-bounded witnesses} if every concurrent program that is not robust against $\mm$ has an $\ell$-$\pw$-bounded witness.
In such cases, the robustness of a program $\ConcProg$ against $\mm$ simplifies to searching for a violation over graphs of bounded pathwidth only.
The next lemma implies that the class of memory models admitting this approach is broad.
\looseness=-1

\begin{restatable}{lemma}{lemprefixborderline}\label{LEM:PREFIX_BORDERLINE}
Every prefix-closed memory model admits $(\pwSC+1)$-$\pw$-bounded witnesses.
\end{restatable}

\section{Execution Graphs as Words}\label{SEC:ALGEBRA}

\cref{SEC:PW} reduces robustness to the search for a witness of pathwidth at most $\ell$, with $\ell=\pwSC+1$ when $\mm$ is prefix-closed.
In this section we turn that search into a search over words.
We develop a graph algebra whose operations build execution graphs~(\cref{SUBSEC:ALGEBRA_BASE}), and show that its words with $k+1$ colors denote exactly the graphs of pathwidth at most $k$~(\cref{SUBSEC:ALGEBRA_WORDS}).

\subsection{A Graph Algebra}\label{SUBSEC:ALGEBRA_BASE}

We now introduce an algebra whose operations build execution graphs.
It is similar in spirit to the HR-algebra of Courcelle~\cite{Courcelle1990}, which builds graphs with distinguished vertices.
Our algebra is simpler, as it has no operation that glues two graphs together.
So, its terms are words rather than trees, and they correspond to path decompositions rather than tree decompositions.

\Paragraph{Free execution graphs.}
A \emph{free} execution graph is a tuple $G = \tuple{\Events, \seqbef, \rf, \mb}$, where $\Events\subseteq\EventDom$ is a finite set of events and $\seqbef$, $\rf$, and $\mb$ are arbitrary binary relations over $\Events$, i.e., with none of the constraints of \cref{SUBSEC:PRELIM_EXEC}.
For example, in a free execution graph $\seqbef$ may be cyclic.
Path decompositions and pathwidth~(\cref{SUBSEC:PW_PRELIMS}) apply to free execution graphs unchanged.
Identifiers only serve to distinguish events that carry the same label, so we consider two free execution graphs equal if they differ only by a renaming of identifiers.

\Paragraph{CFE graphs.}
Given $k\in\Nats$, we take the set $\IntSet{k}$ as our colors.
A \emph{colored free execution (CFE) graph} is a tuple $G_c = \tuple{\Events, \seqbef, \rf, \mb, \chi}$, where $\tuple{\Events, \seqbef, \rf, \mb}$ is a free execution graph and $\chi\colon \Events \rightharpoonup \IntSet{k}$ is a partial injective function.
We call $\event \in \dom{\chi}$ a \emph{colored} event, and $\event \in \Events \setminus \dom{\chi}$ an \emph{uncolored} event.
We call $c \in \img{\chi}$ an \emph{active} color, and $c \in \IntSet{k} \setminus \img{\chi}$ an \emph{inactive} color.
We write $\bot$ for the empty CFE graph, in which $\Events$, all three relations, and $\chi$ are empty.

\Paragraph{Operations.}
We define three operations on a CFE graph $G_c = \tuple{\Events, \seqbef, \rf, \mb, \chi}$, analogous to the steps of very nice path decompositions~(\cref{SUBSEC:PW_PRELIMS}).
In each case, $G'_c = \tuple{\Events', \seqbef', \rf', \mb', \chi'}$ denotes the resulting graph, and any component not explicitly mentioned is unchanged.
\begin{compactitem}
  \item $\CFEIntro(\llab, c)$, where $\llab \in \LabelDom$ and $c \in \IntSet{k} \setminus \img{\chi}$:~it introduces a new event carrying the label $\llab$ and the color $c$.
  Formally, letting $\event \triangleq \tuple{|\Events|+1, \llab}$, we set $\Events' \triangleq \Events \cup \{\event\}$ and $\chi' \triangleq \chi \cup \{\event \mapsto c\}$. 
  \item $\CFEForget(c)$, where $c \in \img{\chi}$:~it uncolors the event with color $c$, which makes $c$ inactive.
  Formally, letting $\event$ be the unique event with $\chi(\event) = c$, $\chi' \triangleq \chi \setminus \{\event \mapsto c\}$.
  \item $\CFEIntroR{\rsym}(c_1, c_2)$, where $\rsym \in \{\seqbef, \rf, \mb\}$, $c_1, c_2 \in \img{\chi}$, and $\tuple{\chi^{-1}(c_1),\chi^{-1}(c_2)}\notin \rsym$:~it adds a $\rsym$-edge between the events with colors $c_1$ and $c_2$.
  Formally, letting $\event_i \triangleq \chi^{-1}(c_i)$, we set $\rsym' \triangleq \rsym \cup \{\tuple{\event_1, \event_2}\}$.
  We write $\CFEIntrosb$, $\CFEIntrorf$, and $\CFEIntromb$ when the relation of $\CFEIntroR{\rsym}$ is fixed.
\end{compactitem}

Each operation $\mathrm{op}$ has \emph{preconditions} for it to be valid:~$c$ must be inactive for $\CFEIntro(\llab, c)$, and the colors it names must be active for $\CFEForget(c)$ and $\CFEIntroR{\rsym}(c_1, c_2)$, where the edge must moreover be new.
When $G_c$ satisfies the precondition of $\mathrm{op}$, we write $\mathrm{op}(G_c)$ for the resulting graph.

\subsection{Words Representing Path Decompositions}\label{SUBSEC:ALGEBRA_WORDS}

We use the operations above to define words that define bounded-pathwidth execution graphs.

\Paragraph{The alphabet.}
Given $k\in\Nats$, let
\[
\begin{aligned}
  \Ops(k) \triangleq{}
    & \{\CFEIntro(\llab, c) \mid \llab \in \LabelDom,\ c \in \IntSet{k}\} \  \cup \  \{\CFEForget(c) \mid c \in \IntSet{k}\} \\
    {} \cup \ & \{\CFEIntroR{\rsym}(c_1, c_2) \mid \rsym \in \{\seqbef, \rf, \mb\},\ c_1, c_2 \in \IntSet{k}\}
\end{aligned}
\]
be the set of operation symbols with $k$ colors.
The size of $\Ops(k)$ is $k\cdot|\LabelDom| + k + 3\cdot k^2$.
We write $\varepsilon$ for the empty word and, for $w \in \Ops(k)^*$, $w_i$ for the prefix of $w$ of length $i$, so $w_0 = \varepsilon$.

\Paragraph{Well-formed and closed words.}
We define well-formedness of $w \in \Ops(k)^*$, and the CFE graph $G_c^w$ \emph{induced} by a well-formed $w$, by induction on $w$.
The empty word is well-formed, with $G_c^{\varepsilon} \triangleq \bot$.
The word $w\cdot \mathrm{op}$ is well-formed if $w$ is well-formed and $G_c^{w}$ satisfies the precondition of $\mathrm{op}$; then, $G_c^{w\,\mathrm{op}} \triangleq \mathrm{op}(G_c^{w})$.
We write $\chi^w$ for the coloring of $G_c^w$, and $G^w$ for the free execution graph obtained by discarding $\chi^w$.
A CFE graph is \emph{closed} if it has no colored event, and a word $w$ is \emph{closed} if it is well-formed and $G_c^w$ is closed, i.e., $\dom{\chi^w} = \emptyset$.

\tikzset{
  evx/.style    = {inner sep=2pt, font=\footnotesize, fill=colorLocX!15},
  evsb/.style   = {kattrans, colorSB},
  evrf/.style   = {kattrans, colorRF, dashed},
  evmb/.style   = {kattrans, colorMB, dotted},
  bagev/.style  = {inner sep=0.5pt, font=\scriptsize, minimum size=3mm},
  bagbox/.style = {draw, rounded corners=2pt},
}

\begin{figure}
\centering
\begin{tabular*}{\textwidth}{@{}p{.22\textwidth}@{\extracolsep{\fill}}p{.72\textwidth}@{}}
\centering
$\vcenter{\hbox{%
\begin{tikzpicture}[katpic, katlab/.append style={font=\footnotesize}]
  \node[font=\footnotesize] at (0,0.46)   {$t_0$};
  \node[font=\footnotesize] at (1.55,0.46) {$t_1$};
  \node[evx] (e1) at (0,0)      {$\event_1\colon\wt(x,1)$};
  \node[evx] (e3) at (0,-0.8)  {$\event_3\colon\rd(x,2)$};
  \node[evx] (e2) at (1.55,0)   {$\event_2\colon\wt(x,2)$};
  \draw[evsb] (e1) -- node[katlab, left=1pt]{$\seqbef$} (e3);
  \draw[evmb] ([xshift=-2.5mm]e1.north east) to[out=60, in=120] node[katlab, above=0pt]{$\mb$} ([xshift=2.5mm]e2.north west);
  \draw[evrf, bend left] (e2) to node[katlab, below right=1pt, pos=.45]{$\rf$} (e3.0);
\end{tikzpicture}%
}}$ &
\centering
$\vcenter{\hbox{%
\begin{tikzpicture}[katpic]
  \foreach \i/\evs in {0/{}, 1/{1}, 2/{1,2}, 3/{1,2}, 4/{1,2,3}, 5/{1,2,3},
                       6/{1,2,3}, 7/{2,3}, 8/{3}, 9/{}} {
    \pgfmathsetmacro{\bx}{0.99*\i+0.15*(\i>1)}
    \draw[bagbox] (\bx-0.29,0.24) rectangle (\bx+0.29,-0.96);
    \foreach \e in \evs
      \node[bagev] (b\i-\e) at (\bx-0.12,-0.36*\e+0.36) {$\event_\e$};
    \node[font=\tiny, anchor=south] at (\bx,0.30) {$\Bag^{w_\i}$};
  }
  \draw[evmb, shorten >=1pt] ([xshift=2.5mm]b3-1.center) -- ([xshift=2.5mm]b3-2.center);
  \draw[evsb, shorten >=1pt] ([xshift=2.5mm]b5-1.center) -- ([xshift=2.5mm]b5-3.center);
  \draw[evrf, shorten >=1pt] ([xshift=2.5mm]b6-2.center) -- ([xshift=2.5mm]b6-3.center);
  \foreach \i/\op/\args in {
      1/\CFEIntro/{(\lx{\wt(x,1)},1)}, 2/\CFEIntro/{(\lx{\wt(x,2)},2)},
      3/\CFEIntromb/{(1,2)},         4/\CFEIntro/{(\lx{\rd(x,2)},3)},
      5/\CFEIntrosb/{(1,3)},         6/\CFEIntrorf/{(2,3)},
      7/\CFEForget/{(1)}, 8/\CFEForget/{(2)}, 9/\CFEForget/{(3)}} {
    \pgfmathsetmacro{\bx}{0.99*\i+0.15*(\i>1)}
    \node[font=\scriptsize, anchor=base] at (\bx,-1.30) {$\op$};
    \node[font=\tiny, anchor=base] at (\bx,-1.56) {$\args$};
  }
\end{tikzpicture}%
}}$
\tabularnewline[-9.4pt] %
\centering \subcaption{An execution graph $G$ with two threads.}\label{SUBFIG:WORD_GRAPH} &
\centering \subcaption{A very nice path decomposition of $G$, with the $i$-th letter of the word $w$ below $\Bag^{w_i}$.
Each step that introduces an edge draws it inside the bag.}\label{SUBFIG:WORD_VNICE}
\tabularnewline
\end{tabular*}
\caption{A graph $G$ (\subref{SUBFIG:WORD_GRAPH}), and a very nice path decomposition of $G$, along with the word $w\in\Ops(3)^*$ it yields (\subref{SUBFIG:WORD_VNICE}).
}
\label{FIG:WORD}
\end{figure}
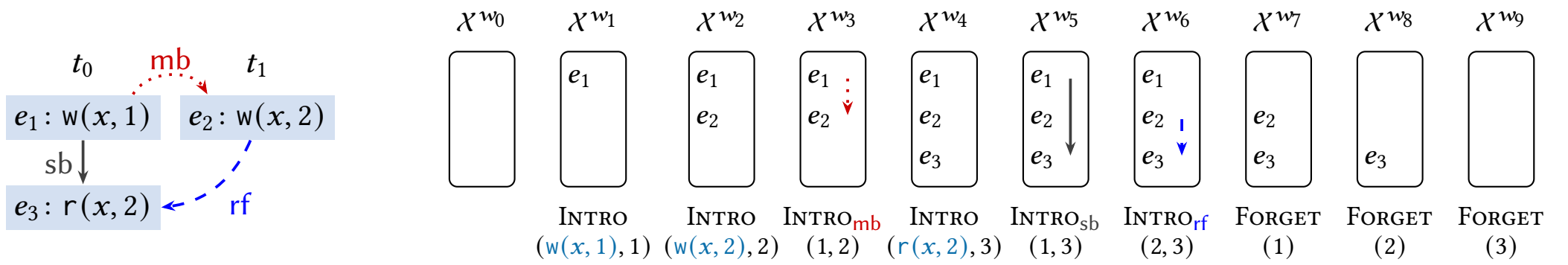

\Paragraph{Words and pathwidth.}
A closed word $w \in \Ops(k+1)^*$ of length $n-1$ defines a very nice path decomposition $\Bag_1,\ldots,\Bag_n$ of $G^w$, where $\Bag_{i+1}$ contains the events that are colored at the end of the prefix $w_i$ of $w$.
This is illustrated in \cref{FIG:WORD}, and made formal in the following lemma.

\begin{restatable}{lemma}{lempathwidth}\label{LEM:PATHWIDTH}
For any closed $w\in \Ops(k+1)^*$, the sequence $\Bag^{w_0}, \Bag^{w_1}, \ldots, \Bag^{w_{|w|}}$, where $\Bag^{w_i} \triangleq \dom{\chi^{w_i}}$, is a very nice path decomposition of $G^w$ of width at most $k$, in which each bag $\Bag^{w_i}$ with $i\geq 1$ is labeled by the step corresponding to the $i$-th operation of $w$.
Conversely, for any free execution graph $G$ and very nice path decomposition $\Bag_1,\Bag_2,\ldots, \Bag_n$ of $G$ of width at most $k$, there is a closed $w\in \Ops(k+1)^*$ of length $n-1$ such that $G^w=G$ and $\Bag^{w_i} = \Bag_{i+1}$ for every $i\in\{0,\ldots,n-1\}$.
\end{restatable}

A direct corollary of \cref{LEM:PATHWIDTH} and \cref{LEM:NICE_PW} is that a free execution graph $G$ has pathwidth at most $k$ iff there is a closed word $w\in \Ops(k+1)^*$ such that $G = G^w$.
This is the analogue of a result of~\cite{Courcelle92} for the HR-algebra, where the graphs denoted by terms with $k+1$ sources are exactly the graphs of treewidth at most $k$.
Here, the colors play the role of the sources.

\Paragraph{Automata over $\Ops(k+1)$.}
In the following sections, we build automata that read words over $\Ops(k+1)$, and we use each such automaton to describe a set of free execution graphs.
A free execution graph belongs to this set if it is $G^{w}$ for some closed word $w$ and the automaton accepts $w$.
By \cref{LEM:PATHWIDTH}, every such $G^{w}$ has pathwidth at most $k$, and every free execution graph of pathwidth at most $k$ equals $G^{w}$ for some closed $w\in\Ops(k+1)^*$.
Taking $k = \ell$ thus makes the words over $\Ops(\ell+1)$ a search space for $\ell$-$\pw$-bounded witnesses.

\section{An Automaton for Recognizing Non-Robustness Witnesses}\label{SEC:AUT}

A witness for the non-$\mm$-robustness of $\ConcProg$ is an execution graph $G$ with $\tuple{G.\Events,G.\seqbef}\in\GLang{\ConcProg}$, $G\models\mm$ and $G\not\models\scmm$.
For each condition, we build an automaton that accepts all words $w\in\Ops(k+1)^*$ for which $G^{w}$ satisfies it:~\cref{SUBSEC:AUT_WF} performs some preparatory checks on the well-formedness of $w$ and $G^{w}$, \cref{SUBSEC:AUT_PROG} checks that $G^{w}$ represents the execution of $\ConcProg$, and \cref{SUBSEC:AUT_MM} that $G^{w}$ is consistent/inconsistent in a $\KAT$ memory model.
Throughout, we bound the size of each automaton by an exponential in the input parameters, while each transition can be written down in polynomial time.
For the most part, we give a high-level description of these automata, aiming at readability and intuitiveness.
Their formal construction appears in \cref{SEC:APP_AUT}.

\subsection{Recognizing Well-Formedness}\label{SUBSEC:AUT_WF}

Free execution graphs may not be execution graphs; for instance, $\rf$ may relate events on different locations, or $\seqbef$ may contain cycles.
The following lemma takes care of such ill-formed graphs.

\begin{restatable}{lemma}{lematwf}\label{LEM:AUT_WF}
  For every $k\in\Nats$, there is a deterministic automaton $\mathcal{A}^{k}_{\textsc{wf}}$ with $2^{O(k^2+k\log|\LabelDom|+|\ThreadDom|+|\LocationDom|)}$ states that accepts $w\in\Ops(k+1)^{*}$ iff $w$ is closed and $G^{w}$ is an execution graph.
\end{restatable}

The automaton $\mathcal{A}^{k}_{\textsc{wf}}$ is the product of four deterministic automata: $\mathcal{A}^{k}_{\textsc{form}}$ checks that $w$ itself is closed, and $\mathcal{A}^{k}_{\seqbef}$, $\mathcal{A}^{k}_{\rf}$ and $\mathcal{A}^{k}_{\mb}$ check that the respective relation is well-formed.

\Paragraph{Closedness of the word.}
A word is closed if each of its operations meets its preconditions, and no color is active at the end.
All conditions depend only on the active colors and on the edges between their events.
For example, $\CFEIntroR{\seqbef}(c_1, c_2)$ requires that $c_1$ and $c_2$ are active colors (i.e., they color events of the graph), and that no $\seqbef$-edge has been drawn between $c_1$ and $c_2$ since these were last introduced.
A state of $\mathcal{A}^{k}_{\textsc{form}}$ is, thus, a pair $\tuple{\ActSet,\EdgSet}$, where $\ActSet\subseteq \IntSet{k+1}$ is the set of active colors and $\EdgSet\subseteq\{\seqbef,\rf,\mb\}\times\ActSet\times\ActSet$ records the edges drawn between their events.
Following the previous example,  $\CFEIntroR{\seqbef}(c_1, c_2)$ can be executed if $c_1,c_2\in \ActSet$ and $\tuple{\seqbef,c_1,c_2}\not\in\EdgSet$.

\Paragraph{Well-formedness of $\seqbef$.}
The relation $\seqbef$ is well-formed iff, for each thread,
\begin{enumerate*}[label=(\roman*)]
\item\label[wfcond]{ITEM:SB_IN} all events have exactly one incoming $\seqbef$-edge, except one, which has none,
\item\label[wfcond]{ITEM:SB_OUT} all events have exactly one outgoing $\seqbef$-edge, except one, which has none, and
\item\label[wfcond]{ITEM:SB_CONN} all events are connected.
\end{enumerate*}
Note that $\seqbef$ may not be built incrementally, in general, which introduces some bookkeeping for checking the above conditions.
We call an event of a thread with no incoming or outgoing $\seqbef$-edge a \emph{first} or \emph{last} event, respectively. 
For \cref{ITEM:SB_IN,ITEM:SB_OUT}, it suffices to record which colored events already have an incoming and an outgoing edge, and for which threads a first and a last event has already been forgotten.
For \cref{ITEM:SB_CONN}, we record which colored events are connected.

\begin{figure}
\centering
\newcommand{\sbExBase}{%
  \useasboundingbox (0.42,0.32) rectangle (2.46,-1.52);
  \node[evnode] (e1) at (0.6,-0.2)       {$\event_1$};
  \node[evnode] (e2) at (1.3,-0.2)     {$\event_2$};
  \node[evnode] (e5) at (1.3,-0.7)  {$\event_5$};
  \node[evnode] (e6) at (1.3,-1.2)  {$\event_6$};
}
\newcommand{\sbExRight}[3]{%
  \node[evnode, #1] (e3) at (2,-0.2)    {$\event_3$};
  \node[evnode, #2] (e4) at (2,-0.7) {$\event_4$};
  \node[evnode, #3] (e7) at (2,-1.2) {$\event_7$};
  \draw[evsb] (e1) -- (e2);
  \draw[evsb] (e2) -- (e3);
  \draw[evsb] (e4) -- (e5);
  \draw[evsb] (e5.west) to[bend right=60] (e6.west);
  \draw[evsb] (e6) -- (e7);
}
\tikzset{
  evnode/.append style = {rounded corners=4pt},
  sbExCa/.style = {fill=colorCa, fill opacity=0.3, text opacity=1},
  sbExCb/.style = {fill=colorCb, fill opacity=0.3, text opacity=1},
  sbExCc/.style = {fill=colorCc, fill opacity=0.3, text opacity=1},
}
\newcommand{\sbExColor}[2]{\node[font=\scriptsize, right=0.12cm of #1] {$#2$};}
\newcommand{\sbExBox}[2]{\draw[rounded corners=5pt, dashed] (1.68,#1) rectangle (2.32,#2);}
\newcommand{\sbExStep}[1]{%
  \begin{tikzpicture}[katpic, baseline=(current bounding box.north)]
    \useasboundingbox (0,0.92) rectangle (0.8,-0.92);
    \draw[->, thick] (0,0) -- node[above, font=\scriptsize] {$#1$} (0.8,0);
  \end{tikzpicture}%
}
\newcommand{\sbExSets}[6]{%
  \makebox[\linewidth][c]{\hspace{0.6cm}{\scriptsize\renewcommand{\arraystretch}{1.3}$\begin{array}{@{}r@{\;}l@{\quad}r@{\;}l@{}}
    \InSet & = #1 & \OutSet & = #2 \\
    \FstSet & = #3 & \LstSet & = #4 \\
    \RepMap & \multicolumn{3}{@{}l@{}}{= #5} \\
    \ThrMap & \multicolumn{3}{@{}l@{}}{= #6}
  \end{array}$}}%
}
\begin{tabular}[t]{@{}>{\centering\arraybackslash}p{3.5cm}@{}}
\begin{tikzpicture}[katpic, baseline=(current bounding box.north)]
  \sbExBase
  \sbExRight{sbExCa}{sbExCb}{sbExCc}
  \sbExBox{0.08}{-1.48}
  \sbExColor{e3}{\cA} \sbExColor{e4}{\cB} \sbExColor{e7}{\cC}
\end{tikzpicture}
\\\noalign{\vskip 0.0ex}
\sbExSets{\{\cA,\cC\}}{\{\cB\}}{\{t_1\}}{\emptyset}{\{\cA{\to}\cA,\cB{\to}\cB,\cC{\to}\cB\}}{\{\cA{\to}t_1,\cB{\to}t_1,\cC{\to}t_1\}}
\end{tabular}\hspace{0.8em}%
\sbExStep{\CFEIntrosb(\cA,\cB)}\hspace{0.8em}%
\begin{tabular}[t]{@{}>{\centering\arraybackslash}p{3.5cm}@{}}
\begin{tikzpicture}[katpic, baseline=(current bounding box.north)]
  \sbExBase
  \sbExRight{sbExCa}{sbExCb}{sbExCc}
  \draw[evsb] (e3) -- (e4);
  \sbExBox{0.08}{-1.48}
  \sbExColor{e3}{\cA} \sbExColor{e4}{\cB} \sbExColor{e7}{\cC}
\end{tikzpicture}
\\\noalign{\vskip 0.0ex}
\sbExSets{\{\cA,\cB,\cC\}}{\{\cA,\cB\}}{\{t_1\}}{\emptyset}{\{\cA{\to}\cA,\cB{\to}\cA,\cC{\to}\cA\}}{\{\cA{\to}t_1,\cB{\to}t_1,\cC{\to}t_1\}}
\end{tabular}\hspace{0.8em}%
\sbExStep{\CFEForget(\cC)}\hspace{0.8em}%
\begin{tabular}[t]{@{}>{\centering\arraybackslash}p{3.5cm}@{}}
\begin{tikzpicture}[katpic, baseline=(current bounding box.north)]
  \sbExBase
  \sbExRight{sbExCa}{sbExCb}{}
  \draw[evsb] (e3) -- (e4);
  \sbExBox{0.08}{-0.98}
  \sbExColor{e3}{\cA} \sbExColor{e4}{\cB}
\end{tikzpicture}
\\\noalign{\vskip 0.0ex}
\sbExSets{\{\cA,\cB\}}{\{\cA,\cB\}}{\{t_1\}}{\{t_1\}}{\{\cA{\to}\cA,\cB{\to}\cA\}}{\{\cA{\to}t_1,\cB{\to}t_1\}}
\end{tabular}
\caption{Three consecutive states $\tuple{\InSet,\OutSet,\FstSet,\LstSet,\RepMap,\ThrMap}$ of $\mathcal{A}^{k}_{\seqbef}$, and the operations read between them.
Above each state, we draw $G_c^{w}$ restricted to $\seqbef$, where $w$ is the word read so far; all its events belong to thread $t_1$.
The events in the dashed region are those that currently have a color (i.e., the bag of the corresponding path decomposition), which is written next to each of them; the others have been forgotten.}
\label{FIG:AUT_SB_EXAMPLE}
\end{figure}
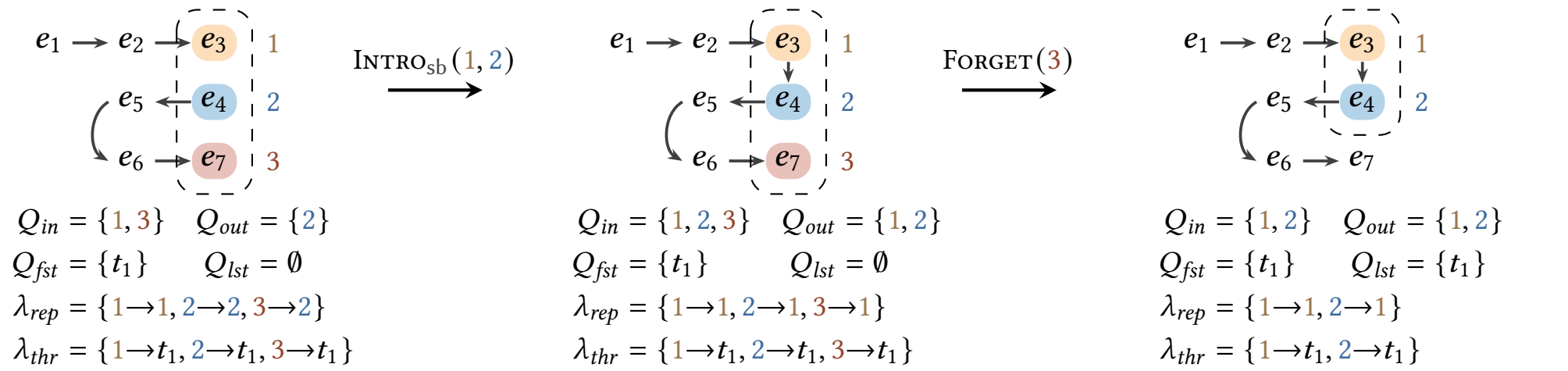

A state of $\mathcal{A}^{k}_{\seqbef}$ is, thus, a tuple $\tuple{\InSet,\OutSet,\FstSet,\LstSet,\RepMap,\ThrMap}$.
Here, $\InSet,\OutSet\subseteq\IntSet{k+1}$ are the active colors whose events have an incoming, respectively outgoing, $\seqbef$-edge, and $\FstSet,\LstSet\subseteq\ThreadDom$ are the threads that have an uncolored first, respectively last, event.
This information suffices to verify \cref{ITEM:SB_IN,ITEM:SB_OUT}.
Moreover, $\RepMap\colon \IntSet{k+1}\rightharpoonup\IntSet{k+1}$ represents the connected components induced by $\seqbef$, where each color is mapped to the smallest color in the same component, and
$\ThrMap\colon \IntSet{k+1}\rightharpoonup\ThreadDom$ maps each active color to the thread of its event.
This information suffices to verify \cref{ITEM:SB_CONN}:~$\ThrMap$ allows to gather all colors of a thread, and $\RepMap$ to check that all such colors belong to the same component.
\cref{FIG:AUT_SB_EXAMPLE} illustrates $\mathcal{A}^{k}_{\seqbef}$ on an example.

\Paragraph{Well-formedness of $\rf$ and $\mb$.}
A state of $\mathcal{A}^{k}_{\rf}$ is a pair $\tuple{\LabMap,\SrcSet}$, where  $\LabMap\colon \IntSet{k+1}\rightharpoonup\LabelDom$ maps each active color to the label of its event, and $\SrcSet\subseteq \IntSet{k+1}$ is the set of active colors representing reading events without yet an $\rf$-edge.
$\mathcal{A}^{k}_{\rf}$ checks that every new edge $\CFEIntroR{\rf}(c_1, c_2)$ is valid (e.g., $c_1, c_2$ represent events on the same location, and the event of $c_2$ does not have an incoming $\rf$-edge).
Moreover, $\CFEForget(c)$ requires $c\notin\SrcSet$, so that every reading event receives its incoming $\rf$-edge before its color is forgotten.
The well-formedness requirements of $\mb$ are identical to those of $\seqbef$, with locations replacing threads and $\rd$ events excluded.
$\mathcal{A}^{k}_{\mb}$ is, likewise, similar to $\mathcal{A}^{k}_{\seqbef}$.

\subsection{Recognizing Executions of $\ConcProg$}\label{SUBSEC:AUT_PROG}

The following lemma ensures that the guessed graph $G^w$ is valid for the program $\ConcProg$.

\begin{restatable}{lemma}{lemautprog}\label{LEM:AUT_PROG}
  For every $k\in\Nats$ and concurrent program $\ConcProg$, there exists a word automaton $\mathcal{A}^{k}_{\ConcProg}$ with $O(|\ConcProg|^{k+1})\, 2^{O(k)}$ states such that, for every $w\in\Lang{\mathcal{A}^{k}_{\textsc{wf}}}$, $\mathcal{A}^{k}_{\ConcProg}$ accepts $w$ iff $\tuple{G^{w}.\Events, G^{w}.\seqbef}\in\GLang{\ConcProg}$.
\end{restatable}

Fix a concurrent program $\ConcProg = \tuple{\Prog_{t_1}, \dots, \Prog_{t_{|\ThreadDom|}}}$.
When an event of a thread $t$ is introduced, $\mathcal{A}^{k}_{\ConcProg}$ guesses a transition of $\Prog_{t}$ that emits the label of that event.
This guess is nondeterministic, as several transitions may emit the same label.
A state of $\mathcal{A}^{k}_{\ConcProg}$ is, thus, a pair $\tuple{\SrcSet, \TrMap}$, where $\TrMap\colon \IntSet{k+1} \rightharpoonup \bigcup_{t \in \ThreadDom} \Trans_{t}$ maps each active color to the transition guessed for its event, and $\SrcSet \subseteq \IntSet{k+1}$ is the set of active colors whose events have no incoming $\seqbef$-edge yet.
When a $\seqbef$-edge is drawn from $c_1$ to $c_2$, $\tgt(\TrMap(c_1))$ must reach $\src(\TrMap(c_2))$ by silent transitions.
Similarly, when a color $c\in\SrcSet$ is forgotten, its event is the first of its thread $t$, so $\src(\TrMap(c))$ must be reachable from the initial state of $\Prog_{t}$ by silent transitions.
\cref{FIG:AUT_PROG_EXAMPLE} shows two steps of $\mathcal{A}^{k}_{\ConcProg}$.

\begin{figure}
\centering
\tikzset{
  evnode/.append style = {rounded corners=4pt},
  pExCa/.style = {fill=colorCa, fill opacity=0.2, text opacity=1},
  pExCb/.style = {fill=colorCb, fill opacity=0.2, text opacity=1},
  pExCc/.style = {fill=colorCc, fill opacity=0.2, text opacity=1},
}

\newcommand{\pExBase}[1]{%
  \useasboundingbox (-0.48,0.6) rectangle (1.48,-1.02);
  \node[font=\footnotesize] at (0,0.45)   {$t_0$};
  \node[font=\footnotesize] at (1.0,0.45) {$t_1$};
  \node[evnode, #1]    (e1) at (0,0)       {$\event_1$};
  \node[evnode, pExCb] (e2) at (0,-0.62)   {$\event_2$};
  \node[evnode]        (e3) at (1.0,0)     {$\event_3$};
  \node[evnode, pExCc] (e4) at (1.0,-0.62) {$\event_4$};
  \draw[evsb] (e3) -- (e4);
  \node[font=\scriptsize, left=0.12cm of e2]  {$\cB$};
  \node[font=\scriptsize, right=0.12cm of e4] {$\cC$};
}
\newcommand{\pExColorA}{\node[font=\scriptsize, left=0.12cm of e1] {$\cA$};}
\newcommand{\pExRegionL}{\draw[rounded corners=5pt, dashed] (-0.32,0.28) -- (0.32,0.28) -- (0.32,-0.34) -- (1.32,-0.34) -- (1.32,-0.9) -- (-0.32,-0.9) -- cycle;}
\newcommand{\pExRegionB}{\draw[rounded corners=5pt, dashed] (-0.32,-0.34) rectangle (1.32,-0.9);}
\newcommand{\pExStep}[1]{%
  \begin{tikzpicture}[katpic, baseline=(current bounding box.north)]
    \useasboundingbox (0,0.91) rectangle (0.8,-0.91);
    \draw[->, thick] (0,0) -- node[above, font=\scriptsize] {$#1$} (0.8,0);
  \end{tikzpicture}%
}
\newcommand{\pExSets}[2]{%
  \makebox[\linewidth][c]{\hspace{0.25cm}{\scriptsize\renewcommand{\arraystretch}{1.3}$\begin{array}{@{}r@{\;}l@{}}
    \SrcSet & = #1 \\
    \TrMap & = #2
  \end{array}$}}%
}
\begin{tabular}[t]{@{}>{\centering\arraybackslash}p{3.45cm}@{}}
\begin{tikzpicture}[katpic, baseline=(current bounding box.north)]
  \pExBase{pExCa}
  \pExColorA
  \pExRegionL
\end{tikzpicture}
\\\noalign{\vskip 0.0ex}
\pExSets{\{\cA,\cB\}}{\{\cA{\to}\delta_{0,1},\cB{\to}\delta_{0,2},\cC{\to}\delta_{1,2}\}}
\end{tabular}\hspace{0.4em}%
\pExStep{\CFEIntrosb(\cA,\cB)}\hspace{0.4em}%
\begin{tabular}[t]{@{}>{\centering\arraybackslash}p{3.45cm}@{}}
\begin{tikzpicture}[katpic, baseline=(current bounding box.north)]
  \pExBase{pExCa}
  \draw[evsb] (e1) -- (e2);
  \pExColorA
  \pExRegionL
\end{tikzpicture}
\\\noalign{\vskip 0.0ex}
\pExSets{\{\cA\}}{\{\cA{\to}\delta_{0,1},\cB{\to}\delta_{0,2},\cC{\to}\delta_{1,2}\}}
\end{tabular}\hspace{0.4em}%
\pExStep{\CFEForget(\cA)}\hspace{0.4em}%
\begin{tabular}[t]{@{}>{\centering\arraybackslash}p{3.45cm}@{}}
\begin{tikzpicture}[katpic, baseline=(current bounding box.north)]
  \pExBase{}
  \draw[evsb] (e1) -- (e2);
  \pExRegionB
\end{tikzpicture}
\\\noalign{\vskip 0.0ex}
\pExSets{\emptyset}{\{\cB{\to}\delta_{0,2},\cC{\to}\delta_{1,2}\}}
\end{tabular}
\caption{Three consecutive states $\tuple{\SrcSet,\TrMap}$ of $\mathcal{A}^{k}_{\ConcProg}$, for the program $\ConcProg$ of \cref{SUBFIG:ROB_PROG}, and the operations read between them.
Above each state, we draw $G_c^{w}.\seqbef$, where $w$ is the word read so far; its events are $\event_1,\dots,\event_4$ of \cref{SUBFIG:ROB_GRAPH}.
The events in the dashed region are those that currently have a color, which is written next to each of them; the others have been forgotten.
We write $\delta_{i,j}$ for the $j$-th transition of $\Prog_{t_i}$.}
\label{FIG:AUT_PROG_EXAMPLE}
\end{figure}

\subsection{Recognizing $\mm$-consistency}\label{SUBSEC:AUT_MM}

Finally, here we construct automata for checking memory model (in)consistency.
The construction is more involved than the automata we have seen so far, and one of the main technical contributions of this work.
We thus present it in more detail.

\begin{restatable}{lemma}{lemautmmyes}\label{LEM:AUT_MM_YES}
  For every $k\in\Nats$ and memory model $\mm$, there is a deterministic word automaton $\mathcal{A}^{k}_{\mm}$ with $2^{O(k^2\cdot|\mm|^2)}$ states such that, for every $w\in\Lang{\mathcal{A}^{k}_{\textsc{wf}}}$, $\mathcal{A}^{k}_{\mm}$ accepts $w$ iff $G^{w}\models\mm$.
  Complementing $\mathcal{A}^{k}_{\mm}$ yields a deterministic word automaton $\mathcal{A}^{k}_{\lnot\mm}$ with the same states that, for every $w\in\Lang{\mathcal{A}^{k}_{\textsc{wf}}}$, accepts $w$ iff $G^{w}\not\models\mm$.
\end{restatable}

Fix a memory model $\mm$, and let $\mathcal{K}_{\mm}=\tuple{\States,\TestDom\cup\RelBase,\Trans,\InitState,\States_F}$ be a $\KAT$-NFA for $\mm$.
Recall that $G\models\mm$ iff no word accepted by $\mathcal{K}_{\mm}$ has a cyclic realization in $G$.
So, $\mathcal{A}^{k}_{\mm}$ has to detect such cyclic realizations in $G^{w}$ while reading $w$.
The difficulty is that a realization may be arbitrarily long and visit (or even start at) events that are forgotten by the end of $w$, where the violation happens.
The key insight underpinning \cref{LEM:AUT_MM_YES} is that it is possible and sufficient to summarize each realization in its (colored) endpoints, by keeping track of the states of $\mathcal{K}_{\mm}$ that it connects.

Throughout, $w$ is well-formed and $G_c^{w}=\tuple{\Events,\seqbef,\rf,\mb,\chi}$, and we write $\event_c\triangleq\chi^{-1}(c)$ for the event of an active color $c$.
A \emph{walk} from $\event$ to $\event'$ is a pair of a word $u\in (\TestDom\cup\RelBase)^*$ and a realization $\event_0,\event_1,\ldots,\event_m$ of $u$ in $G^{w}$ from $\event_0=\event$ to $\event_m=\event'$.
Unlike a realization, a walk thus fixes the word it realizes.
Its \emph{interior} consists of the events $\event_1,\ldots,\event_{m-1}$.
We say that the walk \emph{leads} $\mathcal{K}_{\mm}$ from $q$ to $q'$ if $\mathcal{K}_{\mm}$ has a run on $u$ from $q$ to $q'$.
In these terms, $G^{w}\not\models\mm$ iff some walk from an event to itself leads $\mathcal{K}_{\mm}$ from $\InitState$ to a state of $\States_F$.
We call such a walk a \emph{violation}.

Our construction rests on the observation that uncolored events never gain new edges, so a walk can only enter or leave uncolored events by passing through some colored event.
The automaton summarizes a walk between two colored events with uncolored interior by storing colors of its endpoints and the states of $\mathcal{K}_{\mm}$ it connects.
This is recorded in the following tuples.

A tuple $\tuple{c_1,c_2,q_1,q_2}\in\IntSet{k+1}\times\IntSet{k+1}\times\States\times\States$, with $c_1,c_2\in\img{\chi}$, is a \emph{path summary} of $w$ if some walk from $\event_{c_1}$ to $\event_{c_2}$ with uncolored interior leads $\mathcal{K}_{\mm}$ from $q_1$ to $q_2$; and  a \emph{cycle summary} of $w$ if, for some uncolored event $\event$, some walk from $\event$ to $\event_{c_1}$ leads $\mathcal{K}_{\mm}$ from $\InitState$ to $q_1$, and some walk from $\event_{c_2}$ to $\event$ leads $\mathcal{K}_{\mm}$ from $q_2$ to a state of $\States_F$, both with uncolored interior.

We write $\SumPath[w]$ and $\SumCyc[w]$ for the sets of path and cycle summaries of $w$.
Intuitively, we store cycle summaries because a violation may still be formed by adding a walk from $\event_{c_1}$ to $\event_{c_2}$ that leads $\mathcal{K}_{\mm}$ from $q_1$ to $q_2$.
This violation cannot be cut into path summaries, as its first event may not be colored.
In particular, when $c_1=c_2$ and $q_1=q_2$, the empty walk at $\event_{c_1}$ is such a walk, so a cycle summary $\tuple{c,c,q,q}$ witnesses a violation~(see \cref{FIG:CONS_MM_EXAMPLE}).

\begin{figure}
\centering
\tikzset{
  evnode/.append style = {rounded corners=4pt},
  mmQCa/.style = {fill=colorCa, fill opacity=0.2, text opacity=1},
  mmQCb/.style = {fill=colorCb, fill opacity=0.2, text opacity=1},
  mmQWalk/.style = {kattrans, decorate, decoration={snake, amplitude=1.2pt, segment length=4pt, post length=3pt}},
  mmQFrame/.style = {rounded corners=5pt, draw=black!35},
  mmQName/.style = {font=\footnotesize, anchor=south west, inner sep=2pt},
}
\begin{tabular*}{\textwidth}{@{}p{.5\textwidth}@{\extracolsep{\fill}}p{.45\textwidth}@{}}
\centering
$\vcenter{\hbox{%
\begin{tikzpicture}[katpic]
  \node[evnode]        (e)  at (0,0)       {$\event$};
  \node[evnode, mmQCa] (c1) at (1.25,0.36)  {$\event_1$};
  \node[evnode, mmQCb] (c2) at (1.25,-0.36) {$\event_2$};
  \node[font=\scriptsize, right=0.00cm of c1] {$\cA$};
  \node[font=\scriptsize, right=0.00cm of c2] {$\cB$};
  \draw[mmQWalk, bend left=20] (e) to node[katlab, above left=0pt, font=\scriptsize] {$u_1$} (c1);
  \draw[mmQWalk, bend left=20] (c2) to node[katlab, below left=0pt, font=\scriptsize] {$u_2$} (e);
  \draw[mmQFrame] (-0.45,0.78) rectangle (2.0,-0.78);
  \node[mmQName] at (-0.45,0.78) {$G^{w}_c$};

  \node[katnode, font=\scriptsize] (qi) at (2.7,0)   {$\InitState$};
  \node[katnode, font=\scriptsize] (q1) at (3.75,0)  {$q_1$};
  \node[katnode, font=\scriptsize] (q2) at (4.35,0)  {$q_2$};
  \node[katacc, font=\scriptsize]  (qf) at (5.4,0)   {$q_F$};
  \draw[katstart] (2.7,0.6) -- (qi);
  \draw[mmQWalk] (qi) -- node[katlab, below=1pt, font=\scriptsize] {$u_1$} (q1);
  \draw[mmQWalk] (q2) -- node[katlab, below=1pt, font=\scriptsize] {$u_2$} (qf);
  \draw[mmQFrame] (2.25,0.78) rectangle (5.85,-0.78);
  \node[mmQName] at (2.25,0.78) {$\mathcal{K}_{\mm}$};
\end{tikzpicture}%
}}$ &
\centering
$\vcenter{\hbox{%
\begin{tikzpicture}[katpic]
  \node[evnode]        (e) at (0,0)   {$\event$};
  \node[evnode, mmQCa] (c) at (1.25,0) {$\event_1$};
  \node[font=\scriptsize, right=0.00cm of c] {$\cA$};
  \draw[mmQWalk, bend left=35] (e) to node[katlab, above=1pt, font=\scriptsize] {$u_1$} (c);
  \draw[mmQWalk, bend left=35] (c) to node[katlab, below=1pt, font=\scriptsize] {$u_2$} (e);
  \draw[mmQFrame] (-0.45,0.78) rectangle (2.0,-0.78);
  \node[mmQName] at (-0.45,0.78) {$G^{w}_c$};

  \node[katnode, font=\scriptsize] (qi) at (2.7,0)  {$\InitState$}; 
  \node[katnode, font=\scriptsize] (q)  at (3.8,0)  {$q$};
  \node[katacc, font=\scriptsize]  (qf) at (4.9,0)  {$q_F$};
  \draw[katstart] (2.7,0.6) -- (qi);
  \draw[mmQWalk] (qi) -- node[katlab, below=1pt, font=\scriptsize] {$u_1$} (q);
  \draw[mmQWalk] (q)  -- node[katlab, below=1pt, font=\scriptsize] {$u_2$} (qf);
  \draw[mmQFrame] (2.25,0.78) rectangle (5.35,-0.78);
  \node[mmQName] at (2.25,0.78) {$\mathcal{K}_{\mm}$};
\end{tikzpicture}%
}}$
\tabularnewline[-9.4pt] %
\centering \subcaption{$\tuple{\cA,\cB,q_1,q_2}\in \SumCyc$.}\label{SUBFIG:CONS_MM_EXAMPLE_1} &
\centering \subcaption{$\tuple{\cA,\cA,q,q}\in \SumCyc$: a violation.}\label{SUBFIG:CONS_MM_EXAMPLE_2}
\tabularnewline
\end{tabular*}
\caption{
(\subref{SUBFIG:CONS_MM_EXAMPLE_1}) shows an illustration of a cycle summary of $\SumCyc$.
In $G_c^w$, there exists an uncolored event $e$ and walks $e\rightsquigarrow e_1$ and $e_2\rightsquigarrow e$ with words $u_1$ and $u_2$, respectively.
In $\mathcal{K}_{\mm}$, $u_1$ takes $\InitState$ to $q_1$, and $u_2$ takes $q_2$ to $q_F\in \States_F$.
In (\subref{SUBFIG:CONS_MM_EXAMPLE_2}), the two walks end and start, respectively, on the same endpoint (event $e_1$), therefore closing a cycle on $e$ with the word $u_1 u_2$.
In $\mathcal{K}_{\mm}$, $u_1 u_2$ takes $\InitState$ to $q_F\in \States_F$ via $q$, thereby witnessing a violation.}
\label{FIG:CONS_MM_EXAMPLE}
\end{figure}

\SubParagraph{States.}
A state of $\mathcal{A}^{k}_{\mm}$ is a pair $\tuple{\SumPath,\SumCyc}$ of sets $\SumPath,\SumCyc\subseteq \IntSet{k+1}\times \IntSet{k+1}\times\States\times\States$, which hold path summaries and cycle summaries, respectively.
If $\mathcal{A}^{k}_{\mm}$ has a run on a well-formed word $w$, then the state it reaches after reading $w$ is $\tuple{\SumPath[w],\SumCyc[w]}$.
Each of $\SumPath$ and $\SumCyc$ takes $2^{(k+1)^2\cdot|\States|^2}$ values, where $|\States|\leq 2|\mm|+2$ by \cref{LEM:KAT_COMPILE}.

The automaton detects violations when it forgets an event.
Consider a violation, and the operation $\CFEForget(c)$ that forgets the last of the events in the violation.
Just before this operation, every event of the violation other than $\event_c$ is uncolored, so its first event is either $\event_c$ or an uncolored event.
When the automaton forgets $\event_c$, it lets walks pass through $\event_c$.
Cutting the violation at a visit to $\event_c$, then, yields the two walks of a cycle summary $\tuple{c,c,q,q}$, for some $q\in\States$.
The automaton rejects as soon as such a tuple appears, before it discards the summaries that mention $c$.
The initial state and the unique accepting state are both $\tuple{\emptyset,\emptyset}$.
\cref{FIG:MM_RUN} shows a run of $\mathcal{A}^{k}_{\mm}$ that detects a violation.

\begin{figure}
\centering
\tikzset{
  evnode/.append style = {rounded corners=4pt},
  mrExCa/.style = {fill=colorCa, fill opacity=0.2, text opacity=1},
  mrExCb/.style = {fill=colorCb, fill opacity=0.2, text opacity=1},
}
\newcommand{\mrExBase}[1]{%
  \useasboundingbox (-0.2,0.32) rectangle (1.75,-0.94);
  \node[evnode]         (e1) at (0,0)       {$\event_1$};
  \node[evnode]         (e2) at (0,-0.62)   {$\event_2$};
  \node[evnode, #1]     (e3) at (1.3,0)     {$\event_3$};
  \node[evnode, mrExCb] (e4) at (1.3,-0.62) {$\event_4$};
  \draw[evsb] (e1) -- (e2);
  \draw[evrf] (e2) -- (e3);
  \draw[evrf] (e4) -- (e1);
  \node[font=\scriptsize, right=0.12cm of e4] {$\cB$};
}
\newcommand{\mrExColorA}{\node[font=\scriptsize, right=0.12cm of e3] {$\cA$};}
\newcommand{\mrExLast}{\draw[evsb] (e3) -- (e4);}
\newcommand{\mrExRegionBoth}{\draw[rounded corners=5pt, dashed] (0.98,0.28) rectangle (1.62,-0.9);}
\newcommand{\mrExRegionBottom}{\draw[rounded corners=5pt, dashed] (0.98,-0.34) rectangle (1.62,-0.9);}
\newcommand{\mrExStep}[2][0]{%
  \begin{tikzpicture}[katpic, baseline=(current bounding box.north)]
    \useasboundingbox (0,0.63) rectangle (0.5+#1,-0.63);
    \draw[->, thick] (-0.25+2*#1,0) -- node[above, font=\scriptsize] {$#2$} (0.75+2*#1,0);
  \end{tikzpicture}%
}
\newcommand{\mrExSets}[4]{%
  {\scriptsize\renewcommand{\arraystretch}{1.3}$\begin{array}{@{}r@{\;}l@{}}
    #1 & = \{#2\} \\
    #3 & = \{#4\}
  \end{array}$}%
}
\newcommand{\mrExNL}{\\ & \phantom{=\{}}
\newcommand{\mrExViol}[1]{{\setlength{\fboxsep}{1pt}\colorbox{colorMB!20}{$#1$}}}
\begin{tabular*}{\textwidth}{@{}p{.092\textwidth}@{\extracolsep{\fill}}p{.885\textwidth}@{}}
\centering
$\vcenter{\hbox{%
\begin{tikzpicture}[katpic]
  \node[katnode] (q0) at (0,0)      {$q_0$};
  \node[katacc]  (q1) at (0,-1.1)   {$q_1$};
  \draw[katstart] (0,0.6) -- (q0);
  \draw[kattrans] (q0) -- node[katlab, left=1pt, font=\scriptsize]{$\Ksb,\Krf$} (q1);
  \draw (q1) edge[loop below] node[katlab, below=0pt, font=\scriptsize]{$\Ksb,\Krf$} (q1);
\end{tikzpicture}%
}}$ &
\centering
$\vcenter{\hbox{%
\begin{tabular}{@{}c@{}}%
\begin{tabular}[t]{@{}c@{}}
\begin{tikzpicture}[katpic, baseline=(current bounding box.north)]
  \mrExBase{mrExCa}
  \mrExColorA
  \mrExRegionBoth
\end{tikzpicture}
\\\noalign{\vskip 0.0ex}
\mrExSets{\SumPath}{\tuple{\cB,\cA,q_0,q_1},\mrExNL\tuple{\cB,\cA,q_1,q_1}}{\SumCyc}{\tuple{\cA,\cB,q_1,q_0},\mrExNL\tuple{\cA,\cB,q_1,q_1}}
\end{tabular}\hspace{0pt}%
\mrExStep[0.25]{\CFEIntrosb(\cA,\cB)}\hspace{0pt}%
\begin{tabular}[t]{@{}c@{}}
\begin{tikzpicture}[katpic, baseline=(current bounding box.north)]
  \mrExBase{mrExCa}
  \mrExLast
  \mrExColorA
  \mrExRegionBoth
\end{tikzpicture}
\\\noalign{\vskip 0.0ex}
\mrExSets{\SumPath}{\tuple{\cB,\cA,q_0,q_1},\tuple{\cB,\cA,q_1,q_1},\mrExNL\tuple{\cA,\cB,q_0,q_1},\tuple{\cA,\cB,q_1,q_1}}{\SumCyc}{\tuple{\cA,\cB,q_1,q_0},\tuple{\cA,\cB,q_1,q_1}}
\end{tabular}\hspace{0pt}%
\mrExStep{\CFEForget(\cA)}\hspace{0pt}%
\begin{tabular}[t]{@{}c@{}}
\begin{tikzpicture}[katpic, baseline=(current bounding box.north)]
  \mrExBase{}
  \mrExLast
  \mrExRegionBottom
\end{tikzpicture}
\\\noalign{\vskip 0.0ex}
\mrExSets{\SumPath''}{\tuple{\cB,\cA,q_0,q_1},\tuple{\cB,\cA,q_1,q_1},\tuple{\cA,\cB,q_0,q_1},\mrExNL\tuple{\cA,\cB,q_1,q_1},\tuple{\cB,\cB,q_0,q_1}}{\SumCyc''}{\tuple{\cA,\cB,q_1,q_0},\tuple{\cA,\cB,q_1,q_1},\tuple{\cA,\cA,q_0,q_1},\mrExNL\tuple{\cA,\cB,q_0,q_0},\tuple{\cA,\cB,q_0,q_1},\tuple{\cB,\cA,q_1,q_1},\mrExNL\tuple{\cB,\cB,q_1,q_0},\mrExViol{\tuple{\cB,\cB,q_1,q_1}}}
\end{tabular}%
\end{tabular}%
}}$
\tabularnewline[-9.4pt] %
\centering \subcaption{$\mathcal{K}_{\mm}$.}\label{SUBFIG:MM_RUN_K} &
\centering \subcaption{A run of $\mathcal{A}^{2}_{\mm}$.}\label{SUBFIG:MM_RUN_STEPS}
\tabularnewline
\end{tabular*}
\caption{(\subref{SUBFIG:MM_RUN_K}) shows a $\KAT$-NFA $\mathcal{K}_{\mm}$ for $\mm=(\seqbef\cup\rf)^{+}$, and (\subref{SUBFIG:MM_RUN_STEPS}) shows three consecutive states $\tuple{\SumPath,\SumCyc}$ of $\mathcal{A}^{2}_{\mm}$, and the operations read between them.
Above each state, we draw $G_c^{w}$, where $w$ is the word read so far.
The events in the dashed region are those currently with a color, which is written next to each of them; the others have been forgotten.
We omit from $\SumPath$ the tuples $\tuple{c,c,q,q}$ of the empty walks.
The last panel shows the sets computed by $\textsc{tc}$ when reading $\CFEForget(\cA)$, which contain the violation $\mrExViol{\tuple{\cB,\cB,q_1,q_1}}$, so the run rejects.}
\label{FIG:MM_RUN}
\end{figure}

\SubParagraph{Auxiliary functions.}
The transitions of $\mathcal{A}^{k}_{\mm}$ are based on some auxiliary computations that we describe here.
When a new event or edge is introduced, the only new walks with uncolored interior have empty interior: the empty walk and single tests at the new event, or a single step along the new edge.
For $c_1,c_2\in \IntSet{k+1}$ and a relation symbol or test $l$, the path summaries of a step labeled $l$ from $\event_{c_1}$ to $\event_{c_2}$ are $\textsc{step}(c_1,l,c_2)\triangleq\{\tuple{c_1,c_2,q_1,q_2}\mid\tuple{q_1,l,q_2}\in\Trans\}$.
We then let
\[
\begin{aligned}
  \textsc{addTests}(c,\SumPath,\llab) &\triangleq \SumPath\cup\{\tuple{c,c,q,q}\mid q\in\States\}\cup\textstyle\bigcup\{\textsc{step}(c,\ttest,c)\mid\ttest\in\TestDom,\ \llab\in\ttest\},\\
  \textsc{addEdges}(c_1,c_2,\SumPath,\rsym) &\triangleq \SumPath\cup\textsc{step}(c_1,\rsym,c_2)\cup\textsc{step}(c_2,\rsym^{-1},c_1).
\end{aligned}
\]
\textsc{addTests} adds the empty walk at $\event_c$ and the tests that its label $\llab$ satisfies.
\textsc{addEdges} adds the new edge in both directions, as $\mathcal{K}_{\mm}$ may traverse it backwards.

Forgetting a color $c$ has two effects:~walks may now pass through $\event_c$, and $\event_c$ may be the first event of a violation.
We define an operation $\textsc{tc}(c,\SumPath,\SumCyc)$ that returns $\tuple{\SumPath'',\SumCyc''}$, where $\SumPath''$ is the smallest set that contains $\SumPath$ and satisfies the following rule.
\begin{enumerate}[label=(\roman*),nosep]
  \item\label[tcrule]{ITEM:TC_COMP} If $\tuple{c_1,c,q_1,q}\in \SumPath''$ and $\tuple{c,c_2,q,q_2}\in \SumPath''$, then $\tuple{c_1,c_2,q_1,q_2}\in \SumPath''$.
  This composes two path summaries at $\event_c$.
\end{enumerate}
Similarly, $\SumCyc''$ is the smallest set that contains $\SumCyc$ and satisfies the following rules.
\begin{enumerate}[label=(\roman*),nosep,resume]
  \item\label[tcrule]{ITEM:TC_EXT_IN} If $\tuple{c,c_2,q,q_2}\in \SumCyc''$ and $\tuple{c,c_1,q,q_1}\in \SumPath''$, then $\tuple{c_1,c_2,q_1,q_2}\in \SumCyc''$.
  This extends the walk into $\event_c$ of a cycle summary by a walk from $\event_c$ to $\event_{c_1}$.
  \item\label[tcrule]{ITEM:TC_EXT_OUT} If $\tuple{c_1,c,q_1,q}\in \SumCyc''$ and $\tuple{c_2,c,q_2,q}\in \SumPath''$, then $\tuple{c_1,c_2,q_1,q_2}\in \SumCyc''$.
  This extends the walk out of $\event_c$ of a cycle summary backwards, by a walk from $\event_{c_2}$ to $\event_c$.
  \item\label[tcrule]{ITEM:TC_START} If $\tuple{c,c_1,\InitState,q_1}\in \SumPath''$ and $\tuple{c_2,c,q_2,q_F}\in \SumPath''$ for some $q_F\in\States_F$, then $\tuple{c_1,c_2,q_1,q_2}\in \SumCyc''$.
  This starts a new cycle summary, whose walks start and end at $\event_c$.
\end{enumerate}
We write $\textsc{violationFree}(\SumCyc'')$ when $\SumCyc''$ contains no tuple of the form $\tuple{c',c',q,q}$.
Finally, the operation $\textsc{erase}(c,\SumPath'',\allowbreak\SumCyc'')$ removes every tuple that mentions $c$ from $\SumPath''$ and $\SumCyc''$.

\SubParagraph{Transition rules.}
$\CFEIntro(\llab,c)$ adds the path summaries of the new walks at $\event_c$, namely its empty walk and the tests that $\llab$ satisfies.
$\CFEIntroR{\rsym}(c_1,c_2)$ adds the path summaries of a single step along the new edge, in both directions.
Both leave the cycle summaries unchanged, as these only change when uncoloring an event.
$\CFEForget(c)$ first lets walks pass through $\event_c$, by computing $\textsc{tc}$.
It rejects if this reveals a violation; if not, it discards the summaries that mention $c$, as $c$ is no longer active.
\[
\abovedisplayskip=5pt plus 2pt minus 2pt \abovedisplayshortskip=\abovedisplayskip \belowdisplayskip=\abovedisplayskip \belowdisplayshortskip=\abovedisplayskip
\mbox{\small$
\inferrule{
  \mathrm{op} = \CFEIntro(\llab, c) \\\\
  \SumPath' = \textsc{addTests}(c, \SumPath, \llab)
}{
  \tuple{\SumPath, \SumCyc} \xrightarrow{\mathrm{op}} \tuple{\SumPath', \SumCyc}
}\hspace{0.4em}
\inferrule{
  \mathrm{op} = \CFEIntroR{\rsym}(c_1, c_2) \\\\ \rsym\in\{\seqbef,\rf,\mb\} \\\\
  \SumPath' = \textsc{addEdges}(c_1,c_2, \SumPath, \rsym)
}{
  \tuple{\SumPath, \SumCyc} \xrightarrow{\mathrm{op}} \tuple{\SumPath', \SumCyc}
}\hspace{0.4em}
\inferrule{
  \mathrm{op} = \CFEForget(c) \\\\
  \tuple{\SumPath'', \SumCyc''} = \textsc{tc}(c, \SumPath, \SumCyc) \\\\
  \textsc{violationFree}(\SumCyc'') \\\\
  \tuple{\SumPath', \SumCyc'} = \textsc{erase}(c, \SumPath'', \SumCyc'')
}{
  \tuple{\SumPath, \SumCyc} \xrightarrow{\mathrm{op}} \tuple{\SumPath', \SumCyc'}
}
$}
\]

\subsection{The Complexity of Robustness}\label{SUBSEC:AUT_ROB}

Let $\ConcProg$ be a concurrent program, given by the instruction sequences $S_t$ for $t \in \ThreadDom$.
Let $N$ be the total size of the input, which consists of these sequences, together with explicit representations of $\ThreadDom$, $\LocationDom$, $\RegisterDom$, $\ValueDom$, $\AnnDom$, and of $\mm$ as a $\KAT$ expression.

\Paragraph{Exploring the product.}
Consider any KAT-definable memory model $\mm$ that admits $\ell$-$\pw$-bounded witnesses.
\cref{LEM:PATHWIDTH}, together with \cref{LEM:AUT_WF,LEM:AUT_PROG,LEM:AUT_MM_YES}, implies that
\[
\ConcProg \text{ is robust against } \mm \qquad \iff \qquad 
\Lang{\mathcal{A}^{\ell}_{\textsc{wf}}}\cap \Lang{\mathcal{A}^{\ell}_{\ConcProg}}\cap \Lang{\mathcal{A}^{\ell}_{\mm}} \cap\Lang{\mathcal{A}^{\ell}_{\lnot\scmm}} = \emptyset\ ,
\]
with a word $w$ of the intersection on the right-hand side representing a violation witness $G^w$.
Computing an explicit representation of the above automata (let alone their product) is prohibitive, as their size depends exponentially on $N$.
However, 
\begin{enumerate*}[label=(\roman*)]
    \item if their intersection is not empty, then it contains a word $w$ of at most exponential length, and 
    \item each automaton transition can be verified in polynomial space.
\end{enumerate*}
These two facts lead to an $\NPSPACE$ decision procedure, by guessing and simulating on the fly the synchronous product of the above automata.
Since $\NPSPACE=\PSPACE$, due to Savitch's well-known theorem~\cite{Arora09}, we thus obtain the following.

\begin{restatable}{theorem}{thmrobustnesspspaceexplicit}\label{THE:ROB_PSPACE}
Let $\mm$ be a KAT-definable memory model that admits $\ell$-$\pw$-bounded witnesses, for some $\ell$ bounded by a polynomial in $N$.
Given instruction sequences $S_t$ for $t \in \ThreadDom$ and explicit domains $\ThreadDom$, $\LocationDom$, $\RegisterDom$, $\ValueDom$, $\AnnDom$, deciding whether the induced program $\ConcProg$ is robust against $\mm$ is in $\PSPACE$.
\end{restatable}

\Paragraph{Prefix-closed models.}
Recall from \cref{LEM:PREFIX_BORDERLINE} that every prefix-closed memory model admits $(\pwSC+1)$-$\pw$-bounded witnesses.
Since $\pwSC+1\leq N+1$, \cref{THE:ROB_PSPACE} applies to every prefix-closed memory model.
Moreover, the $\KAT$-NFA $\mathcal{K}_{\mm}$ of \cref{LEM:KAT_COMPILE} can be computed in polynomial time.
This is because expanding the derived relations of $\mm$ yields an expression of size polynomial in $|\mm|$ and $|\ThreadDom|+|\LocationDom|+|\AnnDom|$, and hence in $N$.
We thus obtain \cref{THE:PSPACE_PREFIX}.

\section{A Sequential Monitor}\label{SEC:MON}

The procedure of \cref{THE:ROB_PSPACE} guesses, for every event, a transition of the program that emits it, which is not efficient in practice.
This is required for the generality of \cref{THE:ROB_PSPACE}, which does not assume prefix-closedness of the memory model, but can be simplified under this assumption.
Indeed, prefix-closedness allows us to consider the events of each thread in $\seqbef$ order.
This allows us to replace $\mathcal{A}^{k}_{\ConcProg}$ of \cref{LEM:AUT_PROG} by an automaton that is essentially a realization of the standard operational semantics of $\ConcProg$, in the language of our graph algebra (\cref{SEC:ALGEBRA}) as opposed to the language of event labels.
This treatment requires adapting our definitions on pathwidth.

\Paragraph{Canonical pathwidth.}
A word $w\in\Ops(k)^{*}$ is \emph{canonical} if, for every edge $\event_1\LTo{\seqbef}\event_2$ of $G^{w}$, the $\CFEIntro$ operation that introduced $\event_1$ precedes that of $\event_2$ in $w$.
Similarly, a path decomposition of an execution graph is \emph{canonical} if, for every edge $\event_1\LTo{\seqbef}\event_2$, the first bag containing $\event_1$ comes before the first bag containing $\event_2$.
The \emph{canonical pathwidth} $\cpw(G)$ of $G$ is the smallest width of a canonical path decomposition of $G$.
Thus, canonicity restricts the set of path decompositions of $G$, and the set of words $w$ that yield $G^w$, implying $\pw(G)\leq\cpw(G)$.
\cref{LEM:PATHWIDTH} has a direct analogue:~canonical words over $\Ops(k+1)$ describe exactly the graphs of canonical pathwidth at most $k$.
The bound of \cref{LEM:PW_SC_REDUCED} still holds for $\cpw$, because the path decomposition in its proof is canonical.

\begin{restatable}{lemma}{lempwscreducedcan}\label{LEM:PW_SC_REDUCED_CAN}
    For every execution graph $G$, if $G \models \scmm$ then $\pw(G)\leq\cpw(G)\leq\pwSC$.
\end{restatable}

\Paragraph{Canonical-pathwidth-bounded witnesses.}
We define \emph{$\ell$-$\cpw$-bounded} witnesses as in \cref{SUBSEC:PW_MM}, with $\cpw$ in place of $\pw$.
Prefix-closed memory models admit $\ell$-$\cpw$-bounded witnesses, but for a slightly larger $\ell$ compared to $\pw$-bounded, as the following lemma states.
Intuitively, this is because canonicity forces the last event $e$ of a witness $G$ that causes the violation to be introduced in a bag towards the end of the path decomposition of $G$, and thus its neighboring events (in terms of $\rf$ and $\mb$ edges) must also be carried through until that bag.

\begin{restatable}{lemma}{lemprefixborderlinecan}\label{LEM:PREFIX_BORDERLINE_CAN}
    Every prefix-closed memory model admits $(\pwSC+3)$-$\cpw$-bounded witnesses, and $(\pwSC+2)$-$\cpw$-bounded witnesses if every RMW reads from its $\mb$-predecessor.
\end{restatable}

\Paragraph{The $\mathcal{A}^{k}_{S}$ automaton.}
We now replace $\mathcal{A}^{k}_{\ConcProg}$ by an automaton $\mathcal{A}^{k}_{S}$, with the following guarantees.

\begin{restatable}{lemma}{lemautprogpl}\label{LEM:AUT_PROG_PL}
    Consider a concurrent program $\ConcProg$ given as a collection of instruction sequences $\tuple{S_t}_{t\in \ThreadDom}$, and their respective LTSs $\tuple{\Prog_t}_{t\in \ThreadDom}$.
    For every $k\in \Nats$, there exists a word automaton $\mathcal{A}^{k}_{S}$ with $O\bigl(n^{|\ThreadDom|}\cdot (|\ValueDom|+1)^{|\RegisterDom|}\bigr)\cdot 2^{O((k+|\ThreadDom|)\log k)}$ states, where $n = \max_{t \in \ThreadDom} n_t$, such that, for every $w\in\Lang{\mathcal{A}^{k}_{\textsc{wf}}}$, $\mathcal{A}^{k}_{S}$ accepts $w$ iff $w$ is canonical and $\tuple{G^{w}.\Events,G^{w}.\seqbef} \in \GLang{\ConcProg}$.
\end{restatable}

Intuitively, $\mathcal{A}^{k}_{S}$ realizes the standard operational semantics of $\ConcProg$, but instead of emitting event labels (e.g., $\rd_a(t,x,v)$), it emits the respective operation in $\Ops(k+1)$ (e.g., $\CFEIntro(\rd_a(t,x,v),c)$, for some color $c$).
It also performs some bookkeeping in order to introduce $\seqbef$ edges (i.e., operations $\CFEIntrosb(c_1,c_2)$) in the order that the corresponding events were executed.
An edge $e_1\xrightarrow{\seqbef}e_2$ may be introduced long after $e_1$ and $e_2$ were introduced.
Until then, we call the edge \emph{pending}, and neither of the colors of $e_1$ and $e_2$ may be forgotten.
We make this intuition precise next.

\SubParagraph{States.}
A state of $\mathcal{A}^{k}_{S}$ is a tuple $\tuple{\PC, \Phi, \PndSet, \gamma}$.
Here, $\PC\colon \ThreadDom\to\Nats$ and $\Phi\colon \RegisterDom\to\ValueDom\cup\{\bot\}$ are the program counters and the store, as in the standard program semantics (\cref{SUBSEC:APP_PRELIM_PROG}).
$\PndSet\subseteq \IntSet{k+1}\times \IntSet{k+1}$ is the set of pending $\seqbef$ edges.
Since the word is canonical, the $\seqbef$-predecessor of a new event of thread $t$ is the most recent event of $t$.
$\PndSet$ records these edges until they are drawn, and thereby ensures that the word draws exactly the $\seqbef$-edges of the program order.
Finally, $\gamma\colon \ThreadDom\to \IntSet{k+1}\cup\{\bot\}$ maps each thread to the color of its last event, or to $\bot$ if that event is uncolored or does not exist.
This information is used to create the pending edge of the next event of $t$.
In the initial state, every program counter is $1$, every register holds $\bot$, $\PndSet$ is empty, and $\gamma$ maps every thread to $\bot$.
All states are accepting.

\SubParagraph{Transition rules.}
We give a high-level description of the transition rules of $\mathcal{A}^{k}_{S}$, together with a few examples.
The full set of rules appears in \cref{SUBSEC:APP_MON_PROG}.

Silent transitions (corresponding, e.g., to the execution of a local assignment $\reg := \expr$) behave identically to the standard operational semantics (\cref{SUBSEC:APP_PRELIM_PROG}), leaving the bookkeeping components $\PndSet$ and $\gamma$ unchanged.
The operation $\CFEIntro(\llab,c)$ is emitted upon executing the instruction at $\PC(t)$ for $t=\tid(\llab)$.
It adds the pending edge $\tuple{\gamma(t),c}$ if $\gamma(t)\neq\bot$, which we write $\PndSet'=\textsc{pend}(\PndSet,\gamma,t,c)$, and sets $\gamma(t)$ to $c$.
For example, the rule for a read takes the value $v$ from the label, and stores it in the register:
\[
\abovedisplayskip=5pt plus 2pt minus 2pt \abovedisplayshortskip=\abovedisplayskip \belowdisplayskip=\abovedisplayskip \belowdisplayshortskip=\abovedisplayskip
\mbox{\small$
  \inferrule{
    \mathrm{op} = \CFEIntro(\llab,c) \quad \llab = \rd_a(t, x, v) \quad
    S_t(\PC(t)) = \reg :=_a x \\\\
    \PC' = \PC[t \mapsto \PC(t)+1] \quad
    \Phi' = \Phi[\reg \mapsto v] \quad \PndSet' = \textsc{pend}(\PndSet,\gamma,t,c) \quad \gamma' = \gamma[t\mapsto c]
  }{
    \tuple{\PC, \Phi, \PndSet, \gamma} \xrightarrow{\mathrm{op}} \tuple{\PC', \Phi', \PndSet', \gamma'}
  }
$}
\]
The remaining rules handle the other operations.
$\CFEIntrosb(c_1,c_2)$ requires that the $\seqbef$ edge is pending, and removes it from $\PndSet$.
$\CFEForget(c)$, shown for $c\notin\img{\gamma}$ (else $\gamma$ is reset), requires that $c$ is not an endpoint of a pending edge, which could no longer be drawn.
Finally, $\CFEIntrorf$ and $\CFEIntromb$ simply self-loop, as they do not affect the execution of each thread in isolation.
\[
\abovedisplayskip=5pt plus 2pt minus 2pt \abovedisplayshortskip=\abovedisplayskip \belowdisplayskip=\abovedisplayskip \belowdisplayshortskip=\abovedisplayskip
\makebox[0pt][c]{\small$
\inferrule{
  \mathrm{op} = \CFEIntrosb(c_1,c_2) \\\\
  \tuple{c_1,c_2}\in \PndSet \\\\
  \PndSet' = \PndSet \setminus \{\tuple{c_1,c_2}\}
}{
  \tuple{\PC, \Phi, \PndSet, \gamma} \xrightarrow{\mathrm{op}} \tuple{\PC, \Phi, \PndSet', \gamma}
}\hspace{0.4em}
\inferrule{
  \mathrm{op} = \CFEForget(c) \\\\
  \tuple{c,\cdot},\tuple{\cdot,c}\notin\PndSet \\\\
  c\notin\img{\gamma}
}{
  \tuple{\PC, \Phi, \PndSet, \gamma} \xrightarrow{\mathrm{op}} \tuple{\PC, \Phi, \PndSet, \gamma}
}\hspace{0.4em}
\inferrule{
  \mathrm{op} = \CFEIntroR{\rsym}(c_1,c_2) \\\\ \rsym\in\{\rf,\mb\}
}{
  \tuple{\PC, \Phi, \PndSet, \gamma} \xrightarrow{\mathrm{op}} \tuple{\PC, \Phi, \PndSet, \gamma}
}
$}
\]

\Paragraph{Searching for witnesses.}
Let $\mm$ be a prefix-closed memory model, and let $\ell\triangleq\pwSC+3$.
By \cref{LEM:PREFIX_BORDERLINE_CAN}, if $\ConcProg$ is not robust against $\mm$, then it has a witness $G$ with $\cpw(G)\leq\ell$.
By the canonical analogue of \cref{LEM:PATHWIDTH}, $G=G^{w}$ for some closed canonical $w\in\Ops(\ell+1)^{*}$.
Together with \cref{LEM:AUT_WF,LEM:AUT_PROG_PL,LEM:AUT_MM_YES}, this gives
\[
\ConcProg \text{ is robust against } \mm \qquad \iff \qquad
\Lang{\mathcal{A}^{\ell}_{\textsc{wf}}}\cap \Lang{\mathcal{A}^{\ell}_{S}}\cap \Lang{\mathcal{A}^{\ell}_{\mm}} \cap\Lang{\mathcal{A}^{\ell}_{\lnot\scmm}} = \emptyset\ ,
\]
which is the characterization of \cref{SUBSEC:AUT_ROB}, with $\mathcal{A}^{\ell}_{S}$ in place of $\mathcal{A}^{\ell}_{\ConcProg}$.
A state of $\mathcal{A}^{\ell}_{S}$ also has polynomial size, and its transitions can be verified in polynomial space, leading to the $\PSPACE$ bound.
As $\mathcal{A}^{\ell}_{S}$ executes the instructions of each thread in program order,
the above automata can also be executed on the fly in a dynamic analysis setting, as, e.g., in \cite{Burckhardt2008,Bouajjani2013,Margalit2025}.

\section{Undecidability of Robustness for Relaxed}\label{SEC:RELAXED}

In this section we prove \cref{THE:RELAXED}, thereby showing that the prefix-closedness condition of \cref{THE:PSPACE_PREFIX} cannot be dropped, in general.

\Paragraph{The $\rlxmm$ memory model.}
$\rlxmm$ enforces $\scmm$ per location.
Formally, let
\[
\poloc\triangleq\textstyle\bigcup_{x\in\LocationDom}\identity{\locx{\LabelDom}};\po;\identity{\locx{\LabelDom}}.
\]
Then $\rlxmm\triangleq(\poloc\cup\rf\cup\mo\cup\fr)^{+}$.
Since each of these relations only connects events on the same location, every forbidden cycle stays on one location; thus $\rlxmm$ allows $(\po\cup \rf)$-cycles.

\Paragraph{A tight undecidability of reachability.}
State reachability under $\mm$ asks, given a concurrent program $\ConcProg$ and a state $s_t$ of each $\Prog_t$, whether some $G\in\ExecutionGraphsOf{\ConcProg}{\mm}$ \emph{brings} each thread $t$ to $s_t$, i.e., $\Prog_t$ has a run on the word of $t$ in $G$ that ends in $s_t$.
Equivalently, we may add to each $\Prog_t$ a transition from $s_t$ that emits a special event $\mathbf{term}$, which accesses no location, and ask whether some $G$ \emph{reaches} $\mathbf{term}$, i.e., every thread has a $\mathbf{term}$ event in $G$.
We prove the following theorem.

\begin{restatable}{theorem}{therelaxedreach}\label{THE:RELAXED_REACH}
 State reachability under $\rlxmm$ is undecidable, already over programs with two threads and two memory locations.
\end{restatable}
Note that, over a single location, we have $\poloc=\po$, and thus $\rlxmm$ coincides with $\scmm$.
Similarly, over a single thread, $\rlxmm$ coincides with $\scmm$.
This is because $\po$ is total, so every $\rf$-, $\mo$- or $\fr$-edge goes $\po$-forward, as it would otherwise close a $\poloc$-cycle, and so every cycle of $\po\cup\rf\cup\mo\cup\fr$ is a $\po$-cycle.
Since reachability for $\scmm$ is decidable ($\PSPACE$-complete~\cite{Kozen1977,Sistla1985}), \cref{THE:RELAXED_REACH} gives a tight dichotomy on the decidability of reachability for $\rlxmm$ with respect to each parameter\footnote{
The undecidability of reachability for $\rlxmm$ also follows from a proof in~\cite{Atig2010}, but that proof is on four locations and thus does not establish the dichotomy we obtain with \cref{THE:RELAXED_REACH}.
}.

\Paragraph{Undecidability of robustness.}
We now prove \cref{THE:RELAXED} via a reduction from state reachability, and relying on \cref{THE:RELAXED_REACH}.
We remark that such a reduction is not always possible, e.g., $\ramm$ has undecidable reachability~\cite{Abdulla19,Conrado2026} but decidable robustness~\cite{Lahav2019} (also, \cref{THE:PSPACE_PREFIX}).
Intuitively, such a reduction is possible for $\rlxmm$ because the model allows for unconstrained $(\po\cup \rf)$-cycles\footnote{
Requiring $\acyclic(\po\cup\rf)$ strengthens $\rlxmm$ to $\arlxmm$~\cite{Chakraborty2024}, which is prefix-closed and thus robustness is decidable due to \cref{THE:PSPACE_PREFIX}.
It can also be obtained as a fragment of $\rcmm$.
}.

\begin{wrapfigure}[5]{r}{0.30\linewidth}
\centering
{\small\begin{tabular}[c]{@{}l@{\hspace{1em}}l@{}}
\textbf{Thread} $t_1$ & \textbf{Thread} $t_2$\\ \hline
$\mathbf{if}\ x=1\ \mathbf{then}$ & $\mathbf{if}\ y=1\ \mathbf{then}$\\
\quad $\Prog_{t_1}$ & \quad $\Prog_{t_2}$\\
\quad $y:=1$ & \quad $x:=1$
\end{tabular}}
\vspace{-1em}
\end{wrapfigure}

Consider a two-threaded program $\ConcProg=\tuple{\Prog_{t_1}, \Prog_{t_2}}$ in which $\mathbf{term}$ is the last statement of each thread, and the reachability problem of determining whether $\mathbf{term}$ is reachable via a $\rlxmm$ execution.
Let $\ConcProg'$ extend $\ConcProg$ with two fresh locations $x$ and $y$, as shown on the right.
Note that each thread can execute its \texttt{if} block iff the other thread executes its last write.
This results in a $(\po\cup \rf)$-cycle, which makes the execution impossible under $\scmm$.
On the other hand, $\rlxmm$ allows this cycle (in practice, this could be due to speculation).
We thus have that $\ConcProg'$ is not robust against $\rlxmm$ iff $\ConcProg$ reaches $\mathbf{term}$ on both threads.
Intuitively, the lack of prefix-closedness implies that the smallest pathwidth of a witness might be arbitrarily large.
Here, it is the pathwidth of the execution that brings $\ConcProg$ to $\mathbf{term}$, plus a constant, and the former is unbounded.

\begin{restatable}{lemma}{lemrelaxedenc}\label{LEM:RELAXED_ENC}
  For every concurrent program $\ConcProg$ with two threads, $\ConcProg'$ is robust against $\rlxmm$ iff no $G\in\ExecutionGraphsOf{\ConcProg}{\rlxmm}$ reaches $\mathbf{term}$.
\end{restatable}

\section{Experiments}\label{SEC:EXP}

\newcommand{\expViol}[1]{\textcolor{red!80!black}{$\boldsymbol{\times}$}\,\num{#1}}
\newcommand{\expRob}[1]{\textcolor{green!50!black}{$\checkmark$}\,\num{#1}}
\newcommand{\expTO}{\textsc{to}}
\newcommand{\expMO}{\textsc{mo}}
\newcommand{\expNA}{\textsc{na}}
\renewcommand{\arraystretch}{0.97}

Here we report on an implementation of our algorithm behind \cref{THE:PSPACE_PREFIX} in a prototype tool $\ToolName$, and an experimental evaluation on a range of memory models.

\Paragraph{Implementation.}
$\ToolName$ is implemented inside the \textsc{Spin} model checker~\cite{Holzmann97}.
Its inputs are a program (as in \cref{SUBSEC:PRELIM_PROG}) and a KAT-defined memory model.
Both are compiled into an instrumented \textsc{Promela} program that implements the sequential monitor of \cref{SEC:MON}.
$\ToolName$ further detects and removes color symmetries, thereby collapsing monitor states that are equal up to recoloring.
Robustness violation witnesses are reported through \textsc{Spin} trails.

\Paragraph{Benchmarks and memory models.}
We experiment with a range of memory models, with their $\KAT$ definitions as listed in \cref{SEC:APP_MODELS}, on benchmarks across four categories:
\begin{compactenum}
    \item Classic litmus tests from the literature, whose accesses all have the same access mode, checked against $\tsomm$, $\psomm$, $\sramm$, $\ramm$, $\wramm$\footnote{We add \Ax{write-coherence} to $\wramm$ in order to avoid trivial violations where $\mo$ contradicts $\hb$.} and $\psimm$, as these models do not mix accesses.
    \item Litmus tests with accesses of mixed modes, checked against $\rcelevenmm$ and $\rcmm$.
    \item Litmus tests from the $\ocamlmm$ memory model repository~\cite{ocaml-memory-model}, checked against $\ocamlmm$.
    \item The benchmarks of \textsc{Rocker}~\cite{Lahav2019}, checked against $\tsomm$ and $\ramm$, as in the corresponding paper.
\end{compactenum}

\Paragraph{Setup.}
We run $\ToolName$ using \textsc{Spin} 6.5.2 and \texttt{gcc -O2}, single-threaded.
We use an Intel Xeon Platinum 8168 machine with 192 GB of RAM, running Debian.
For the more demanding benchmarks, we enable state hashing in \textsc{Spin}, which stores 64-bit hashes instead of concrete states.
We report the running time for each benchmark in seconds.

\begin{table}
\centering
\scriptsize
\setlength{\tabcolsep}{1.8pt}
\captionsetup{skip=3pt,position=top}
\begin{minipage}[t]{.5\linewidth}
\centering
\caption{Litmus tests without mixed access modes.}
\label{TAB:EXP_LITMUS}
\begin{tabular}{@{}l r r *{6}{r}@{}}
\toprule
Test & $|\ThreadDom|$ & $|\LocationDom|$ & \multicolumn{1}{c}{$\tsomm$} & \multicolumn{1}{c}{$\psomm$} & \multicolumn{1}{c}{$\sramm$} & \multicolumn{1}{c}{$\ramm$} & \multicolumn{1}{c}{$\wramm$} & \multicolumn{1}{c}{$\psimm$} \\
\midrule
SB & 2 & 2 & \expViol{8.6} & \expViol{8.6} & \expViol{8.5} & \expViol{8.5} & \expViol{8.5} & \expViol{8.5} \\
MP & 2 & 2 & \expRob{8.6}  & \expViol{8.5} & \expRob{8.5}  & \expRob{8.5}  & \expRob{8.5}  & \expRob{8.5}  \\
LB & 2 & 2 & \expRob{8.5}  & \expRob{8.6}  & \expRob{8.5}  & \expRob{8.6}  & \expRob{8.5}  & \expRob{8.5}  \\
S & 2 & 2 & \expRob{9.1}  & \expViol{9.0} & \expRob{9.0}  & \expRob{9.0}  & \expRob{9.0}  & \expRob{9.0}  \\
R & 2 & 2 & \expViol{9.0} & \expViol{9.0} & \expViol{9.0} & \expViol{9.1} & \expViol{9.0} & \expRob{9.1}  \\
2+2W & 2 & 2 & \expRob{9.5}  & \expViol{9.5} & \expRob{9.4}  & \expViol{9.5} & \expViol{9.5} & \expRob{9.5}  \\
CoRR & 2 & 1 & \expRob{6.7}  & \expRob{6.7}  & \expRob{6.6}  & \expRob{6.6}  & \expRob{6.7}  & \expRob{6.7}  \\
CoWR & 2 & 1 & \expRob{7.2}  & \expRob{7.1}  & \expRob{7.1}  & \expRob{7.1}  & \expViol{7.1} & \expRob{7.1}  \\
CoRW1 & 1 & 1 & \expRob{5.5}  & \expRob{5.5}  & \expRob{5.5}  & \expRob{5.4}  & \expRob{5.5}  & \expRob{5.5}  \\
CoRW2 & 2 & 1 & \expRob{7.2}  & \expRob{7.1}  & \expRob{7.1}  & \expRob{7.1}  & \expRob{7.1}  & \expRob{7.1}  \\
CoWW & 1 & 1 & \expRob{5.9}  & \expRob{5.9}  & \expRob{5.9}  & \expRob{5.9}  & \expRob{5.8}  & \expRob{5.9}  \\
WRC & 3 & 2 & \expRob{9.5}  & \expRob{9.5}  & \expRob{9.4}  & \expRob{9.4}  & \expRob{9.4}  & \expRob{9.4}  \\
ISA2 & 3 & 3 & \expRob{12.9}  & \expViol{12.0} & \expRob{12.1}  & \expRob{12.1}  & \expRob{12.5}  & \expRob{12.6}  \\
WWC & 3 & 2 & \expRob{10.0}  & \expRob{10.0}  & \expRob{9.8}  & \expRob{9.8}  & \expRob{9.9}  & \expRob{9.9}  \\
RWC & 3 & 2 & \expViol{9.4} & \expViol{9.4} & \expViol{9.3} & \expViol{9.4} & \expViol{9.4} & \expViol{9.4} \\
W+RWC & 3 & 3 & \expViol{12.4} & \expViol{12.3} & \expViol{11.9} & \expViol{11.9} & \expViol{12.2} & \expViol{12.1} \\
WRW+2W & 3 & 2 & \expRob{10.8}  & \expViol{10.3} & \expRob{10.5}  & \expViol{10.2} & \expViol{10.3} & \expRob{10.6}  \\
WRW+WR & 3 & 2 & \expViol{9.8} & \expViol{9.8} & \expViol{9.8} & \expViol{9.8} & \expViol{9.8} & \expRob{10.0}  \\
WRR+2W & 3 & 2 & \expRob{10.3}  & \expViol{9.8} & \expViol{9.9} & \expViol{9.9} & \expViol{9.9} & \expRob{10.1}  \\
IRIW & 4 & 2 & \expRob{12.3}  & \expRob{12.1}  & \expViol{10.9} & \expViol{11.0} & \expViol{11.0} & \expViol{11.0} \\
3.SB & 3 & 3 & \expViol{12.7} & \expViol{12.8} & \expViol{12.1} & \expViol{12.0} & \expViol{12.4} & \expViol{12.5} \\
3.LB & 3 & 3 & \expRob{12.3}  & \expRob{12.3}  & \expRob{11.9}  & \expRob{11.9}  & \expRob{12.1}  & \expRob{12.2}  \\
3.2W & 3 & 3 & \expRob{19.6}  & \expViol{15.7} & \expRob{15.3}  & \expViol{14.0} & \expViol{15.7} & \expRob{18.9}  \\
\bottomrule
\end{tabular}

\end{minipage}\hfill
\begin{minipage}[t]{.485\linewidth}
\centering
\caption{Litmus tests with mixed access modes.}
\label{TAB:EXP_MIXED}
\begin{tabular}{@{}l r r r r@{}}
\toprule
Test & $|\ThreadDom|$ & $|\LocationDom|$ & \multicolumn{1}{c}{$\rcelevenmm$} & \multicolumn{1}{c}{$\rcmm$} \\
\midrule
MP+rel+acq & 2 & 2 & \expRob{11.0}  & \expRob{8.7}  \\
MP+rel+rlx & 2 & 2 & \expViol{11.0} & \expViol{8.6} \\
MP+rlx+acq & 2 & 2 & \expViol{10.9} & \expViol{8.6} \\
MP+relseq+acq & 2 & 2 & \expRob{12.5}  & \expViol{9.9} \\
S+rel+acq & 2 & 2 & \expRob{11.4}  & \expRob{9.1}  \\
WRC+rel+acq & 3 & 2 & \expRob{11.9}  & \expRob{9.5}  \\
ISA2+rel+acq-rel+acq & 3 & 3 & \expRob{15.9}  & \expRob{12.3}  \\
WWC+rel+acq & 3 & 2 & \expRob{12.5}  & \expRob{9.9}  \\
SB+ra & 2 & 2 & \expViol{11.0} & \expViol{8.6} \\
R+ra & 2 & 2 & \expViol{11.4} & \expViol{9.1} \\
2+2W+ra & 2 & 2 & \expViol{11.9} & \expViol{9.5} \\
IRIW+ra & 4 & 2 & \expViol{13.6} & \expViol{11.1} \\
SB+sc & 2 & 2 & \expRob{12.0}  & \expNA \\
SB+sc+rlx & 2 & 2 & \expViol{11.2} & \expNA \\
2+2W+sc & 2 & 2 & \expRob{12.2}  & \expNA \\
RWC+sc & 3 & 2 & \expRob{17.1}  & \expNA \\
IRIW+sc & 4 & 2 & \expRob{54.5}  & \expNA \\
IRIW-acq-sc & 4 & 2 & \expViol{18.0} & \expNA \\
\bottomrule
\end{tabular}

\end{minipage}

\vspace{-6pt}

\begin{minipage}[b]{.5\linewidth}
\centering
\addtocounter{table}{1}
\caption{Benchmarks of Rocker.}
\label{TAB:EXP_PROGRAMS}
\begin{tabular}{@{}l r r r r@{}}
\toprule
Program & $|\ThreadDom|$ & $|\LocationDom|$ & \multicolumn{1}{c}{$\tsomm$} & \multicolumn{1}{c}{$\ramm$} \\
\midrule
barrier & 2 & 2 & \expRob{9.0} & \expRob{9.0} \\
chase-lev-sc & 3 & 2 & \expViol{842.8} & \expViol{398.4} \\
chase-lev-tso & 3 & 3 & \expTO$^\dagger$ & \expViol{734.5} \\
cilk-the-wsq-sc & 2 & 3 & \expViol{40.8} & \expViol{37.0} \\
dekker-sc & 2 & 3 & \expViol{38.2} & \expViol{37.4} \\
lamport2-sc & 2 & 4 & \expViol{35.8} & \expViol{35.4} \\
lamport2-tso & 2 & 5 & \expTO$^\dagger$ & \expViol{}$^\dagger$ \\
peterson-sc & 2 & 3 & \expViol{18.8} & \expViol{18.2} \\
peterson-tso & 2 & 4 & \expRob{}$^\dagger$ & \expViol{24.6} \\
peterson-ra-bratosz & 2 & 3 & \expViol{19.4} & \expViol{19.0} \\
\bottomrule
\end{tabular}
\par\vspace{0pt}
\end{minipage}\hfill
\begin{minipage}[b]{.485\linewidth}
\centering
\addtocounter{table}{-2}
\caption{Litmus tests for $\ocamlmm$.}
\label{TAB:EXP_OCAML}
\begin{tabular}{@{}l r r r@{}}
\toprule
Test & $|\ThreadDom|$ & $|\LocationDom|$ & \multicolumn{1}{c}{$\ocamlmm$} \\
\midrule
CoRR+W+ponns & 2 & 2 & \expViol{8.6} \\
IRIWaa & 4 & 2 & \expRob{11.7}  \\
IRIWan & 4 & 2 & \expViol{11.0} \\
MP-broken & 2 & 3 & \expViol{11.8} \\
MPco & 3 & 2 & \expRob{10.0}  \\
MPco2 & 2 & 2 & \expRob{9.0}  \\
R-ocaml & 2 & 2 & \expRob{9.0}  \\
SB+extrareads & 2 & 2 & \expRob{11.0}  \\
SB & 2 & 2 & \expRob{11.0}  \\
SBcoh & 2 & 1 & \expRob{13.4}  \\
corw & 2 & 1 & \expViol{7.1} \\
wat & 1 & 1 & \expRob{5.5}  \\
\bottomrule
\end{tabular}
\par\vspace{0pt}
\addtocounter{table}{1}
\end{minipage}
\end{table}

\Paragraph{Results.}
\cref{TAB:EXP_LITMUS} lists the results on the first benchmark category,
with \textcolor{green!50!black}{$\checkmark$} and \textcolor{red!80!black}{$\boldsymbol{\times}$} reporting a robust and non-robust program, respectively.
The verdicts are consistent with the relative strength of the models, with more violations reported on weaker models.
SB is the canonical example of weak behaviors and is non-robust against all models of \cref{TAB:EXP_LITMUS}.
MP exhibits a standard message-passing pattern that is robust against all models that preserve the order of writes;
$\psomm$ does not, as it reorders writes to different locations.
LB has a weak behavior involving $(\po\cup \rf)$-cycles, which are disallowed by all models, and thus is found robust against all.
R separates $\psimm$ from $\sramm$, intuitively because in $\psimm$, $\mo$ also propagates causality (similarly for WRW+WR and WRR+2W).
Notably, R is not robust against $\tsomm$, while IRIW is robust against $\tsomm$ but not against $\psimm$,
which shows that the two models are incomparable.

\cref{TAB:EXP_MIXED} lists the results of the second category, with
\expNA{} marking a non-applicable benchmark due to $\msc$ accesses that $\rcmm$ lacks.
The robustness checks agree between $\rcelevenmm$ and $\rcmm$ on every test except MP+relseq+acq.
This is because, in $\rcelevenmm$, an acquire read $\rd(x)$ synchronizes with a release write $\wt_1(x)$ even if $\rd(x)$ reads from a write $\wt_2(x)$ with $\wt_1(x)\LTo{\po}\wt_2(x)$; in contrast, $\rcmm$ mirrors the C++20 change~\cite{Boehm2018ReleaseSeq}, where release sequences go only along RMW chains.

\cref{TAB:EXP_OCAML} lists the results on $\ocamlmm$.
Here, every location is either atomic, which behaves like $\scmm$, or non-atomic, which behaves very weakly, e.g., without read-read coherence.
The robustness checks showcase this split.
E.g., IRIW is robust when both locations are atomic (IRIWaa), but not when one of them is non-atomic (IRIWan), and CoRR+W+ponns exhibits a read-read coherence violation.

Across all three tables, the total runtime is in the order of seconds, with compilation taking up 84\% of the time, on average.
Of the remaining 16\%, about 10\% is spent on generating \textsc{Promela} code from the input, and only 6\% in \textsc{Spin} for robustness checking (the median is 0.12s).

Finally, we discuss \cref{TAB:EXP_PROGRAMS} on the \textsc{Rocker} benchmarks, which are bigger and more challenging.
Here, the relative time for robustness checking increases up to 81\%, dominating the total time.
$\ToolName$ catches all violations reported by \textsc{Rocker} (for $\ramm$) and \textsc{Trencher} (for $\tsomm$) in~\cite[Fig.~7]{Lahav2019}, but manages to prove robust only barrier (against both $\ramm$ and $\tsomm$) and peterson-tso (against $\tsomm$).
($^\dagger$) marks the instances where state hashing was turned on.
\textsc{Rocker} has 15 more benchmarks that are robust and not shown here; $\ToolName$ times out on them  after 1h (with state hashing).

\section{Related Work}\label{SEC:RELWORK}

Here we discuss related literature beyond the work on robustness covered in \cref{SEC:INTRO}.

\Paragraph{Data-race freedom.}
A common syntactic approximation to robustness is data-race freedom (DRF)~\cite{Adve1990,Adve1993,Gharachorloo1992}, where a program free of certain types of races is guaranteed to observe only $\scmm$ behaviors.
DRF holds for various models, such as $\tsomm$~\cite{Owens2010}, C/C++~\cite{Boehm2008,Batty2011,Lahav2017}, Java~\cite{Manson2005}, $\ocamlmm$~\cite{Dolan2018} and $\wasmmm$~\cite{Raad2026}, but it is known to be too conservative, as racy programs can still be robust.

\Paragraph{Fences.}
When a program is not robust, it can be made so by inserting fences that restrict the reordering of its memory accesses~\cite{Shasha1988}.
As fences incur a runtime overhead~\cite{Alglave2017}, much work aims to optimize fence placement for various hardware models~\cite{Alglave2010,Alglave2017,Bouajjani2013,Abdulla2015b}.
For language-level models, \cite{Lahav2016} proposes syntactic criteria for fences that restore $\scmm$ under $\sramm$, and
\cite{Ou2015} proposes a dynamic-analysis approach for strengthening C/C++ memory accesses to recover $\scmm$.
As our robustness checker is parametric in the memory model, it can serve as an exact oracle for fence placement across these settings.
Fences also arise in compilation mappings between memory models~\cite{Sarkar2012,Lahav2017,Podkopaev2019,Reimers2026}, which act conservatively as they must be correct for all programs.

\Paragraph{Treewidth in concurrency.}
Our approach follows the recipe behind Courcelle's theorem~\cite{Courcelle1990}, with $\KAT$ in place of MSO and pathwidth in place of treewidth, which yields automata of single-exponential rather than non-elementary size.
For automata with stacks and queues, Courcelle's result enables decidable under-approximations of reachability, by restricting to behaviors of bounded treewidth~\cite{Torre2007,Madhusudan2011} or split-width~\cite{Cyriac2012}.
The corresponding languages were recently shown to coincide with the MCFLs~\cite{Aiswarya2026}.
Similarly, message-passing communication models are MSO-definable, so model checking is decidable as an under-approximation over behaviors of bounded special treewidth~\cite{DiGiusto2023}.
The notion closest to robustness is synchronizability, which asks whether all behaviors of a message-passing system lie in a well-behaved class.
It is decidable when this class is artificially restricted to bounded special treewidth, via a generic MSO procedure with non-elementary complexity.
Equivalence to synchronous communication is, however, undecidable~\cite{Bollig2021a,Finkel2017}.
For robustness, in contrast, the target class is the natural $\scmm$, whose pathwidth is always bounded (\cref{LEM:PW_SC_REDUCED}), and we decide it exactly and uniformly in $\PSPACE$.
Finally, treewidth and MSO were recently used as a framework for weak-memory verification, partly also tackling the coarser notion of reads-from robustness~\cite{Conrado26}.
\looseness=-1

\section{Conclusion}\label{SEC:CONCLUSION}

We have developed a generic algorithm for deciding robustness against any prefix-closed, $\KAT$-definable memory model, in $\PSPACE$.
Our algorithm unifies a plethora of earlier, individual results on robustness, and also tackles several new memory models.
We have further shown that the prefix-closedness condition cannot be dropped, in general.
A natural next step is to tackle dependency-based models, such as $\powermm$~\cite{Alglave14}, $\armmm$~\cite{Pulte2017}, $\riscmm$~\cite{RISCV2019} and $\immmm$~\cite{Podkopaev2019}.
The challenge here is that executions have richer signatures, which, if represented naively, break the bounded pathwidth of $\scmm$ (\cref{LEM:PW_SC_REDUCED}).
Another step is to handle memory models beyond $\KAT$, e.g., those containing intersections or differences nested within other operators, as in $\lkmm$~\cite{Alglave2018}.
Finally, $\ToolName$ is a prototype that leaves ample room for further engineering optimizations.

\begin{acks}
This work was partially supported by VILLUM FONDEN research grant VIL42117, the Independent Research Foundation of Denmark grant 5251-00076B, and the Swedish Research Council (VR) grant 2023-05242.
\end{acks}

\section*{Disclosure of AI Use}
We have used Claude AI for our Lean formalization.
This was done post-factum, based on our paper proofs.
We have also used Claude AI and ChatGPT in the development of $\ToolName$ and experimental setup in \cref{SEC:EXP}.
All results were cross-checked for correctness.

\bibliographystyle{plainurl}
\bibliography{bibliography}

\pagebreak

\appendix
\renewcommand{\MathparLineskip}{\lineskiplimit=.2cm\lineskip=.2cm}
\section{Details of \cref{SEC:PRELIM}}\label{SEC:APP_PRELIM}

\subsection{Program Semantics}\label{SUBSEC:APP_PRELIM_PROG}

Fix $t\in\ThreadDom$ and an instruction sequence $S_t:\IntSet{n_t}\to\InstructionDom$.
This subsection defines the LTS $\Prog_t$ that $S_t$ induces~(\cref{SUBSEC:PRELIM_PROG}).
The transitions of $\Prog_t$ from a state $\tuple{\pc,\phi}$ with $\pc\in\IntSet{n_t}$ are given by the rules below; a state with $\pc=n_t+1$ has terminated and has no outgoing transition.
In the rules, a target program counter outside $\IntSet{n_t}$ stands for $n_t+1$.
Each rule applies only when the expressions it evaluates are not $\bot$; we leave this premise implicit.
A failed $\mathsf{CAS}$ does not evaluate $\expr_W$.

A failed CAS reads without writing, so it emits a read label.
To let memory models tell this read apart from the read of a read instruction, we assume that, for every annotation $a$ carried by an instruction, $\AnnDom$ contains a \emph{sibling} annotation $\failann{a}$, which no instruction carries and only failed CAS operations emit.
Tests compare annotations up to siblings, i.e., a condition $\ann(\llab)=a$ in a test stands for $\ann(\llab)\in\{a,\failann{a}\}$.
For example, $\{\llab\in\LabelDom\colon\ann(\llab)=\mrlx\}$ also contains the reads of failed $\mrlx$ CAS operations.
So, by default, a failed CAS behaves as a read with the annotation of its instruction.
A model may still single out failed CAS operations with the test $\FailDom\triangleq\{\llab\in\LabelDom\colon\ann(\llab)=\failann{a}\text{ for some }a\in\AnnDom\}$.
This assumption at most doubles $|\AnnDom|$.

The first four rules are silent, and the others emit the label of an event of $t$.
{\renewcommand{\LTransTo}[1]{\xhookrightarrow{\;\text{\footnotesize $#1$}\;}}
\begin{mathpar}
  \inferrule{
    S_t(\pc) = \reg := \expr
  }{
    \tuple{\pc, \phi} \LTransTo{\epsilon} \tuple{\pc+1, \phi[\reg \mapsto \phi(\expr)]}
  }
  \and
  \inferrule{
    S_t(\pc) = \reg := \NDet(\ValueDom') \\\\ v\in \ValueDom'
  }{
    \tuple{\pc, \phi} \LTransTo{\epsilon} \tuple{\pc+1, \phi[\reg \mapsto v]}
  }
  \and
  \inferrule{
    S_t(\pc) = \mathtt{if}\ \expr\ \mathtt{goto}\ m \\\\ \phi(\expr) \neq 0
  }{
    \tuple{\pc, \phi} \LTransTo{\epsilon} \tuple{m, \phi}
  }
  \and
  \inferrule{
    S_t(\pc) = \mathtt{if}\ \expr\ \mathtt{goto}\ m \\\\ \phi(\expr) = 0
  }{
    \tuple{\pc, \phi} \LTransTo{\epsilon} \tuple{\pc+1, \phi}
  }
  \and
  \inferrule{
    S_t(\pc) = x :=_a \expr
  }{
    \tuple{\pc, \phi} \LTransTo{\wt_a(t,x,\phi(\expr))} \tuple{\pc+1, \phi}
  }
  \and
  \inferrule{
    S_t(\pc) = \mathsf{fence}_a
  }{
    \tuple{\pc, \phi} \LTransTo{\fc_a(t)} \tuple{\pc+1, \phi}
  }
  \and
  \inferrule{
    S_t(\pc) = \reg :=_a x \\\\ v\in\ValueDom
  }{
    \tuple{\pc, \phi} \LTransTo{\rd_a(t,x,v)} \tuple{\pc+1, \phi[\reg\mapsto v]}
  }
  \and
  \inferrule{
    S_t(\pc) = \reg :=_a \mathsf{CAS}(x, \expr_R, \expr_W) \\\\ v\in\ValueDom\setminus\{\phi(\expr_R)\}
  }{
    \tuple{\pc, \phi} \LTransTo{\rd_{\failann{a}}(t,x,v)} \tuple{\pc+1, \phi[\reg\mapsto v]}
  }
  \and
  \inferrule{
    S_t(\pc) = \reg :=_a \mathsf{CAS}(x, \expr_R, \expr_W)
  }{
    \tuple{\pc, \phi} \LTransTo{\rmw_a(t,x,\phi(\expr_R),\phi(\expr_W))} \tuple{\pc+1, \phi[\reg\mapsto\phi(\expr_R)]}
  }
\end{mathpar}}%
A read may observe any value $v$, and a failed CAS any value other than $\phi(\expr_R)$; which value is observed is determined by the $\rf$-edge of its event, not by $\Prog_t$.
Let $|S_t|$ be the total number of symbols in the instructions of $S_t$, counting the expressions and the sets $\ValueDom'$ of $\NDet(\ValueDom')$.
We assume that each binary operator on $\ValueDom$ is evaluated in constant time, e.g., by a lookup table.
Then, given two states $\tuple{\pc,\phi}$ and $\tuple{\pc',\phi'}$ and a label or $\epsilon$, whether they form a transition of $\Prog_t$ is decidable in time $O(|S_t|+|\RegisterDom_t|)$.
To check this, we evaluate the at most two expressions of $S_t(\pc)$ under $\phi$, in time $O(|S_t(\pc)|)$.
For $\NDet(\ValueDom')$, we also check $v\in\ValueDom'$, in time $O(|\ValueDom'|)$, where $|\ValueDom'|\leq|S_t|$.
Finally, we compare the label and $\pc'$, and compare $\phi'$ with the updated store register by register, in time $O(|\RegisterDom_t|)$.

\subsection{Compiling $\KAT$ Expressions}\label{SUBSEC:APP_PRELIM_KAT}

In this subsection, we prove \cref{LEM:KAT_COMPILE}.
We build $\mathcal{K}_{\kexp}$ by induction on $\kexp$, as for regular expressions.
Throughout, $\mathcal{K}_{\kexp}$ has a single accepting state $q_f\neq\InitState$, i.e., $\States_F=\{q_f\}$, and no transition enters $\InitState$ or leaves $q_f$.

A transition labeled by the test $\LabelDom$ behaves like a silent transition, as it is realized exactly by staying at any event.

\Paragraph{Constructions.}
The cases of the induction are the following.
\begin{compactitem}
\item \emph{Tests and base relations.}
For a test $\identity{\ttest}$ or a base relation symbol $\rsym$, $\mathcal{K}_{\kexp}$ is the single transition $\InitState\LTo{\ttest}q_f$ or $\InitState\LTo{\rsym}q_f$, respectively.
\item \emph{Union.}
For $\kexp_1\cup \kexp_2$, take $\mathcal{K}_{\kexp_1}$ and $\mathcal{K}_{\kexp_2}$, and merge their initial states and their accepting states.
\item \emph{Composition.}
For $\kexp_1;\kexp_2$, take $\mathcal{K}_{\kexp_1}$ and $\mathcal{K}_{\kexp_2}$, and merge the accepting state of $\mathcal{K}_{\kexp_1}$ with the initial state of $\mathcal{K}_{\kexp_2}$.
\item \emph{Transitive closure.}
For $\kexp^{+}$, take $\mathcal{K}_{\kexp}$, with initial state $p$ and accepting state $p'$, and add a fresh initial state $\InitState$, a fresh accepting state $q_f$, and the transitions $\InitState\LTo{\LabelDom}p$, $p'\LTo{\LabelDom}p$ and $p'\LTo{\LabelDom}q_f$.
\item \emph{Inverse.}
For $\kexp^{-1}$, reverse every transition of $\mathcal{K}_{\kexp}$, swap $\InitState$ and $q_f$, and replace each base relation symbol $\rsym$ by $\rsym^{-1}$, where $(\rsym^{-1})^{-1}=\rsym$.
Tests are unchanged.
\end{compactitem}
Each construction keeps a single initial state and a single accepting state, which are distinct, with no transition entering the former or leaving the latter.
This is what allows us to merge states in the union and the composition, without creating new runs.

\lemkatcompile*
\begin{proof}
    We first show, by induction on $\kexp$, that $\mathcal{K}_{\kexp}$ has at most $2|\kexp|+2$ states and $3|\kexp|$ transitions.
    \begin{compactitem}
    \item \emph{Tests and base relations.}
    $\mathcal{K}_{\kexp}$ has $2\leq 2|\kexp|+2$ states and $1\leq 3|\kexp|$ transitions.
    \item \emph{Union and composition.}
    Both merge states of $\mathcal{K}_{\kexp_1}$ and $\mathcal{K}_{\kexp_2}$, so they have at most $2|\kexp_1|+2|\kexp_2|+4\leq 2|\kexp|+2$ states and $3|\kexp_1|+3|\kexp_2|\leq 3|\kexp|$ transitions, as $|\kexp|=|\kexp_1|+|\kexp_2|+1$.
    \item \emph{Transitive closure.}
    The construction adds two states and three transitions, and $|\kexp^{+}|=|\kexp|+1$.
    \item \emph{Inverse.}
    The construction adds no states and no transitions.
    \end{compactitem}
    Each construction takes time linear in the number of states and transitions it adds, so $\mathcal{K}_{\kexp}$ is constructed in time $O(|\kexp|)$.

    Now fix an execution graph $G$.
    We show by induction on $\kexp$ that $\tuple{\event,\event'}\in G.\kexp$ iff some $w\in\Lang{\mathcal{K}_{\kexp}}$ has a realization in $G$ from $\event$ to $\event'$.
    \begin{compactitem}
    \item \emph{Tests and base relations.}
    This is the definition of a realization.
    \item \emph{Union.}
    An accepting run of $\mathcal{K}_{\kexp_1\cup \kexp_2}$ stays in one of $\mathcal{K}_{\kexp_1}$ and $\mathcal{K}_{\kexp_2}$, as no transition enters the initial states or leaves the accepting states.
    So, $\Lang{\mathcal{K}_{\kexp_1\cup \kexp_2}}=\Lang{\mathcal{K}_{\kexp_1}}\cup\Lang{\mathcal{K}_{\kexp_2}}$, and the claim follows from $G.(\kexp_1\cup\kexp_2)=G.\kexp_1\cup G.\kexp_2$.
    \item \emph{Composition.}
    An accepting run of $\mathcal{K}_{\kexp_1;\kexp_2}$ visits the merged state exactly once, so $\Lang{\mathcal{K}_{\kexp_1;\kexp_2}}=\Lang{\mathcal{K}_{\kexp_1}}\cdot\Lang{\mathcal{K}_{\kexp_2}}$.
    Realizations of the concatenation $w_1w_2$ from $\event$ to $\event'$ are exactly the realizations of $w_1$ from $\event$ to some $\event''$, followed by realizations of $w_2$ from $\event''$ to $\event'$.
    So, the claim follows from $G.(\kexp_1;\kexp_2)=G.\kexp_1;G.\kexp_2$.
    \item \emph{Transitive closure.}
    The words of $\Lang{\mathcal{K}_{\kexp^{+}}}$ are the words $\LabelDom\,w_1\,\LabelDom\,w_2\cdots\LabelDom\,w_k\,\LabelDom$, for $k\geq 1$ and $w_1,\ldots,w_k\in\Lang{\mathcal{K}_{\kexp}}$.
    Each $\LabelDom$ is realized by staying at the current event, so such a word has a realization from $\event$ to $\event'$ iff there is a chain $\event=\event_0,\ldots,\event_k=\event'$ with $\tuple{\event_{i-1},\event_i}\in G.\kexp$ for each $i$, i.e., iff $\tuple{\event,\event'}\in(G.\kexp)^{+}$.
    \item \emph{Inverse.}
    The words of $\Lang{\mathcal{K}_{\kexp^{-1}}}$ are the reversals of the words of $\Lang{\mathcal{K}_{\kexp}}$, with each base relation symbol inverted.
    A realization of such a word from $\event$ to $\event'$, read backwards, is a realization of the original word from $\event'$ to $\event$, and vice versa.
    So, the claim follows from $G.\kexp^{-1}=(G.\kexp)^{-1}$.
    \end{compactitem}
\end{proof}

\section{Memory Models as $\KAT$ Expressions}\label{SEC:APP_MODELS}

In this section, we write several memory models of the literature as $\KAT$ expressions, and show that they are prefix-closed.
\cref{TAB:APP_MODELS} summarizes them.
We first fix some notation and the axioms that several models share, then give a general sufficient condition for prefix-closedness, and finally treat each model in turn.

\Paragraph{Non-atomic accesses.}
For the C11-style models $\ramm$, $\rcmm$, $\rcelevenmm$ and $\wrcelevenmm$, we omit non-atomic accesses, as these do not affect robustness unless they participate in a data race.
Such races can be handled similarly to previous work~\cite{Lahav2019}.
First, we strengthen every non-atomic access to a release/acquire access, and check robustness with our procedure.
A violation means that the original program is not robust.
Otherwise, every consistent execution is $\scmm$-consistent, so the original program has a data race iff some reachable $\scmm$ state is racy, which can be checked in $\PSPACE$~\cite{Kozen1977,Sistla1985}.

\begin{table}[h]
\centering
\footnotesize
\caption{The memory models of this section.}
\label{TAB:APP_MODELS}
\begin{tabular}{@{}l l l l l@{}}
\toprule
Model & Source & Adaptations & Prefix-closed by & Prior $\PSPACE$ robustness \\
\midrule
$\tsomm$ & \cite{Sewell2010} & -- & \Ax{ghb} & \cite{Bouajjani2013} \\
$\psomm$ & \cite{Alglave2012} & -- & \Ax{ghb} & \cite{Derevenetc2015} \\
$\ramm$ & \cite{Lahav2016} & no non-atomics & \Ax{irr-hb} & \cite{Lahav2019} \\
$\sramm$ & \cite{Lahav2016,Lahav2022} & -- & \Ax{strong-write-coherence} & \cite{Beillahi19} (RMW-free) \\
$\wramm$ & \cite{Lahav2022} & with \Ax{write-coherence} & \Ax{irr-hb} & \cite{Beillahi19} (RMW-free) \\
$\lramm$ & \cite{Singh24} & with \Ax{write-coherence}, tagged values & \Ax{irr-hb} & -- \\
$\rcmm$ & \cite{Margalit2021} & no non-atomics & \Ax{irr-hb} & \cite{Margalit2021} \\
$\rcelevenmm$ & \cite{Lahav2017} & no non-atomics, single-event RMWs & \Ax{no-thin-air} & -- \\
$\wrcelevenmm$ & \cite{Kokologiannakis2018} & \begin{tabular}[t]{@{}l@{}}no non-atomics, single-event RMWs,\\ with \Ax{mo-agreement}\end{tabular} & \Ax{no-thin-air} & -- \\
$\ocamlmm$ & \cite{Dolan2018} & -- & \Ax{causality} & -- \\
$\psimm$ & \cite{Raad18} & unary transactions & \Ax{psi} & -- \\
\bottomrule
\end{tabular}
\end{table}

\Paragraph{Notation.}
Each axiom below is an irreflexivity constraint, and a model is the union of its axioms, as in \cref{SUBSEC:PRELIM_MM}.
For an expression $\kexp$, we let $\kexp|_{\mathsf{loc}}\triangleq\bigcup_{x\in\LocationDom}\identity{\locx{\LabelDom}};\kexp;\identity{\locx{\LabelDom}}$ be its restriction to pairs of events on the same location.
For the C11-style models $\rcmm$ and $\rcelevenmm$, the annotations are access modes $\AnnDom\subseteq\{\mrlx,\mrel,\macq,\macqrel,\msc\}$, and we let $\LabelDom_{\sqsupseteq\mrel}$, $\LabelDom_{\sqsupseteq\macq}$ and $\LabelDom_{\msc}$ be the labels annotated with $\mrel$, $\macqrel$ or $\msc$, with $\macq$, $\macqrel$ or $\msc$, and with $\msc$, respectively.
As annotations are compared up to siblings~(\cref{SUBSEC:APP_PRELIM_PROG}), these sets also contain the reads of failed CAS operations with the respective modes, and likewise for the modes of reads, writes and fences in the models below.

\Paragraph{Shared axioms.}
The following axioms are shared by several models.
{\allowdisplaybreaks
\begin{align*}
  & \hb \tag*{\Ax{irr-hb}} \\
  & \mo;\hb \tag*{\Ax{write-coherence}} \\
  & \mo;\hb;\rf^{-1} \tag*{\Ax{read-coherence}} \\
  & \mo;\mo;\rf^{-1} \tag*{\Ax{atomicity}} \\
  & (\hb\cup\mo)^{+} \tag*{\Ax{strong-write-coherence}} \\
  & \textstyle\bigcup_{x\in\LocationDom}\hb;\identity{\locx{\WriteDom}\cup\locx{\RMWDom}};\hb;\identity{\locx{\LabelDom}};\rf^{-1} \tag*{\Ax{weak-read-coherence}} \\
  & \mo;\identity{\RMWDom};\rf^{-1};\rf;\identity{\RMWDom} \tag*{\Ax{weak-atomicity}}
\end{align*}}%
In words, \Ax{irr-hb} forbids $(\po\cup\rf)$-cycles, \Ax{write-coherence} requires $\mo$ to agree with $\hb$, and \Ax{read-coherence} forbids a read from reading a write that is $\mo$-before another write $\hb$-before the read.
\Ax{atomicity} forbids a write $\mo$-between an RMW and the write it reads from, \Ax{strong-write-coherence} requires $\hb\cup\mo$ to be acyclic, \Ax{weak-read-coherence} forbids a write on $x$ $\hb$-between a write and a read on $x$ that reads from it, and \Ax{weak-atomicity} forbids two RMWs from reading the same write.

\Paragraph{On the encoding of $\fr$.}
In this appendix, we distinguish the two definitions of $\fr$ given in \cref{SUBSEC:PRELIM_MM}.
We write $\fr\triangleq(\rf^{-1};\mo)\setminus\id$ for the standard from-read relation, and $\fr_{\KAT}\triangleq\identity{\ReadDom};\rf^{-1};\mo\cup\identity{\RMWDom};\rf^{-1};\mo;\mo$ for its $\KAT$ definition.
The axioms below are usually stated with $\fr$, and their $\KAT$ expressions use $\fr_{\KAT}$ instead.
We now show that this does not change the consistent graphs of the models we consider.
The two relations agree on reads, as $\mo$ only relates writes and RMWs.
So, consider an RMW $\event$ that reads from an event $\event_w$, both on a location $x$.
As $\mo$ totally orders the writes and RMWs on $x$, there are three cases.
\begin{compactitem}
\item $\event_w\LTo{\mb}\event$. Then both $\fr$ and $\fr_{\KAT}$ relate $\event$ to exactly the events $\mo$-after $\event$.
\item $\event_w\LTo{\mo}\event'\LTo{\mo}\event$ for some $\event'$. This violates \Ax{atomicity}, and we have $\event\LTo{\fr}\event'\LTo{\mo}\event$ and $\tuple{\event,\event}\in\fr_{\KAT}$.
\item $\event\LTo{\mo}\event_w$. Then $\event\LTo{\mo}\event_w\LTo{\rf}\event$ is a cycle of $\mo;\rf$.
\end{compactitem}
So, $\fr$ and $\fr_{\KAT}$ coincide in every graph that satisfies \Ax{atomicity} and in which $\mo;\rf$ is irreflexive.
Hence, replacing $\fr$ by $\fr_{\KAT}$ in a model does not change its consistent graphs, as long as the model forbids, with either relation, the violations of \Ax{atomicity} and the cycles of $\mo;\rf$.
For example, $\scmm$ forbids the cycles of $\rf\cup\mo\cup\fr$, and the last two cases yield such a cycle with either relation.
$\ramm$, $\sramm$, $\wramm$, $\lramm$ and $\wrcelevenmm$ do not use $\fr$, and RMWs do not occur in $\ocamlmm$.
For the other models, we show the condition along with their definitions.

\Paragraph{Prefix-closedness.}
All the models of this section forbid $(\po\cup\rf)$-cycles, and this suffices for prefix-closedness.
We say that an expression $\kexp$ uses $\mb$ \emph{only through $\mo$} if every occurrence of $\mb$ and $\mb^{-1}$ in $\kexp$, after expanding the derived relations, is in a subexpression $\mb^{+}$ or $(\mb^{-1})^{+}$, respectively.
This is the case for all the expressions of this section.

\begin{lemma}\label{LEM:PORF_PREFIX}
    Let $\mm$ be a memory model that uses $\mb$ only through $\mo$, and such that $\po\cup\rf$ is acyclic in every $G\models\mm$.
    Then $\mm$ is prefix-closed.
\end{lemma}
\begin{proof}
    Let $G=\tuple{E,\seqbef,\rf,\mb}\models\mm$ be non-empty.
    As $\po\cup\rf$ is acyclic and $E$ is finite and non-empty, some event $\event$ has no outgoing $\po$- or $\rf$-edge, so $\event$ is maximal.
    Let $G'=G\setminus\{\event\}$.
    Moreover, as $\event$ has no outgoing $\seqbef$- or $\rf$-edge, $G'.\seqbef$, $G'.\po$, $G'.\rf$ and $G'.\mo$ are the restrictions of $G.\seqbef$, $G.\po$, $G.\rf$ and $G.\mo$ to $E\setminus\{\event\}$, and so are their inverses.
    Every operator of $\KAT$ expressions is monotone, so, as $\mm$ uses $\mb$ only through $\mo$, $G'.\mm\subseteq G.\mm$.
    Since $G.\mm$ is irreflexive, so is $G'.\mm$, i.e., $G'\models\mm$.
\end{proof}

Note that the condition on $\mb$ is needed, as removing $\event$ may create an $\mb$-edge that skips over it.
In particular, every model that contains \Ax{irr-hb}, or another axiom whose expression contains $\hb=(\po\cup\rf)^{+}$, is prefix-closed by \cref{LEM:PORF_PREFIX}.

\Paragraph{$\tsomm$ and $\psomm$.}
We take $\tsomm$ from x86-TSO~\cite{Sewell2010}, and $\psomm$ from the SPARC hierarchy of~\cite{Alglave2012}.
In both, a full fence ($\mathsf{mfence}$ on x86, $\mathsf{membar}$ on SPARC) is a fence annotated with $\msc$, and we let $\FenceDom_{\msc}\triangleq\{\llab\in\FenceDom\colon\ann(\llab)=\msc\}$.
On x86, a CAS is a locked instruction, which acts as a full fence whether it succeeds or fails.
So, the full barriers of $\tsomm$ are $\mathsf{FB}\triangleq\FenceDom_{\msc}\cup\RMWDom\cup\FailDom$, where $\FailDom$ consists of the reads of failed CAS operations~(\cref{SUBSEC:APP_PRELIM_PROG}).
For $\mm\in\{\tsomm,\psomm\}$, the derived relations are
\[
\begin{array}{r@{\ }l}
  \mathsf{ppo}_{\tsomm} & \triangleq \big(\identity{\ReadDom};\po\cup\po;\identity{\WriteDom}\cup\identity{\mathsf{FB}};\po\cup\po;\identity{\mathsf{FB}}\big)^{+} \\
  \mathsf{ppo}_{\psomm} & \triangleq (\identity{\ReadDom\cup\RMWDom};\po)\cup(\identity{\WriteDom\cup\RMWDom};\po;\identity{\FenceDom_{\msc}};\po) \\
  \mathsf{ghb}_{\mm} & \triangleq \mathsf{ppo}_{\mm}\cup\rfe\cup\mo\cup\fr
\end{array}
\]
In words, $\mathsf{ppo}_{\tsomm}$ preserves the program order between any two events, except from a write to a later read, unless one of the two is a full barrier or a full barrier lies between them.
$\mathsf{ppo}_{\psomm}$ preserves the program order from every read and RMW, but orders a write before later events only through a full fence.
The axioms are
\begin{align*}
  & (\po|_{\mathsf{loc}}\cup\rf\cup\mo\cup\fr)^{+} \tag*{\Ax{uniproc}} \\
  & \mathsf{ghb}_{\mm}^{+} \tag*{\Ax{ghb}}
\end{align*}
So, $\mm\triangleq\Ax{uniproc}\cup\Ax{ghb}$.
As for $\scmm$, \Ax{uniproc} forbids the cycles of $\rf\cup\mo\cup\fr$, and hence the violations of \Ax{atomicity} and the cycles of $\mo;\rf$, with either $\fr$ or $\fr_{\KAT}$.
Both models forbid $(\po\cup\rf)$-cycles.
Indeed, such a cycle contains an $\rfe$-edge, as $\rf$-edges within a thread follow $\po$ by \Ax{uniproc}.
Between two consecutive $\rfe$-edges, the cycle follows a $\po$-path that starts at a read or an RMW, which is in $\mathsf{ppo}_{\mm}$.
So, the cycle yields a cycle of $\mathsf{ghb}_{\mm}$, and $\tsomm$ and $\psomm$ are prefix-closed by \cref{LEM:PORF_PREFIX}.

\Paragraph{$\ramm$.}
Following~\cite{Lahav2016,Lahav2019}, $\ramm\triangleq\Ax{irr-hb}\cup\Ax{write-coherence}\cup\Ax{read-coherence}\cup\Ax{atomicity}$, as in \cref{SUBSEC:PRELIM_MM}.

\Paragraph{$\sramm$.}
Following~\cite{Lahav2016,Lahav2022}, $\sramm\triangleq\Ax{strong-write-coherence}\cup\Ax{read-coherence}\cup\Ax{atomicity}$.
Without RMWs, $\sramm$ coincides with Causal Convergence~\cite{Lahav2022}.
As $\hb\subseteq(\hb\cup\mo)^{+}$, it is prefix-closed by \cref{LEM:PORF_PREFIX}.

\Paragraph{$\wramm$.}
$\wramm$~\cite{Lahav2022} is given by \Ax{irr-hb}, \Ax{weak-read-coherence} and \Ax{weak-atomicity}, and, without RMWs, it coincides with Causal Consistency~\cite{Lahav2022}.
These axioms do not constrain $\mo$, apart from the distinctness of two RMWs in \Ax{weak-atomicity}, which holds for either order of the two.
In our setting, however, $\mb$ is part of every execution graph, and $\scmm$ does constrain it.
So, robustness against these three axioms alone is nearly trivial: if some $\scmm$-consistent graph of $\ConcProg$ has two writes $\event_1\LTo{\po}\event_2$ on the same location, then swapping $\event_1$ and $\event_2$ in $\mb$ preserves the three axioms, but creates the $\scmm$-cycle $\event_1\LTo{\po}\event_2\LTo{\mo}\event_1$, so $\ConcProg$ is not robust.
We therefore let
\[
  \wramm\triangleq\Ax{irr-hb}\cup\Ax{write-coherence}\cup\Ax{weak-read-coherence}\cup\Ax{weak-atomicity}.
\]
Adding \Ax{write-coherence} does not change which $\tuple{\Events,\seqbef,\rf}$ are allowed.
Indeed, given a graph that satisfies the three axioms, we obtain one that also satisfies \Ax{write-coherence} by replacing, for each location $x$, the $\mb$-edges on $x$ by the successor relation of a total order on the writes and RMWs on $x$ that extends $\hb$; such an order exists as $\hb$ is acyclic by \Ax{irr-hb}.
The other three axioms still hold, as \Ax{irr-hb} and \Ax{weak-read-coherence} do not mention $\mo$, and \Ax{weak-atomicity} holds for either order.
So, \Ax{write-coherence} only rules out choices of $\mb$ that contradict $\hb$, which would otherwise yield trivial violations, such as $\mo$-edges from a write to a $\po$-earlier write on the same location.

\Paragraph{$\lramm$.}
$\lramm$~\cite{Singh24} strengthens $\wramm$ with the irreflexivity of
\[
  (\hb|_{\mathsf{loc}}\setminus\rf);\identity{\ReadDom\cup\RMWDom};\hb;\rf^{-1}.
\]
In words, a read $\event_r$ may not read from a write $\event_w$ on $x$ if some read $\event_r'$ on $x$, with $\event_w\LTo{\hb}\event_r'\LTo{\hb}\event_r$, reads from a write other than $\event_w$.
As for $\wramm$, we include \Ax{write-coherence}, and the argument above applies verbatim, as the new axiom does not mention $\mo$.
The difference with $\rf$ cannot be expressed directly.
Instead, we extend the value domain to $\ThreadDom\times\ValueDom$, so that every value also records the thread that wrote it, and we modify the LTSs accordingly (for programs given as code, the rules of \cref{SUBSEC:APP_PRELIM_PROG}).
A write of thread $t$ writes $\tuple{t,v}$ instead of $v$, and a read may observe $\tuple{t',v}$ for any thread $t'$, but only uses $v$.
So, the label of a read tells us which thread it reads from.
This does not affect robustness, and the number of labels only grows by a factor of $|\ThreadDom|$, so the bounds of \cref{SEC:AUT} still apply.

For $t\in\ThreadDom$ and $x\in\LocationDom$, let $\ReadDom_x^{\neq t}$ be the read and RMW labels on $x$ whose read value records a thread other than $t$.
The new axiom is
\begin{align*}
  & \textstyle\bigcup_{t\in\ThreadDom}\bigcup_{x\in\LocationDom}\identity{(\locx{\WriteDom}\cup\locx{\RMWDom})\cap\LabelDom^{t}};\hb;\identity{\ReadDom_x^{\neq t}};\hb;\rf^{-1} \tag*{\Ax{local-read-coherence}}
\end{align*}
and $\lramm\triangleq\wramm\cup\Ax{local-read-coherence}$.
We show that a graph $G$ satisfies the definition of~\cite{Singh24} together with \Ax{write-coherence} iff $\lramm$ holds in the corresponding graph of the modified program, where each read observes the thread of its $\rf$-predecessor.
If $G$ violates $\wramm$, then both sides fail.
So, assume that $G\models\wramm$.
Suppose that $G$ violates the axiom of~\cite{Singh24}, i.e., there are $\event_w$, $\event_r'$ and $\event_r$ as above, where $\event_r'$ reads from some $\event_w'\neq \event_w$.
Both $\event_w$ and $\event_w'$ write to $x$.
If $\event_w\LTo{\hb}\event_w'$, then $\event_w\LTo{\hb|_{\mathsf{loc}}}\event_w'\LTo{\hb}\event_r$, as $\event_w'\LTo{\rf}\event_r'\LTo{\hb}\event_r$, and this violates \Ax{weak-read-coherence}, since $\event_r$ reads from $\event_w$.
Similarly, if $\event_w'\LTo{\hb}\event_w$, then $\event_w'\LTo{\hb|_{\mathsf{loc}}}\event_w\LTo{\hb}\event_r'$, which violates \Ax{weak-read-coherence}, since $\event_r'$ reads from $\event_w'$.
So, $\event_w$ and $\event_w'$ are $\hb$-incomparable, and hence they belong to different threads, as $\po\subseteq\hb$.
So, in the corresponding graph, the label of $\event_r'$ is in $\ReadDom_x^{\neq t}$ for the thread $t$ of $\event_w$, and \Ax{local-read-coherence} has a cycle through $\event_w$.
Conversely, a cycle of \Ax{local-read-coherence} yields $\event_w$, $\event_r'$ and $\event_r$ as above, where $\event_r'$ reads from a thread other than the one of $\event_w$, and so from a write other than $\event_w$.

\Paragraph{$\rcmm$.}
$\rcmm$~\cite{Margalit2021} has the access modes $\AnnDom=\{\mrlx,\mrel,\macq,\macqrel\}$.
Reads are $\mrlx$ or $\macq$, writes are $\mrlx$ or $\mrel$, and RMWs and fences have any mode, except that fences are not $\mrlx$.
The derived relations are
\[
  \sw\triangleq\identity{\LabelDom_{\sqsupseteq\mrel}};(\identity{\FenceDom};\po)^{?};\rf^{+};(\po;\identity{\FenceDom})^{?};\identity{\LabelDom_{\sqsupseteq\macq}}
  \qquad\text{and}\qquad
  \hb_{\rcmm}\triangleq(\po\cup\sw)^{+}.
\]
The axioms are
\begin{align*}
  & \mo;\rf^{?};\hb_{\rcmm}^{?} \tag*{\Ax{rc-write-coherence}} \\
  & \fr;\rf^{?};\hb_{\rcmm} \tag*{\Ax{rc-read-coherence}}
\end{align*}
So, $\rcmm\triangleq\Ax{irr-hb}\cup\Ax{atomicity}\cup\Ax{rc-write-coherence}\cup\Ax{rc-read-coherence}$.
It forbids the violations of \Ax{atomicity} directly, and the cycles of $\mo;\rf$ by \Ax{rc-write-coherence}, and neither axiom involves $\fr$.
Its release/acquire fragment, with only $\macq$ reads, $\mrel$ writes and $\macqrel$ RMWs, and no fences, is $\ramm$~\cite{Margalit2021}.

\Paragraph{$\rcelevenmm$.}
We consider the fragment of $\rcelevenmm$~\cite{Lahav2017} without non-atomic accesses, as their races make the whole program undefined, which is not a constraint on single graphs.
We also omit the initialization events of~\cite{Lahav2017}, which precede all other events in $\po$, as our graphs have none.
The modes are $\AnnDom=\{\mrlx,\macq,\mrel,\macqrel,\msc\}$, where reads are $\mrlx$, $\macq$ or $\msc$, writes are $\mrlx$, $\mrel$ or $\msc$, and fences are not $\mrlx$.
Lahav et al.~\cite{Lahav2017} represent an RMW by a read and a write related by $\mathsf{rmw}$, with five allowed pairs of modes, which correspond to our five modes.
With a single event, $\rf;\mathsf{rmw}$ becomes $\rf;\identity{\RMWDom}$, and their atomicity axiom becomes \Ax{atomicity}.
Moreover, the $\mathsf{rmw}$-edge from the read to the write of an RMW is in $\hb_{\rcelevenmm}$, and it disappears, so \Ax{coherence} also requires $\eco$ to be irreflexive.
The derived relations are
\[
\begin{array}{r@{\ }l}
  \po|_{\neq\mathsf{loc}} & \triangleq \textstyle\bigcup_{x\neq y}(\identity{\locx{\LabelDom}};\po;\identity{\LabelDom_{y}})\ \cup\ (\identity{\FenceDom};\po)\ \cup\ (\po;\identity{\FenceDom}) \\
  \mathsf{rs} & \triangleq \identity{\WriteDom\cup\RMWDom};\po|_{\mathsf{loc}}^{?};\identity{\WriteDom\cup\RMWDom};(\rf;\identity{\RMWDom})^{*} \\
  \sw & \triangleq \identity{\LabelDom_{\sqsupseteq\mrel}};(\identity{\FenceDom};\po)^{?};\mathsf{rs};\rf;(\po;\identity{\FenceDom})^{?};\identity{\LabelDom_{\sqsupseteq\macq}} \\
  \hb_{\rcelevenmm} & \triangleq (\po\cup\sw)^{+} \\
  \eco & \triangleq (\rf\cup\mo\cup\fr)^{+} \\
  \mathsf{scb} & \triangleq \po\cup(\po|_{\neq\mathsf{loc}};\hb_{\rcelevenmm};\po|_{\neq\mathsf{loc}})\cup\hb_{\rcelevenmm}|_{\mathsf{loc}}\cup\mo\cup\fr \\
  \mathsf{psc} & \triangleq \bigl(\identity{\LabelDom_{\msc}}\cup(\identity{\LabelDom_{\msc}\cap\FenceDom};\hb_{\rcelevenmm}^{?})\bigr);\mathsf{scb};\bigl(\identity{\LabelDom_{\msc}}\cup(\hb_{\rcelevenmm}^{?};\identity{\LabelDom_{\msc}\cap\FenceDom})\bigr) \\
  & \phantom{\triangleq}\ \cup\ \bigl(\identity{\LabelDom_{\msc}\cap\FenceDom};\bigl(\hb_{\rcelevenmm}\cup(\hb_{\rcelevenmm};\eco;\hb_{\rcelevenmm})\bigr);\identity{\LabelDom_{\msc}\cap\FenceDom}\bigr)
\end{array}
\]
where $\po|_{\neq\mathsf{loc}}$ equals $\po\setminus\po|_{\mathsf{loc}}$ of~\cite{Lahav2017}, since a fence has no location.
The axioms are
\begin{align*}
  & (\hb_{\rcelevenmm};\eco^{?})\cup\eco \tag*{\Ax{coherence}} \\
  & \mathsf{psc}^{+} \tag*{\Ax{sc}} \\
  & (\po\cup\rf)^{+} \tag*{\Ax{no-thin-air}}
\end{align*}
So, $\rcelevenmm\triangleq\Ax{coherence}\cup\Ax{atomicity}\cup\Ax{sc}\cup\Ax{no-thin-air}$, which is prefix-closed by \cref{LEM:PORF_PREFIX}.
It forbids the violations of \Ax{atomicity} directly, and the cycles of $\mo;\rf$ by \Ax{coherence}, as $\mo;\rf\subseteq\eco$ with either $\fr$ or $\fr_{\KAT}$.

\Paragraph{$\wrcelevenmm$.}
$\wrcelevenmm$~\cite{Kokologiannakis2018} weakens $\rcelevenmm$ by not recording a modification order: every use of $\mo$ is replaced by a relation derived from $\hb_{\rcelevenmm}$ and $\rf$.
We make the same adaptations as for $\rcelevenmm$, and take $\mathsf{rs}$, $\sw$ and $\hb_{\rcelevenmm}$ as there.
The derived relations are
\[
\begin{array}{r@{\ }l}
  \mo_{w} & \triangleq \textstyle\bigcup_{x\in\LocationDom}\identity{\locx{\WriteDom}\cup\locx{\RMWDom}};\bigl(\hb_{\rcelevenmm}\cup(\identity{\locx{\LabelDom}};\rf)\bigr)^{+};\identity{\locx{\WriteDom}\cup\locx{\RMWDom}} \\
  \fr_{w} & \triangleq \identity{\ReadDom};\rf^{-1};\mo_{w}\ \cup\ \textstyle\bigcup_{t\in\ThreadDom}\identity{\RMWDom\cap\LabelDom^{t}};\rf^{-1};\mo_{w};\identity{\LabelDom\setminus\LabelDom^{t}} \\
  \eco_{w} & \triangleq (\rf\cup\mo_{w}\cup\fr_{w})^{+}
\end{array}
\]
and $\mathsf{scb}_{w}$ and $\mathsf{psc}_{w}$ are $\mathsf{scb}$ and $\mathsf{psc}$ with $\mo$, $\fr$ and $\eco$ replaced by $\mo_{w}$, $\fr_{w}$ and $\eco_{w}$.
Note that $\mo_{w}$ only follows $\rf$-edges on $x$, as in~\cite{Kokologiannakis2018}, and is thus in general smaller than the restriction of $(\po\cup\rf)^{+}$ to the writes and RMWs on $x$.
The from-read relation of~\cite{Kokologiannakis2018} is $(\rf^{-1};\mo_{w})\setminus\id$.
For an RMW $\event$ of thread $t$, $\fr_{w}$ omits its edges to all events of $t$, and not only to $\event$ itself, which $\KAT$ cannot single out.
The omitted edges do not matter.
An edge from $\event$ to a $\po$-later event is also an $\mo_{w}$-edge, and both relations occur in the same places in the axioms below.
An edge from $\event$ to a $\po$-earlier event $\event'$ yields $\event_w\LTo{\mo_{w}}\event'\LTo{\mo_{w}}\event$ for the event $\event_w$ that $\event$ reads from, which \Ax{w-atomicity} forbids.
The axioms are
\begin{align*}
  & (\hb_{\rcelevenmm};\eco_{w}^{?})\cup\eco_{w} \tag*{\Ax{w-coherence}} \\
  & \mo_{w};\mo_{w};\rf^{-1} \tag*{\Ax{w-atomicity}} \\
  & \mathsf{psc}_{w}^{+} \tag*{\Ax{w-sc}} \\
  & \mo;\mo_{w} \tag*{\Ax{mo-agreement}}
\end{align*}
So,
\[
  \begin{array}{r@{\ }l}
  \wrcelevenmm\triangleq & \Ax{w-coherence}\cup\Ax{w-atomicity}\cup\Ax{w-sc} \\
  & \cup\ \Ax{no-thin-air}\cup\Ax{weak-atomicity}\cup\Ax{mo-agreement},
  \end{array}
\]
which is prefix-closed by \cref{LEM:PORF_PREFIX}.
As for $\wramm$, the axioms of~\cite{Kokologiannakis2018} do not constrain $\mo$, and we add \Ax{mo-agreement} so that $\mo$ extends $\mo_{w}$.
This does not change which $\tuple{\Events,\seqbef,\rf}$ are allowed.
Indeed, $\hb_{\rcelevenmm}\subseteq(\po\cup\rf)^{+}$, so $\mo_{w}$ is acyclic by \Ax{no-thin-air}, and we can replace, for each location $x$, the $\mb$-edges on $x$ by the successor relation of a total order on the writes and RMWs on $x$ that extends $\mo_{w}$.
The other axioms still hold, as only \Ax{weak-atomicity} mentions $\mo$, and it holds for either order.

\Paragraph{$\ocamlmm$.}
We take the $\ocamlmm$ memory model of~\cite{Dolan2018}, which has no RMWs.
We fix a set $\LocationDom_{\mathsf{at}}\subseteq\LocationDom$ of atomic locations, and we let $\LabelDom_{\mathsf{at}}$ be the labels on atomic locations.
The derived relation is
\[
  \hb_{\ocamlmm}\triangleq\bigl(\po\cup(\identity{\LabelDom_{\mathsf{at}}};(\mo\cup\rf))\bigr)^{+},
\]
where we omit the initial writes of~\cite{Dolan2018}, as our graphs have none.
The axioms are
\begin{align*}
  & \bigl(\hb_{\ocamlmm}\cup\rf\cup(\identity{\LabelDom_{\mathsf{at}}};\fr)\bigr)^{+} \tag*{\Ax{causality}} \\
  & \hb_{\ocamlmm};\mo \tag*{\Ax{CoWW}} \\
  & \hb_{\ocamlmm};\fr \tag*{\Ax{CoWR}}
\end{align*}
So, $\ocamlmm\triangleq\Ax{causality}\cup\Ax{CoWW}\cup\Ax{CoWR}$.
\Ax{causality} contains $(\po\cup\rf)^{+}$, and so $\ocamlmm$ is prefix-closed by \cref{LEM:PORF_PREFIX}.

\Paragraph{$\psimm$.}
We consider Parallel Snapshot Isolation~\cite{Raad18} with unary transactions, where every event is its own transaction.
Then, lifting a relation to transactions leaves it unchanged, and by~\cite[Lemma~1]{Raad18}, $\psimm$ has the derived relation $\hb_{\psimm}\triangleq(\po\cup\rf\cup\mo)^{+}$ and the single axiom
\begin{align*}
  & (\hb_{\psimm}|_{\mathsf{loc}}\cup\mo\cup\fr)^{+} \tag*{\Ax{psi}}
\end{align*}
So, $\psimm\triangleq\Ax{psi}$.
As $\rf\subseteq\hb_{\psimm}|_{\mathsf{loc}}$, \Ax{psi} forbids the cycles of $\rf\cup\mo\cup\fr$, and hence, as for $\scmm$, the violations of \Ax{atomicity} and the cycles of $\mo;\rf$, with either $\fr$ or $\fr_{\KAT}$.
Every $(\po\cup\rf)$-cycle contains an $\rf$-edge, and hence an event $\event$ with a location, for which $\tuple{\event,\event}\in\hb_{\psimm}|_{\mathsf{loc}}$.
So, $\psimm$ forbids $(\po\cup\rf)$-cycles, and it is prefix-closed by \cref{LEM:PORF_PREFIX}.

\section{Proofs of \cref{SEC:PW}}\label{SEC:APP_PW}

\lempwscreduced*

\begin{proof}
This proof uses the same construction as \cite{Conrado26}. 
By definition of $\scmm$~(\cref{SUBSEC:PRELIM_MM}), $G\models\scmm$ iff $\po\cup\rf\cup\mo\cup\fr$ is acyclic in $G$.
So, we can fix a linear extension $\sigma=\event_1,\event_2,\dots,\event_n$ of it.
We write $\event\prec\event'$ when $\event$ precedes $\event'$ in $\sigma$.

For $i\in\IntSet{n}$, the bag $\Bag_i$ contains $\event_i$, together with, for each thread $t\in\ThreadDom$, the last event of $t$ before $\event_i$ in $\sigma$, and, for each location $x\in\LocationDom$, the last write or RMW on $x$ before $\event_i$ in $\sigma$, whenever these exist.
So, $|\Bag_i|\leq 1+|\ThreadDom|+|\LocationDom|$, and the width is at most $|\ThreadDom|+|\LocationDom|$.
It remains to show that $\Bag_1,\Bag_2,\dots,\Bag_n$ is a path decomposition of $G$, i.e., that every event lies in a contiguous range of bags, and that the endpoints of every edge share a bag.

First, we show that every event lies in a contiguous range of bags.
Consider an event $\event_i$.
It first appears in $\Bag_i$.
It then stays in the bags as the last event of its thread, up to the bag of the next event of its thread.
If $\event_i$ is a write or RMW, it also stays as the last write or RMW on its location, up to the bag of the next write or RMW on that location, or up to $\Bag_n$ if there is none.
Both ranges start at $\Bag_i$, so their union is contiguous.

Next, we show that the endpoints of every edge share a bag.
Let $\event\LTo{\seqbef}\event'$, with $\event'=\event_{i'}$.
Then $\event$ is the event of its thread that comes right before $\event'$ in $\po$.
Since $\sigma$ extends $\po$, $\event$ is also the last event of its thread before $\event'$ in $\sigma$.
So, $\event\in\Bag_{i'}$, and both endpoints lie in $\Bag_{i'}$.
The case of $\mb$-edges is the same, with $\mo$ in place of $\po$, and the writes and RMWs on a location in place of the events of a thread.

Finally, for $\rf$-edges, let $\event\LTo{\rf}\event'$, with $\event'=\event_{i'}$ and $x\triangleq\lloc(\event)$.
We show that $\event$ is the last write or RMW on $x$ before $\event'$ in $\sigma$, so that both endpoints lie in $\Bag_{i'}$.
Suppose, towards a contradiction, that some write or RMW $\event''$ on $x$ lies in between, i.e., $\event\prec\event''\prec\event'$.
Since $\mo$ totally orders the writes and RMWs on $x$, and $\sigma$ extends $\mo$, we have $\event\LTo{\mo}\event''$.
If $\event'$ is a read, then $\event'\LTo{\fr}\event''$, and since $\sigma$ extends $\fr$, we get $\event'\prec\event''$.
This contradicts $\event''\prec\event'$.
Otherwise, $\event'$ is an RMW on $x$, and for the same reason we have $\event''\LTo{\mo}\event'$.
So, $\event\LTo{\mo}\event''\LTo{\mo}\event'$, and hence $\event'\LTo{\fr}\event'$.
This is a cyclic realization of $\fr$, which contradicts $G\models\scmm$.
\end{proof}

\lemprefixborderline*
\begin{proof}
Fix a prefix-closed memory model $\mm$, and a program $\ConcProg$ that is not robust against $\mm$.
Let $G$ be a witness for the non-$\mm$-robustness of $\ConcProg$ with the fewest events.
We show that $\pw(G)\leq\pwSC+1$.

Since $G$ is a witness, $G\not\models\scmm$, so $G$ is non-empty.
By prefix-closedness, some maximal event $\event$ of $G$ satisfies $G\setminus\{\event\}\in\ExecutionGraphsOf{\ConcProg}{\mm}$.
Since $G\setminus\{\event\}$ has fewer events than $G$, it is not a witness.
So, $G\setminus\{\event\}\models\scmm$, and $\pw(G\setminus\{\event\})\leq\pwSC$ by \cref{LEM:PW_SC_REDUCED}.
It remains to add $\event$ back.
Let $H$ be the graph obtained from $G$ by restricting its events and its three relations to $G.\Events\setminus\{\event\}$.
Note that $H$ and $G\setminus\{\event\}$ agree on $\seqbef$ and $\rf$.
On $\mb$, the graph $G\setminus\{\event\}$ may additionally contain the edge that bridges over $\event$, due to the transitive reduction.
So, every edge of $H$ is an edge of $G\setminus\{\event\}$, every path decomposition of $G\setminus\{\event\}$ is one of $H$, and $\pw(H)\leq\pwSC$.
Finally, adding $\event$ to every bag of a path decomposition of $H$ gives a path decomposition of $G$.
Hence, $\pw(G)\leq\pw(H)+1\leq\pwSC+1$.
\end{proof}

\section{Proofs of \cref{SEC:ALGEBRA}}\label{SEC:APP_ALGEBRA}

\lempathwidth*
\begin{proof}
    For the first part, let $w$ be closed, and consider the bags $\Bag^{w_i} = \dom{\chi^{w_i}}$.
    Each bag holds at most $k+1$ events, as $\chi^{w_i}$ is injective, and $\Bag^{w_0}=\Bag^{w_{|w|}}=\emptyset$, as $w_0=\varepsilon$ and $w$ is closed.
    We label $\Bag^{w_i}$ by the step corresponding to the $i$-th operation of $w$.
    An operation $\CFEIntro(\llab,c)$ adds a new event $\event$ to the bag, which is the step Introduce $\event$.
    An operation $\CFEForget(c)$ removes $\chi^{-1}(c)$ from the bag, which is the step Forget $\chi^{-1}(c)$.
    An operation $\CFEIntroR{\rsym}(c_1,c_2)$ leaves the bag unchanged and adds a new edge $\chi^{-1}(c_1)\LTo{\rsym}\chi^{-1}(c_2)$ whose endpoints are both in the bag, which is the step Introduce of that edge.
    Every event and every edge of $G^w$ is added by exactly one operation, so each of them is introduced exactly once.
    Finally, an event stays in the bags from its $\CFEIntro$ until its color is forgotten, after which nothing colors it again, so it occupies a contiguous range of bags.

    For the converse, let $\Bag_1, \ldots, \Bag_n$ be a very nice path decomposition of $G$ of width at most $k$.
    We translate its steps in order, keeping an injective map $\chi$ from the current bag to $\IntSet{k+1}$.
    Introducing $\event$ becomes $\CFEIntro(\llab(\event), c)$ for some inactive $c$, which exists because the bag \emph{before} the step has at most $k$ events.
    Forgetting $\event$ becomes $\CFEForget(\chi(\event))$.
    Introducing an edge $\event_1\LTo{\rsym}\event_2$ becomes $\CFEIntroR{\rsym}(\chi(\event_1), \chi(\event_2))$.
    The preconditions hold by construction, where each drawn edge is new as every edge of $G$ is introduced exactly once, and $\Bag^{w_i}=\Bag_{i+1}$ for every $i$.
    As every event and every edge of $G$ is introduced exactly once, $G^w=G$, up to the renaming of identifiers of \cref{SUBSEC:ALGEBRA_BASE}.
    Finally, $w$ is closed, as $\Bag_n$ is empty.
\end{proof}

\section{Proofs of \cref{SEC:AUT}}\label{SEC:APP_AUT}

This section follows the structure of \cref{SEC:AUT}.
For each automaton, we first describe the state it reaches after reading a prefix of the word, in the form of an invariant.
We then derive the correctness of the automaton from this invariant.

\subsection{Proofs of \cref{SUBSEC:AUT_WF}}\label{SUBSEC:APP_AUT_WF}

In this subsection, we prove \cref{LEM:AUT_WF}, i.e., that $\mathcal{A}^{k}_{\textsc{wf}}$ recognizes the words that induce execution graphs.
We treat its four components in turn, and then their product.

\Paragraph{Closedness of the word.}
We start with $\mathcal{A}^{k}_{\textsc{form}}$, whose states $\tuple{\ActSet,\EdgSet}$ record the active colors and the edges drawn between their events.
Its initial and unique accepting state is $\tuple{\emptyset,\emptyset}$, and its rules are the following.
\begin{mathparpagebreakable}
\inferrule{
  \mathrm{op} = \CFEIntro(\llab,c)\quad c \not\in \ActSet
}{
  \tuple{\ActSet,\EdgSet} \xrightarrow{\mathrm{op}} \tuple{\ActSet \cup \{c\},\EdgSet}
}
\and
\inferrule{
  \mathrm{op} = \CFEForget(c)\quad c \in \ActSet \\\\
  \EdgSet' = \{\tuple{\rsym,c_1,c_2}\in\EdgSet \mid c\notin\{c_1,c_2\}\}
}{
  \tuple{\ActSet,\EdgSet} \xrightarrow{\mathrm{op}} \tuple{\ActSet \setminus \{c\},\EdgSet'}
}
\and
\inferrule{
  \mathrm{op} = \CFEIntroR{\rsym}(c_1,c_2)\quad c_1,c_2 \in \ActSet \quad \tuple{\rsym,c_1,c_2}\notin\EdgSet
}{
  \tuple{\ActSet,\EdgSet} \xrightarrow{\mathrm{op}} \tuple{\ActSet,\EdgSet\cup\{\tuple{\rsym,c_1,c_2}\}}
}
\end{mathparpagebreakable}
\begin{restatable}{lemma}{lemautform}\label{LEM:AUT_FORM}
  For every $k\in\Nats$, there is a deterministic word automaton $\mathcal{A}^{k}_{\textsc{form}}$ with $2^{O(k^2)}$ states that accepts $w\in\Ops(k+1)^{*}$ iff $w$ is closed.
\end{restatable}

\begin{proof}
    A state is a subset $\ActSet$ of $\IntSet{k+1}$ together with a subset $\EdgSet$ of $\{\seqbef,\rf,\mb\}\times\IntSet{k+1}^2$, so $\mathcal{A}^{k}_{\textsc{form}}$ has at most $2^{k+1}\cdot 2^{3(k+1)^2}=2^{O(k^2)}$ states.
    At most one rule applies to each pair of a state and a letter, so it is deterministic.
    By induction on $|w|$, if $\mathcal{A}^{k}_{\textsc{form}}$ has a run on $w$, then $w$ is well-formed and the state it reaches is $\tuple{\img{\chi^{w}},\EdgSet}$, where $\tuple{\rsym,c_1,c_2}\in\EdgSet$ iff $\tuple{\chi^{-1}(c_1),\chi^{-1}(c_2)}\in G^{w}.\rsym$.
    For $\CFEIntro$, the new event has no edges.
    For $\CFEIntroR{\rsym}$, the drawn edge is added on both sides.
    For $\CFEForget(c)$, the edges that mention $c$ are discarded, and the other active colors keep their events.
    So, each rule is the precondition of its operation in \cref{SUBSEC:ALGEBRA_BASE} read on the state, and a run exists exactly when every operation of $w$ meets its precondition, i.e., when $w$ is well-formed.
    The unique accepting state is $\tuple{\emptyset,\emptyset}$, which by the same induction says $\dom{\chi^{w}}=\emptyset$, i.e., that $w$ is closed.
\end{proof}

\Paragraph{Well-formedness of $\seqbef$.}
The automaton $\mathcal{A}^{k}_{\seqbef}$ is more involved.
Throughout, $w$ is well-formed, $G_c^{w}=\tuple{E,\seqbef,\rf,\mb,\chi}$, and we call an event colored when it lies in $\dom{\chi}$.
The \emph{components} of $\seqbef$ are the connected components of the graph on $E$ whose edges are the $\seqbef$-edges taken in either direction, so that two events are connected iff they lie in the same component.
We first note that the $\seqbef$-edges of an uncolored event are final, as $\CFEIntrosb$ only draws edges between colored events, and a forgotten event is never colored again.

\SubParagraph{Transition rules.}
When an edge joins two components, their colors take the smaller of the two representatives, and when a representative is forgotten, the colors it represents take the smallest among them.
Formally, $\textsc{merge}(c_1,c_2,\RepMap)$ maps every $c$ with $\RepMap(c)\in\{\RepMap(c_1),\RepMap(c_2)\}$ to $\min\{\RepMap(c_1),\RepMap(c_2)\}$, and every other $c\in\dom{\RepMap}$ to $\RepMap(c)$.
Similarly, $\textsc{drop}(c,\RepMap)$ is defined on $\dom{\RepMap}\setminus\{c\}$, maps every $c'$ with $\RepMap(c')=c$ to $\min\{c''\neq c\mid \RepMap(c'')=c\}$, and every other $c'$ to $\RepMap(c')$.
The initial state is $\tuple{\emptyset,\emptyset,\emptyset,\emptyset,\emptyset,\emptyset}$, and every state is accepting.
In each rule below, the components that are not mentioned are unchanged.
\begin{mathparpagebreakable}
\inferrule{
  \mathrm{op} = \CFEIntro(\llab,c) \quad
  \RepMap' = \RepMap \cup \{c \mapsto c\} \quad
  \ThrMap' = \ThrMap \cup \{c \mapsto \tid(\llab)\}
}{
  \tuple{\InSet,\OutSet,\FstSet,\LstSet,\RepMap,\ThrMap} \xrightarrow{\mathrm{op}} \tuple{\InSet',\OutSet',\FstSet',\LstSet',\RepMap',\ThrMap'}
}
\and
\inferrule{
  \mathrm{op} = \CFEForget(c) \quad t=\ThrMap(c) \quad
  c\notin\InSet \Rightarrow t\notin\FstSet \quad
  c\notin\OutSet \Rightarrow t\notin\LstSet \\\\
  \FstSet' = \FstSet\cup\{t\mid c\notin\InSet\} \quad
  \LstSet' = \LstSet\cup\{t\mid c\notin\OutSet\}\quad
  \InSet'=\InSet\setminus\{c\} \quad
  \OutSet'=\OutSet\setminus\{c\} \\\\
  \RepMap'=\textsc{drop}(c,\RepMap) \quad
  \ThrMap'=\ThrMap\setminus\{c\mapsto t\}
}{
  \tuple{\InSet,\OutSet,\FstSet,\LstSet,\RepMap,\ThrMap} \xrightarrow{\mathrm{op}} \tuple{\InSet',\OutSet',\FstSet',\LstSet',\RepMap',\ThrMap'}
}
\and
\inferrule{
  \mathrm{op} = \CFEIntrosb(c_1,c_2)\quad
  \ThrMap(c_1)=\ThrMap(c_2) \quad
  c_1\notin\OutSet \quad c_2\notin\InSet \quad
  \RepMap(c_1)\neq\RepMap(c_2) \\\\
  \OutSet'=\OutSet\cup\{c_1\} \quad
  \InSet'=\InSet\cup\{c_2\} \quad
  \RepMap'=\textsc{merge}(c_1,c_2,\RepMap)
}{
  \tuple{\InSet,\OutSet,\FstSet,\LstSet,\RepMap,\ThrMap} \xrightarrow{\mathrm{op}} \tuple{\InSet',\OutSet',\FstSet',\LstSet',\RepMap',\ThrMap'}
}
\and
\inferrule{
  \mathrm{op} = \CFEIntroR{\rsym}(c_1,c_2) \quad \rsym \in \{\rf,\mb\}
}{
  \tuple{\InSet,\OutSet,\FstSet,\LstSet,\RepMap,\ThrMap} \xrightarrow{\mathrm{op}} \tuple{\InSet,\OutSet,\FstSet,\LstSet,\RepMap,\ThrMap}
}
\end{mathparpagebreakable}
\SubParagraph{Invariant.}
The following lemma describes the state that $\mathcal{A}^{k}_{\seqbef}$ reaches after reading $w$.

\begin{restatable}{lemma}{lemappautsbinv}\label{LEM:APP_AUT_SB_INV}
    Let $w$ be well-formed, and suppose that $\mathcal{A}^{k}_{\seqbef}$ has a run on $w$.
    Then the state
    \[\tuple{\InSet,\OutSet,\FstSet,\LstSet,\RepMap,\ThrMap}\]
    that it reaches satisfies the following.
    \begin{compactenum}
        \item\label{ITEM:INV_DEG} for every active color $c$, $c\in\InSet$ iff $\chi^{-1}(c)$ has an incoming $\seqbef$-edge, and $c\in\OutSet$ iff it has an outgoing one, and $\InSet,\OutSet\subseteq\img{\chi^{w}}$;
        \item\label{ITEM:INV_ENDS} $t\in\FstSet$ iff $t$ has an uncolored first event, and $t\in\LstSet$ iff $t$ has an uncolored last event;
        \item\label{ITEM:INV_REP} $\RepMap(c)$ is the smallest active color whose event lies in the component of $\chi^{-1}(c)$;
        \item\label{ITEM:INV_LAMBDA} $\dom{\ThrMap}=\dom{\RepMap}=\img{\chi^{w}}$ and $\ThrMap(c)=\tid(\chi^{-1}(c))$.
    \end{compactenum}
\end{restatable}

\begin{proof}
    By induction on $|w|$.
    The empty word gives the initial state and $G_c^{\varepsilon}=\bot$, for which all four claims hold vacuously.
    For the inductive step, let $\mathrm{op}$ be the last operation of $w$ and assume the claims for the preceding prefix.

    \SubParagraph{$\CFEIntro(\llab,c)$.}
    The introduced event is colored by $c$ and has no edges, so it forms a component of its own, of which $c$ is the only active color.
    The rule maps $c$ to $\tid(\llab)$ in $\ThrMap$ and to itself in $\RepMap$, and leaves $\InSet$ and $\OutSet$ unchanged, which do not contain $c$ by \cref{ITEM:INV_DEG}, as $c$ was inactive.
    This gives \cref{ITEM:INV_LAMBDA,ITEM:INV_DEG,ITEM:INV_REP}, and \cref{ITEM:INV_ENDS} is immediate, as the uncolored events are unchanged.

    \SubParagraph{$\CFEIntrosb(c_1,c_2)$.}
    Write $\event_i=\chi^{-1}(c_i)$.
    The new edge gives $\event_1$ an outgoing and $\event_2$ an incoming $\seqbef$-edge, which the rule records, giving \cref{ITEM:INV_DEG}.
    By \cref{ITEM:INV_REP}, the premise $\RepMap(c_1)\neq\RepMap(c_2)$ says that $\event_1$ and $\event_2$ lie in different components, which the new edge joins into one.
    The smallest active color of the joined component is the smaller of the two representatives, and the other components are unchanged, so $\textsc{merge}$ gives \cref{ITEM:INV_REP}.
    \Cref{ITEM:INV_LAMBDA} is immediate, and so is \cref{ITEM:INV_ENDS}, as both endpoints of the new edge are colored.

    \SubParagraph{$\CFEIntroR{\rsym}(c_1,c_2)$ with $\rsym\in\{\rf,\mb\}$.}
    The relation $\seqbef$, the coloring and the state are unchanged.

    \SubParagraph{$\CFEForget(c)$.}
    Let $\event=\chi^{-1}(c)$ and $t=\ThrMap(c)$.
    The event $\event$ becomes uncolored, and the edges and components are unchanged.
    So, \cref{ITEM:INV_LAMBDA,ITEM:INV_DEG} follow from the removal of $c$ from $\ThrMap$, $\InSet$ and $\OutSet$.
    For \cref{ITEM:INV_REP}, the colors whose smallest active color changes are those with $\RepMap(c')=c$, i.e., the other active colors of the component of $\event$, whose new smallest active color is the smallest among them, as computed by $\textsc{drop}$.
    For \cref{ITEM:INV_ENDS}, the only new uncolored event is $\event$, whose edges are final.
    By \cref{ITEM:INV_DEG}, $\event$ is a first event of $t$ iff $c\notin\InSet$, in which case the rule adds $t$ to $\FstSet$, and symmetrically for $\LstSet$.
\end{proof}

We now use this invariant to prove that $\mathcal{A}^{k}_{\seqbef}$ is correct.

\begin{restatable}{lemma}{lemautsb}\label{LEM:AUT_SB}
  For every $k\in\Nats$, there is a deterministic word automaton $\mathcal{A}^{k}_{\seqbef}$ with $2^{O(k\log(k\cdot|\ThreadDom|)+|\ThreadDom|)}$ states that accepts a closed $w$ iff $G^{w}.\seqbef$ is well-formed.
\end{restatable}

\begin{proof}
    A state consists of two subsets of $\IntSet{k+1}$, two subsets of $\ThreadDom$, a partial map $\IntSet{k+1}\rightharpoonup\IntSet{k+1}$ and a partial map $\IntSet{k+1}\rightharpoonup\ThreadDom$.
    So, there are at most $2^{2(k+1)}\cdot 2^{2|\ThreadDom|}\cdot(k+2)^{k+1}\cdot(|\ThreadDom|+1)^{k+1}=2^{O(k\log(k\cdot|\ThreadDom|)+|\ThreadDom|)}$ states.
    At most one rule applies to each state and letter, so $\mathcal{A}^{k}_{\seqbef}$ is deterministic.
    Let $w$ be closed.

    \SubParagraph{Soundness.}
    Suppose $\mathcal{A}^{k}_{\seqbef}$ accepts $w$.
    By \cref{ITEM:INV_LAMBDA} and the premise $\ThrMap(c_1)=\ThrMap(c_2)$, every $\seqbef$-edge joins two events of the same thread, so every component lies within some $\Events^{t}$.
    By \cref{ITEM:INV_DEG}, the premises $c_1\notin\OutSet$ and $c_2\notin\InSet$ say that every event has at most one incoming and at most one outgoing $\seqbef$-edge.
    By \cref{ITEM:INV_REP}, the premise $\RepMap(c_1)\neq\RepMap(c_2)$ says that every edge joins two different components, so that every component is acyclic.
    An acyclic graph in which every event has at most one incoming and at most one outgoing edge is a $\seqbef$-path, which has exactly one first event and one last event.
    Since $w$ is closed, every event is forgotten at some point, and the rule for $\CFEForget$ rejects forgetting a first event of a thread that is already in $\FstSet$.
    So, by \cref{ITEM:INV_ENDS}, each thread has at most one first event, and hence at most one component.
    So, each nonempty $\Events^{t}$ is a single $\seqbef$-path, i.e., \cref{ITEM:SB_IN,ITEM:SB_OUT,ITEM:SB_CONN} hold.

    \SubParagraph{Completeness.}
    Conversely, suppose $G^{w}.\seqbef$ is well-formed, i.e., each nonempty $\Events^{t}$ forms a single $\seqbef$-path.
    Consider the $\CFEIntrosb(c_1,c_2)$ that draws an edge $\event_1\LTo{\seqbef}\event_2$.
    Its endpoints belong to the same thread, so $\ThrMap(c_1)=\ThrMap(c_2)$ by \cref{ITEM:INV_LAMBDA}.
    Each event has at most one incoming and at most one outgoing $\seqbef$-edge, and each edge is drawn once, so $c_1\notin\OutSet$ and $c_2\notin\InSet$ by \cref{ITEM:INV_DEG}.
    Since none of the events $\seqbef$-before  $\event_1$ are connected to any of the events $\seqbef$-after $\event_2$, $\RepMap(c_1)\neq\RepMap(c_2)$ by \cref{ITEM:INV_REP}.
    Finally, each thread has at most one first event and one last event, so, by \cref{ITEM:INV_ENDS}, $t\notin\FstSet$ when its first event is forgotten, and $t\notin\LstSet$ when its last event is forgotten.
    So, the run exists, and it is accepting, as every state is.
\end{proof}

\Paragraph{Well-formedness of $\rf$.}
The main text only sketches $\mathcal{A}^{k}_{\rf}$, so we first give its states and transition rules.
By the definition of \cref{SUBSEC:PRELIM_EXEC}, $\rf$ is well-formed iff
\begin{enumerate*}[label=(\roman*)]
\item\label[wfcond]{ITEM:RF_KIND} every $\rf$-edge goes from a write or RMW to a read or RMW,
\item\label[wfcond]{ITEM:RF_MATCH} with the same location and value, and
\item\label[wfcond]{ITEM:RF_ONE} every read and RMW has exactly one incoming $\rf$-edge.
\end{enumerate*}
A state of $\mathcal{A}^{k}_{\rf}$ is a pair $\tuple{\LabMap,\SrcSet}$.
Here, $\LabMap:\IntSet{k+1}\rightharpoonup\LabelDom$ maps each active color to the label of its event, and $\SrcSet\subseteq \IntSet{k+1}$ is the set of active colors whose events are reads or RMWs without an incoming $\rf$-edge yet.
Intuitively, the labels in $\LabMap$ suffice to check \cref{ITEM:RF_KIND,ITEM:RF_MATCH} when an $\rf$-edge is drawn.
The set $\SrcSet$ is used to check \cref{ITEM:RF_ONE}: a read or RMW cannot receive a second incoming $\rf$-edge, and cannot be forgotten before it receives one.
The initial state and the unique accepting state are both $\tuple{\emptyset,\emptyset}$, and the transition rules are the following.
\begin{mathparpagebreakable}
\inferrule{
  \mathrm{op} = \CFEIntro(\llab,c) \quad
  \LabMap' = \LabMap \cup \{c \to \llab\} \\\\
  \SrcSet' = \SrcSet \cup \{c\} \text{ if } \op(\llab) \in \{\rd,\rmw\}, \text{ and } \SrcSet' = \SrcSet \text{ otherwise}
}{
  \tuple{\LabMap, \SrcSet} \xrightarrow{\mathrm{op}} \tuple{\LabMap', \SrcSet'}
}
\and
\inferrule{
  \mathrm{op} = \CFEForget(c) \quad
  c \not\in \SrcSet \\\\
  \LabMap' = \LabMap \setminus \{c \to \LabMap(c)\}
}{
  \tuple{\LabMap, \SrcSet} \xrightarrow{\mathrm{op}} \tuple{\LabMap', \SrcSet}
}
\and
\inferrule{
  \mathrm{op} = \CFEIntrorf(c_1,c_2) \quad
  \op(\LabMap(c_1)) \in \{\wt,\rmw\} \\\\
  c_2 \in \SrcSet \quad
  \lloc(\LabMap(c_1)) = \lloc(\LabMap(c_2)) \\\\
  \val_{\wt}(\LabMap(c_1)) = \val_{\rd}(\LabMap(c_2)) \quad
  \SrcSet' = \SrcSet \setminus \{c_2\}
}{
  \tuple{\LabMap, \SrcSet} \xrightarrow{\mathrm{op}} \tuple{\LabMap, \SrcSet'}
}
\and
\inferrule{
  \mathrm{op} = \CFEIntroR{\rsym}(c_1,c_2) \\\\ \rsym \in \{\seqbef,\mb\}
}{
  \tuple{\LabMap, \SrcSet} \xrightarrow{\mathrm{op}} \tuple{\LabMap, \SrcSet}
}
\end{mathparpagebreakable}
We now prove that $\mathcal{A}^{k}_{\rf}$ is correct.

\begin{restatable}{lemma}{lemautrf}\label{LEM:AUT_RF}
  For every $k\in\Nats$, there is a deterministic word automaton $\mathcal{A}^{k}_{\rf}$ with $2^{O(k\log|\LabelDom|)}$ states that accepts a closed $w$ iff $G^{w}.\rf$ is well-formed.
\end{restatable}

\begin{proof}
    A state of $\mathcal{A}^{k}_{\rf}$ is a partial map $\IntSet{k+1}\rightharpoonup\LabelDom$ together with a subset of $\IntSet{k+1}$.
    So, there are $(|\LabelDom|+1)^{k+1}\cdot 2^{k+1}=2^{O(k\log|\LabelDom|)}$ states.
    Moreover, exactly one rule applies to each state and letter, so $\mathcal{A}^{k}_{\rf}$ is deterministic.

    By induction on $|w|$, if $\mathcal{A}^{k}_{\rf}$ has a run on a well-formed $w$, then $\LabMap(c)=\llab(\chi^{-1}(c))$ for every active color $c$, and $\SrcSet$ is the set of active colors whose events are reads or RMWs without an incoming $\rf$-edge in $G^{w}$.
    The first claim follows directly from the rules for $\CFEIntro$ and $\CFEForget$.
    For the second, a color enters $\SrcSet$ when a read or RMW is introduced, and leaves $\SrcSet$ when its incoming $\rf$-edge is drawn.
    Moreover, the rule for $\CFEForget$ never removes a color of $\SrcSet$.

    For soundness, suppose that $\mathcal{A}^{k}_{\rf}$ accepts a closed $w$.
    The premises of the rule for $\CFEIntrorf(c_1,c_2)$ check that the source is a write or RMW, that the target is a read or RMW without an incoming $\rf$-edge yet, and that both have the same location and value.
    This gives \cref{ITEM:RF_KIND,ITEM:RF_MATCH}.
    For \cref{ITEM:RF_ONE}, a read or RMW cannot receive two incoming $\rf$-edges, as its color leaves $\SrcSet$ after the first one.
    It also cannot receive none.
    Otherwise, its color would stay in $\SrcSet$, and the rule for $\CFEForget$ would prevent it from ever being forgotten, which is impossible as $w$ is closed.

    For completeness, suppose conversely that $G^{w}.\rf$ satisfies \cref{ITEM:RF_KIND,ITEM:RF_MATCH,ITEM:RF_ONE}.
    Consider an operation that draws an $\rf$-edge.
    Its premises on labels hold by \cref{ITEM:RF_KIND,ITEM:RF_MATCH}.
    Its target has no incoming $\rf$-edge yet, as it has exactly one by \cref{ITEM:RF_ONE}, and each edge is drawn only once.
    Moreover, every read or RMW receives its incoming $\rf$-edge before it is forgotten, since this edge can only be drawn while it is colored.
    So, the premise $c\not\in \SrcSet$ of the rule for $\CFEForget$ always holds.
    Finally, the run ends in $\tuple{\emptyset,\emptyset}$, as $w$ is closed.
\end{proof}

\Paragraph{Well-formedness of $\mb$.}
Similarly, the main text only describes $\mathcal{A}^{k}_{\mb}$ as a variant of $\mathcal{A}^{k}_{\seqbef}$, so we first give its rules.
The construction is the same as for $\mathcal{A}^{k}_{\seqbef}$, with locations in place of threads, and restricted to writes and RMWs.
Indeed, $\mb$ is well-formed iff, for each location,
\begin{enumerate*}[label=(\roman*)]
\item\label[wfcond]{ITEM:MB_IN} all writes and RMWs but one have exactly one incoming $\mb$-edge, and that one has none,
\item\label[wfcond]{ITEM:MB_OUT} all writes and RMWs but one have exactly one outgoing $\mb$-edge, and that one has none, and
\item\label[wfcond]{ITEM:MB_CONN} all writes and RMWs are connected.
\end{enumerate*}
A state is a tuple $\tuple{\InSet,\OutSet,\FstSet,\LstSet,\RepMap,\LocMap}$ with $\InSet,\OutSet\subseteq\IntSet{k+1}$, $\FstSet,\LstSet\subseteq\LocationDom$, $\RepMap:\IntSet{k+1}\rightharpoonup\IntSet{k+1}$ and $\LocMap:\IntSet{k+1}\rightharpoonup\LocationDom$.
The components have the same meaning as for $\mathcal{A}^{k}_{\seqbef}$, but for $\mb$, and for the writes and RMWs on each location instead of the events of each thread.
Since reads and fences are not part of any $\mb$-chain, $\RepMap$ and $\LocMap$ are only defined on the colors of writes and RMWs.
So, the rules for $\CFEIntro$ and $\CFEForget$ ignore reads and fences.
The initial state is $\tuple{\emptyset,\emptyset,\emptyset,\emptyset,\emptyset,\emptyset}$, and every state is accepting.
We reuse the functions $\textsc{merge}$ and $\textsc{drop}$ of $\mathcal{A}^{k}_{\seqbef}$, and the transition rules are the following, where the components that are not mentioned are unchanged.
\begin{mathparpagebreakable}
\inferrule{
  \mathrm{op} = \CFEIntro(\llab,c) \quad
  \op(\llab) \in \{\wt,\rmw\} \\\\
  \RepMap' = \RepMap \cup \{c \mapsto c\} \quad
  \LocMap' = \LocMap \cup \{c \mapsto \lloc(\llab)\}
}{
  \tuple{\InSet,\OutSet,\FstSet,\LstSet,\RepMap,\LocMap} \xrightarrow{\mathrm{op}} \tuple{\InSet',\OutSet',\FstSet',\LstSet',\RepMap',\LocMap'}
}
\and
\inferrule{
  \mathrm{op} = \CFEForget(c) \quad c\in\dom{\LocMap} \quad x=\LocMap(c) \quad
  c\notin\InSet \Rightarrow x\notin\FstSet \quad
  c\notin\OutSet \Rightarrow x\notin\LstSet \\\\
  \FstSet' = \FstSet\cup\{x\mid c\notin\InSet\} \quad
  \LstSet' = \LstSet\cup\{x\mid c\notin\OutSet\} \quad
  \InSet'=\InSet\setminus\{c\} \quad
  \OutSet'=\OutSet\setminus\{c\} \\\\
  \RepMap'=\textsc{drop}(c,\RepMap) \quad
  \LocMap'=\LocMap\setminus\{c\mapsto x\}
}{
  \tuple{\InSet,\OutSet,\FstSet,\LstSet,\RepMap,\LocMap} \xrightarrow{\mathrm{op}} \tuple{\InSet',\OutSet',\FstSet',\LstSet',\RepMap',\LocMap'}
}
\and
\inferrule{
  \mathrm{op} = \CFEIntromb(c_1,c_2)\quad
  c_1, c_2 \in \dom{\LocMap} \quad
  \LocMap(c_1)=\LocMap(c_2) \quad
  c_1\notin\OutSet \quad c_2\notin\InSet \\\\
  \RepMap(c_1)\neq\RepMap(c_2) \quad
  \OutSet'=\OutSet\cup\{c_1\} \quad
  \InSet'=\InSet\cup\{c_2\} \quad
  \RepMap'=\textsc{merge}(c_1,c_2,\RepMap)
}{
  \tuple{\InSet,\OutSet,\FstSet,\LstSet,\RepMap,\LocMap} \xrightarrow{\mathrm{op}} \tuple{\InSet',\OutSet',\FstSet',\LstSet',\RepMap',\LocMap'}
}
\and
\inferrule{
  \mathrm{op} = \CFEIntro(\llab,c) \\\\
  \op(\llab) \in \{\rd,\fc\}
}{
  q \xrightarrow{\mathrm{op}} q
}
\and
\inferrule{
  \mathrm{op} = \CFEForget(c) \\\\
  c \not\in \dom{\LocMap}
}{
  q \xrightarrow{\mathrm{op}} q
}
\and
\inferrule{
  \mathrm{op} = \CFEIntroR{\rsym}(c_1,c_2) \\\\ \rsym \in \{\seqbef,\rf\}
}{
  q \xrightarrow{\mathrm{op}} q
}
\end{mathparpagebreakable}
where $q=\tuple{\InSet,\OutSet,\FstSet,\LstSet,\RepMap,\LocMap}$.

The correctness of $\mathcal{A}^{k}_{\mb}$ is thus similar to $\mathcal{A}^{k}_{\seqbef}$.

\begin{restatable}{lemma}{lemautmb}\label{LEM:AUT_MB}
  For every $k\in\Nats$, there is a deterministic word automaton $\mathcal{A}^{k}_{\mb}$ with $2^{O(k\log(k\cdot|\LocationDom|)+|\LocationDom|)}$ states that accepts a closed $w$ iff $G^{w}.\mb$ is well-formed.
\end{restatable}

\begin{proof}
    The number of states is that of \cref{LEM:AUT_SB}, with $\LocationDom$ in place of $\ThreadDom$.
    For correctness, \cref{LEM:APP_AUT_SB_INV} and the proof of \cref{LEM:AUT_SB} carry over, with $\mb$ in place of $\seqbef$, the writes and RMWs on each location in place of the events of each thread.
    It remains to check that reads are handled correctly.
    The premise $c_1,c_2\in\dom{\LocMap}$ of the rule for $\CFEIntromb$ rejects every $\mb$-edge with a read as an endpoint, which well-formedness of $\mb$ also forbids.
    Moreover, the rules for $\CFEIntro$ and $\CFEForget$ leave the state unchanged for reads.
    So, reads never appear in $\InSet$ or $\OutSet$, never add a location to $\FstSet$ or $\LstSet$, and never appear in $\dom{\RepMap}$ or $\dom{\LocMap}$.
\end{proof}

\Paragraph{The product.}
Finally, $\mathcal{A}^{k}_{\textsc{wf}}$ is the product of the four automata above, which proves \cref{LEM:AUT_WF}.

\lematwf*

\begin{proof}
    Let $\mathcal{A}^{k}_{\textsc{wf}}$ be the product of $\mathcal{A}^{k}_{\textsc{form}}$, $\mathcal{A}^{k}_{\seqbef}$, $\mathcal{A}^{k}_{\rf}$ and $\mathcal{A}^{k}_{\mb}$.
    By \cref{LEM:AUT_FORM}, $\mathcal{A}^{k}_{\textsc{form}}$ accepts $w$ iff $w$ is closed.
    For such $w$, by \cref{LEM:AUT_SB,LEM:AUT_RF,LEM:AUT_MB}, the other three automata accept $w$ iff $G^{w}.\seqbef$, $G^{w}.\rf$ and $G^{w}.\mb$ are well-formed.
    So, $\mathcal{A}^{k}_{\textsc{wf}}$ accepts $w$ iff $w$ is closed, and $G^{w}$ is an execution graph.
    Its number of states is the product of those of its components, which gives the stated bound, using $k\log k\leq k^2$ and $|\ThreadDom|,|\LocationDom|\leq|\LabelDom|$.
\end{proof}
\subsection{Proofs of \cref{SUBSEC:AUT_PROG}}\label{SUBSEC:APP_AUT_PROG}

In this subsection, we prove that $\mathcal{A}^{k}_{\ConcProg}$ recognizes the graphs of $\ConcProg$~(\cref{LEM:AUT_PROG}).
Fix $\ConcProg$ as in \cref{SUBSEC:AUT_PROG}, with $\Prog_{t} = \tuple{\States_{t}, \LabelDom^{t}, \Trans_{t}, \InitState_{t}}$ for each $t \in \ThreadDom$.
For states $u$ and $v$ of the same $\Prog_{t}$, we write $\epsilon\textsc{-reach}(u, v)$ when $\Prog_{t}$ has a (possibly empty) path of silent transitions from $u$ to $v$.
The initial state and the unique accepting state of $\mathcal{A}^{k}_{\ConcProg}$ are both $\tuple{\emptyset, \emptyset}$, and its rules are the following.
\begin{mathparpagebreakable}
\inferrule{
  \mathrm{op} = \CFEIntro(\llab,c) \\\\
  \delta\in \Trans_{\tid(\llab)} \quad \llab(\delta) = \llab \\\\
  \SrcSet' = \SrcSet \cup \{c\} \quad \TrMap' = \TrMap \cup \{c \mapsto \delta\}
}{
  \tuple{\SrcSet, \TrMap} \xrightarrow{\mathrm{op}} \tuple{\SrcSet', \TrMap'}
}
\and
\inferrule{
  \mathrm{op} = \CFEForget(c) \quad c \not\in \SrcSet \\\\
  \TrMap' = \TrMap \setminus \{ c \mapsto \TrMap(c) \}
}{
  \tuple{\SrcSet, \TrMap} \xrightarrow{\mathrm{op}} \tuple{\SrcSet, \TrMap'}
}
\and
\inferrule{
  \mathrm{op} = \CFEIntroR{\rsym}(c_1,c_2) \\\\ \rsym \in \{\rf,\mb\}
}{
  \tuple{\SrcSet, \TrMap} \xrightarrow{\mathrm{op}} \tuple{\SrcSet, \TrMap}
}
\and
\inferrule{
  \mathrm{op} = \CFEForget(c) \quad c \in \SrcSet \\\\
  \epsilon\textsc{-reach}(\InitState_{\tid(\llab(\TrMap(c)))},\src(\TrMap(c))) \\\\
  \SrcSet' = \SrcSet \setminus \{c\} \quad \TrMap' = \TrMap \setminus \{ c \mapsto \TrMap(c)\}
}{
  \tuple{\SrcSet, \TrMap} \xrightarrow{\mathrm{op}} \tuple{\SrcSet', \TrMap'}
}
\and
\inferrule{
  \mathrm{op} = \CFEIntrosb(c_1,c_2) \quad c_2\in \SrcSet \\\\
  \epsilon\textsc{-reach}(\tgt(\TrMap(c_1)),\src(\TrMap(c_2))) \\\\
  \SrcSet' = \SrcSet \setminus \{c_2\}
}{
  \tuple{\SrcSet, \TrMap} \xrightarrow{\mathrm{op}} \tuple{\SrcSet', \TrMap}
}
\end{mathparpagebreakable}
Throughout, $w$ is well-formed and $G_c^{w}=\tuple{E,\seqbef,\rf,\mb,\chi}$.
The automaton $\mathcal{A}^{k}_{\ConcProg}$ is nondeterministic, so its state after $w$ depends on the run and not only on $w$.
Given a run $\rho$ of $\mathcal{A}^{k}_{\ConcProg}$ on $w$, we write $\mu_{\rho}(\event)$ for the transition $\delta$ that the $\CFEIntro$ rule chose at the operation introducing $\event\in E$, and $\mu$ when $\rho$ is clear from the context.
The following lemma describes the state reached by such a run, together with the transitions it chose.

\begin{restatable}{lemma}{lemappautproginv}\label{LEM:APP_AUT_PROG_INV}
    Let $\rho$ be a run of $\mathcal{A}^{k}_{\ConcProg}$ on a well-formed $w$, reaching $\tuple{\SrcSet,\TrMap}$, and let $\mu=\mu_{\rho}$.
    Then
    \begin{compactenum}
        \item\label{ITEM:PINV_MU} $\mu(\event)\in\Trans_{\tid(\event)}$ and $\llab(\mu(\event))=\llab(\event)$ for every $\event\in E$;
        \item\label{ITEM:PINV_LAMBDA} $\dom{\TrMap}=\img{\chi}$ and $\TrMap(c)=\mu(\chi^{-1}(c))$;
        \item\label{ITEM:PINV_P} $\SrcSet$ is the set of $c\in\img{\chi}$ such that $\chi^{-1}(c)$ has no incoming $\seqbef$-edge;
        \item\label{ITEM:PINV_EDGE} $\epsilon\textsc{-reach}(\tgt(\mu(\event_1)),\src(\mu(\event_2)))$ for every $\event_1\LTo{\seqbef}\event_2$;
        \item\label{ITEM:PINV_FIRST} $\epsilon\textsc{-reach}(\InitState_{\tid(\event)},\src(\mu(\event)))$ for every uncolored $\event\in E$ with no incoming $\seqbef$-edge.
    \end{compactenum}
\end{restatable}

\begin{proof}
    By induction on $|w|$.
    The empty word gives $\tuple{\emptyset,\emptyset}$ and $G_c^{\varepsilon}=\bot$, for which all five claims hold vacuously.
    For the inductive step, let $\mathrm{op}$ be the last operation of $w$ and assume the claims for the preceding prefix.
    Note that an uncolored event never receives a new $\seqbef$-edge, as $\CFEIntrosb$ only joins colored events.
    So, \cref{ITEM:PINV_EDGE,ITEM:PINV_FIRST} are never invalidated for events that are already uncolored.

    \SubParagraph{$\CFEIntro(\llab,c)$.}
    Let $\event$ be the introduced event and $\delta$ the transition chosen, so $\mu(\event)=\delta$.
    The premises $\delta\in\Trans_{\tid(\llab)}$ and $\llab(\delta)=\llab$ give \cref{ITEM:PINV_MU}, as $\llab(\event)=\llab$.
    The rule sets $\TrMap'=\TrMap\cup\{c\to\delta\}$, giving \cref{ITEM:PINV_LAMBDA}.
    The event $\event$ is colored and has no $\seqbef$-edge, so adding $c$ to $\SrcSet$ gives \cref{ITEM:PINV_P}, and \cref{ITEM:PINV_EDGE,ITEM:PINV_FIRST} are unaffected.

    \SubParagraph{$\CFEIntrosb(c_1,c_2)$.}
    Write $\event_i=\chi^{-1}(c_i)$.
    The operation adds the edge $\event_1\LTo{\seqbef}\event_2$, and the premise $\epsilon\textsc{-reach}(\tgt(\TrMap(c_1)),\src(\TrMap(c_2)))$ is, by \cref{ITEM:PINV_LAMBDA}, the instance of \cref{ITEM:PINV_EDGE} for that edge.
    After the operation $\event_2$ has an incoming $\seqbef$-edge, and the rule removes $c_2$ from $\SrcSet$, giving \cref{ITEM:PINV_P}.
    \Cref{ITEM:PINV_MU,ITEM:PINV_LAMBDA,ITEM:PINV_FIRST} are unaffected, as $\mu$, $\TrMap$ and the coloring are unchanged.

    \SubParagraph{$\CFEForget(c)$.}
    Let $\event=\chi^{-1}(c)$.
    Both rules remove $c$ from $\TrMap$ and from $\SrcSet$, which gives \cref{ITEM:PINV_LAMBDA,ITEM:PINV_P}.
    Since $\event$ is now uncolored, \cref{ITEM:PINV_FIRST} requires $\epsilon\textsc{-reach}(\InitState_{\tid(\event)},\src(\mu(\event)))$ if $\event$ has no incoming $\seqbef$-edge.
    By \cref{ITEM:PINV_P}, this is the case iff $c\in \SrcSet$.
    Then the second rule applies, and its premise is exactly this condition, as $\TrMap(c)=\mu(\event)$ by \cref{ITEM:PINV_LAMBDA}, and $\tid(\llab(\mu(\event)))=\tid(\event)$ by \cref{ITEM:PINV_MU}.
    \Cref{ITEM:PINV_MU,ITEM:PINV_EDGE} are unaffected.

    \SubParagraph{$\CFEIntroR{\rsym}(c_1,c_2)$ with $\rsym\in\{\rf,\mb\}$.}
    The state, $\mu$, the coloring and $\seqbef$ are unchanged, so all five claims carry over.
\end{proof}

We now use this invariant to prove that $\mathcal{A}^{k}_{\ConcProg}$ is correct.

\lemautprog*

\begin{proof}
    A state is a pair of a subset of $\IntSet{k+1}$ and a partial map $\IntSet{k+1}\rightharpoonup\bigcup_{t}\Trans_{t}$.
    By \cref{SUBSEC:PRELIM_PROG}, $|\bigcup_{t}\Trans_{t}|\leq|\ConcProg|$, so there are at most $2^{k+1}\cdot(|\ConcProg|+1)^{k+1}\leq 2^{k+1}\cdot(2|\ConcProg|)^{k+1}=O(|\ConcProg|^{k+1})\,2^{O(k)}$ states.
    Let $w$ be closed, with $G^{w}$ an execution graph.
    Since $G^{w}.\seqbef$ is well-formed, for each $t\in\ThreadDom$ the events of $\Events^{t}$ form a single $\seqbef$-path $\event_1\LTo{\seqbef}\cdots\LTo{\seqbef}\event_n$, and $\ProjectGraph{G^{w}}{\seqbef}{t}=\llab(\event_1)\cdots\llab(\event_n)$.

    \SubParagraph{Soundness.}
    Suppose a run $\rho$ of $\mathcal{A}^{k}_{\ConcProg}$ accepts $w$, and fix $t$ with $\Events^{t}$ as above.
    If $n=0$, the empty run of $\Prog_{t}$ emits the empty word, so assume $n\geq 1$.
    Since $w$ is closed, every event is uncolored at the end of $w$, and $\event_1$ has no incoming $\seqbef$-edge, so \cref{ITEM:PINV_FIRST} gives $\epsilon\textsc{-reach}(\InitState_{t},\src(\mu(\event_1)))$.
    \Cref{ITEM:PINV_EDGE} gives $\epsilon\textsc{-reach}(\tgt(\mu(\event_i)),\src(\mu(\event_{i+1})))$ for every $i<n$.
    Concatenating these $\epsilon$-paths with the transitions $\mu(\event_1),\ldots,\mu(\event_n)$ yields a run of $\Prog_{t}$ from $\InitState_{t}$.
    Its non-silent transitions are exactly $\mu(\event_1),\ldots,\mu(\event_n)$, whose labels are $\llab(\event_1),\ldots,\llab(\event_n)$ by \cref{ITEM:PINV_MU}.
    Hence $\ProjectGraph{G^{w}}{\seqbef}{t}\in\Lang{\Prog_{t}}$ for every $t$, i.e., $\tuple{G^{w}.\Events,G^{w}.\seqbef}\in\GLang{\ConcProg}$.

    \SubParagraph{Completeness.}
    Suppose $\tuple{G^{w}.\Events,G^{w}.\seqbef}\in\GLang{\ConcProg}$, and fix $t$ with $\Events^{t}$ as above.
    Some run of $\Prog_{t}$ emits $\llab(\event_1)\cdots\llab(\event_n)$; let $\delta_1,\ldots,\delta_n$ be its non-silent transitions in order.
    Then $\delta_i\in\Trans_{t}$ and $\llab(\delta_i)=\llab(\event_i)$, the silent transitions before $\delta_1$ give $\epsilon\textsc{-reach}(\InitState_{t},\src(\delta_1))$, and those between $\delta_i$ and $\delta_{i+1}$ give $\epsilon\textsc{-reach}(\tgt(\delta_i),\src(\delta_{i+1}))$.
    Let $\rho$ be the run of $\mathcal{A}^{k}_{\ConcProg}$ on $w$ that chooses $\delta_i$ at the operation introducing $\event_i$, for every thread, so that $\mu(\event_i)=\delta_i$.
    We check that no premise fails along $w$, using \cref{LEM:APP_AUT_PROG_INV} for the prefix preceding each operation.
    The premises of $\CFEIntro$ hold by the choice of $\delta_i$.
    The premise $\epsilon\textsc{-reach}$ of $\CFEIntrosb(c_1,c_2)$ holds because the edge is some $\event_i\LTo{\seqbef}\event_{i+1}$ and $\TrMap(c_j)=\mu(\chi^{-1}(c_j))$ by \cref{ITEM:PINV_LAMBDA}.
    Its premise $c_2\in \SrcSet$ holds by \cref{ITEM:PINV_P}, as $\event_{i+1}$ has exactly one incoming $\seqbef$-edge in $G^{w}$ and each edge is introduced once.
    For the second $\CFEForget$ rule, $c\in \SrcSet$ means that $\chi^{-1}(c)$ has no incoming $\seqbef$-edge, and it takes none after being forgotten; hence it is $\event_1$ of its thread, and the premise holds because $\epsilon\textsc{-reach}(\InitState_{t},\src(\delta_1))$.
    The remaining rules have no premise that can fail.
    Finally, $w$ is closed, so $\TrMap=\emptyset$ by \cref{ITEM:PINV_LAMBDA} and $\SrcSet\subseteq\img{\chi}=\emptyset$ by \cref{ITEM:PINV_P}, and $\rho$ ends in the accepting state $\tuple{\emptyset,\emptyset}$.
\end{proof}

\subsection{Proofs of \cref{SUBSEC:AUT_MM}}\label{SUBSEC:APP_AUT_MM}

In this subsection, we prove that $\mathcal{A}^{k}_{\mm}$ recognizes the graphs that are consistent in $\mm$~(\cref{LEM:AUT_MM_YES}), and derive its complement.
Fix $\mathcal{K}_{\mm}=\tuple{\States,\TestDom\cup\RelBase,\Trans,\InitState,\States_F}$ as in \cref{SUBSEC:AUT_MM}.
Throughout, $w$ is well-formed and $G_c^{w}=\tuple{E,\seqbef,\rf,\mb,\chi}$.
We use the walks, their interiors, and the events $\event_c$ of \cref{SUBSEC:AUT_MM}.
Note that walks compose.
If a walk from $\event$ to $\event'$ leads $\mathcal{K}_{\mm}$ from $q$ to $q'$, and a walk from $\event'$ to $\event''$ leads $\mathcal{K}_{\mm}$ from $q'$ to $q''$, then together they form a walk from $\event$ to $\event''$ that leads $\mathcal{K}_{\mm}$ from $q$ to $q''$.

\begin{definition}\label{DEF:APP_AUT_MM_SUM}
    Let $E'\subseteq E$ be a set of events.
    A tuple $\tuple{c_1,c_2,q_1,q_2}$ with $c_1,c_2\in\img{\chi}$
    \begin{enumerate*}[label=(\roman*)]
        \item\label{ITEM:MMSUM_P} is a \emph{path summary over $E'$} if some walk from $\event_{c_1}$ to $\event_{c_2}$ with interior in $E'$ leads $\mathcal{K}_{\mm}$ from $q_1$ to $q_2$; and
        \item\label{ITEM:MMSUM_Q} is a \emph{cycle summary over $E'$} if, for some $\event\in E'$, some walk from $\event$ to $\event_{c_1}$ with interior in $E'$ leads $\mathcal{K}_{\mm}$ from $\InitState$ to $q_1$, and some walk from $\event_{c_2}$ to $\event$ with interior in $E'$ leads $\mathcal{K}_{\mm}$ from $q_2$ to a state of $\States_F$.
    \end{enumerate*}
    We write $\SumPath[w]$ and $\SumCyc[w]$ for the path and cycle summaries over the uncolored events of $G_c^{w}$.
\end{definition}

Note that $\SumPath[w]$ and $\SumCyc[w]$ are exactly the sets described in \cref{SUBSEC:AUT_MM}.
We define summaries over an arbitrary set $E'$ because $\textsc{tc}(c,\cdot,\cdot)$ lets walks pass through $\event_c$, and so considers the uncolored events together with $\event_c$.
The following lemma shows that $\textsc{tc}$ computes exactly the summaries over this larger set.

\begin{restatable}{lemma}{lemappautmmtc}\label{LEM:APP_AUT_MM_TC}
    Let $c\in\img{\chi}$, let $U$ be the set of uncolored events of $G_c^{w}$, and let $\tuple{\SumPath'',\SumCyc''}=\textsc{tc}(c,\SumPath[w],\SumCyc[w])$.
    Then
    \begin{enumerate*}[label=(\roman*)]
        \item\label{ITEM:TC_P} $\SumPath''$ is the set of path summaries over $U\cup\{\event_c\}$, and
        \item\label{ITEM:TC_Q} $\SumCyc''$ is the set of cycle summaries over $U\cup\{\event_c\}$.
    \end{enumerate*}
\end{restatable}

\begin{proof}
    Let $U'\triangleq U\cup\{\event_c\}$.
    We first show, by induction on the construction of the fixed points, that every tuple of $\SumPath''$ and $\SumCyc''$ is a summary over $U'$.
    The tuples of $\SumPath[w]$ and $\SumCyc[w]$ are summaries over $U$, and hence over $U'$.
    \Cref{ITEM:TC_COMP} composes two walks at $\event_c$, which is in $U'$.
    \Cref{ITEM:TC_EXT_IN} takes the first half of a cycle summary, which ends at $\event_c$, and extends it by a walk from $\event_c$ to $\event_{c_1}$.
    The resulting walk still starts at the anchor $\event$ and has its interior in $U'$.
    \Cref{ITEM:TC_EXT_OUT} does the same for the second half.
    Finally, \cref{ITEM:TC_START} uses $\event_c\in U'$ itself as the anchor.

    Conversely, we show that every summary over $U'$ is in $\SumPath''$, respectively $\SumCyc''$.
    For \cref{ITEM:TC_P}, consider a walk from $\event_{c_1}$ to $\event_{c_2}$ with interior in $U'$.
    We cut it at every occurrence of $\event_c$ in its interior.
    Each piece has its interior in $U$, so its summary is in $\SumPath[w]\subseteq \SumPath''$.
    \Cref{ITEM:TC_COMP} then puts the pieces back together.

    For \cref{ITEM:TC_Q}, consider a cycle summary over $U'$ with anchor $\event$.
    Let $\pi_1$ be its first half, from $\event$ to $\event_{c_1}$, and $\pi_2$ its second half, from $\event_{c_2}$ to $\event$.
    If $\event=\event_c$, then $\pi_1$ and $\pi_2$ are walks with interior in $U'$ that start and end at colored events.
    So, their summaries are in $\SumPath''$ by \cref{ITEM:TC_P}, and \cref{ITEM:TC_START} applies.
    Otherwise, $\event\in U$.
    If $\pi_1$ visits $\event_c$, we cut it at its first occurrence of $\event_c$.
    This gives a walk from $\event$ to $\event_c$ with interior in $U$, followed by a walk from $\event_c$ to $\event_{c_1}$ with interior in $U'$.
    Symmetrically, if $\pi_2$ visits $\event_c$, we cut it at its last occurrence of $\event_c$.
    The two outer pieces share the anchor $\event\in U$ and have their interiors in $U$.
    Hence, they form a cycle summary over $U$, which is in $\SumCyc[w]\subseteq \SumCyc''$.
    The inner pieces are path summaries over $U'$, so they are in $\SumPath''$ by \cref{ITEM:TC_P}.
    \Cref{ITEM:TC_EXT_IN,ITEM:TC_EXT_OUT} then attach them to the cycle summary.
\end{proof}

The following lemma describes the state that $\mathcal{A}^{k}_{\mm}$ reaches after reading $w$.

\begin{restatable}{lemma}{lemappautmminv}\label{LEM:APP_AUT_MM_INV}
    Let $w$ be well-formed and suppose $\mathcal{A}^{k}_{\mm}$ has a run on $w$, reaching $\tuple{\SumPath,\SumCyc}$.
    Then $\SumPath=\SumPath[w]$ and $\SumCyc=\SumCyc[w]$.
\end{restatable}

\begin{proof}
    By induction on $|w|$.
    For the empty word, the state is $\tuple{\emptyset,\emptyset}$ and $G_c^{\varepsilon}=\bot$, so $\SumPath[\varepsilon]=\SumCyc[\varepsilon]=\emptyset$.
    For the inductive step, let $\mathrm{op}$ be the last operation of $w$ and $w'$ the prefix before it, and assume that $\SumPath=\SumPath[w']$ and $\SumCyc=\SumCyc[w']$.

    \SubParagraph{$\CFEIntro(\llab,c)$.}
    The new event $\event_c$ has no edges.
    So, a walk that visits $\event_c$ never leaves it, and consists only of tests satisfied by $\llab$.
    If such a walk has two or more steps, then $\event_c$ is in its interior.
    As $\event_c$ is colored, this walk yields no path summary.
    Hence, the new path summaries come from walks from $\event_c$ to itself with empty interior.
    These are the empty walk, which gives $\tuple{c,c,q,q}$, and the single tests $\ttest$ with $\llab\in\ttest$, which give $\textsc{step}(c,\ttest,c)$.
    This is exactly what $\textsc{addTests}$ adds.
    There are no new cycle summaries, since a half would need an edge to reach $\event_c$ from an uncolored event.

    \SubParagraph{$\CFEIntroR{\rsym}(c_1,c_2)$.}
    The new edge joins two colored events.
    So, a walk with uncolored interior can only use it if the walk consists of this single step.
    Hence, the new path summaries come from the step $\rsym$ from $\event_{c_1}$ to $\event_{c_2}$ and the step $\rsym^{-1}$ back, which is exactly what $\textsc{addEdges}$ adds.
    There are no new cycle summaries.
    This is because every event of a half other than its colored endpoint is uncolored, so a half never uses an edge between two colored events.

    \SubParagraph{$\CFEForget(c)$.}
    The uncolored events of $G_c^{w}$ are those of $G_c^{w'}$ together with $\event_c$.
    By \cref{LEM:APP_AUT_MM_TC}, the function $\textsc{tc}$ of the rule for $\CFEForget(c)$ returns the path and cycle summaries over this set, with respect to the colors of $G_c^{w'}$.
    The summaries with respect to the colors of $G_c^{w}$ are those that do not mention $c$, and these are exactly the ones that $\textsc{erase}$ keeps.
\end{proof}

We now use this invariant to prove that $\mathcal{A}^{k}_{\mm}$ is correct.

\lemautmmyes*

\begin{proof}
    There are $2^{(k+1)^2\cdot|\States|^2}$ values each for $\SumPath$ and $\SumCyc$, and one additional sink state.
    As $|\States|\leq 2|\mm|+2$, $\mathcal{A}^{k}_{\mm}$ has $2^{O(k^2\cdot|\mm|^2)}$ states.
   
    It is deterministic by construction~(\cref{SUBSEC:AUT_MM}).
    Let $w$ be closed, with $G^{w}$ an execution graph.
    For every prefix of $w$ of the form $w'\,\CFEForget(c)$, let $\tuple{\SumPath'',\SumCyc''}=\textsc{tc}(c,\SumPath[w'],\SumCyc[w'])$, and let $U'$ be the set of uncolored events of $G_c^{w'}$ together with $\event_c$.
    If the run of $\mathcal{A}^{k}_{\mm}$ on $w$ reaches the end of $w'$, then its state there is $\tuple{\SumPath[w'],\SumCyc[w']}$ by \cref{LEM:APP_AUT_MM_INV}.
    So, the $\CFEForget(c)$ rule computes exactly $\tuple{\SumPath'',\SumCyc''}$.

    \SubParagraph{Soundness.}
    Suppose that $G^{w}\not\models\mm$.
    We show that $w$ is rejected.
    Consider a cyclic realization $\event_0,\ldots,\event_m$, with $\event_0=\event_m$, of a word on which $\mathcal{K}_{\mm}$ has a run $q_0,\ldots,q_m$ with $q_0=\InitState$ and $q_m\in\States_F$.
    As $w$ is closed, every event of the realization is forgotten at some point of $w$.
    Let $\CFEForget(c)$ be the operation that forgets the last of them, and $w'$ the prefix before it.
    Then every event of the realization is in $U'$.
    Moreover, edges are only drawn between colored events, so no edge between two events of $U'$ is drawn after $w'\,\CFEForget(c)$.
    Hence, the realization already lies in $G^{w'}$.
    If $\event_0=\event_c$, the realization is a walk from $\event_c$ to itself with interior in $U'$.
    So, $\tuple{c,c,q_0,q_m}\in \SumPath''$ by \cref{LEM:APP_AUT_MM_TC}.
    The empty walk at $\event_c$ also gives $\tuple{c,c,q_0,q_0}\in \SumPath''$, and \cref{ITEM:TC_START} yields $\tuple{c,c,q_0,q_0}\in \SumCyc''$.
    Otherwise, let $j$ be the first index with $\event_j=\event_c$.
    Then $\event_0,\ldots,\event_j$ and $\event_j,\ldots,\event_m$ are the two halves of a cycle summary over $U'$ with anchor $\event_0$.
    So, $\tuple{c,c,q_j,q_j}\in \SumCyc''$ by \cref{LEM:APP_AUT_MM_TC}.
    In both cases, $\textsc{violationFree}(\SumCyc'')$ fails at this $\CFEForget$.
    Since $\mathcal{A}^{k}_{\mm}$ is deterministic, its only run on $w$ enters the sink at this point, if it has not done so already.
    Hence, $w$ is rejected.

    \SubParagraph{Completeness.}
    Conversely, suppose that $G^{w}\models\mm$.
    We first show that no $\CFEForget$ along $w$ detects a violation.
    Assume, for contradiction, that $\tuple{c',c',q,q}\in \SumCyc''$ at some prefix $w'\,\CFEForget(c)$.
    By \cref{LEM:APP_AUT_MM_TC}, this tuple is a cycle summary over $U'$.
    In other words, some anchor $\event$ has a walk to $\event_{c'}$ that leads $\mathcal{K}_{\mm}$ from $\InitState$ to $q$, and $\event_{c'}$ has a walk back to $\event$ that leads $\mathcal{K}_{\mm}$ from $q$ to some $q_F\in\States_F$.
    Together, these two walks form a cyclic realization in $G^{w'}$ of a word on which $\mathcal{K}_{\mm}$ has a run from $\InitState$ to $q_F$.
    This word is accepted by $\mathcal{K}_{\mm}$.
    Moreover, operations only add events and edges, so the realization is also a realization in $G^{w}$.
    This contradicts $G^{w}\models\mm$.
    So, the run on $w$ never enters the sink.
    Again by \cref{LEM:APP_AUT_MM_INV}, it ends in $\tuple{\SumPath[w],\SumCyc[w]}$.
    As $w$ is closed and summaries only mention active colors, this state is $\tuple{\emptyset,\emptyset}$, which is accepting.

    \SubParagraph{Complement.}
    Finally, let $\mathcal{A}^{k}_{\lnot\mm}$ be the complement of $\mathcal{A}^{k}_{\mm}$, which is deterministic and has the same states.
    For $w\in\Lang{\mathcal{A}^{k}_{\textsc{wf}}}$, $\mathcal{A}^{k}_{\lnot\mm}$ accepts $w$ iff $\mathcal{A}^{k}_{\mm}$ rejects it, iff $G^{w}\not\models\mm$ by the above.
\end{proof}

\subsection{Proofs of \cref{SUBSEC:AUT_ROB}}\label{SUBSEC:APP_AUT_ROB}

In this subsection, we prove the results of \cref{SUBSEC:AUT_ROB}.
We first define the product automaton $\mathcal{A}^{k}_{\textsc{rob}}$ formally.
We then consider programs that are given explicitly, i.e., as an LTS per thread, for which we can build $\mathcal{A}^{k}_{\textsc{rob}}$ and test it for emptiness.
Finally, we prove that robustness is in $\PSPACE$ for programs given as code.

\Paragraph{The product.}
We state the product for two arbitrary memory models $\mm_1$ and $\mm_2$, as robustness against $\mm$ is the case $\mm_1=\mm$ and $\mm_2=\scmm$.

\begin{lemma}\label{LEM:AUT_MM_ROB}
    Given $k\in\Nats$, a concurrent program $\ConcProg$ and two memory models $\mm_1$ and $\mm_2$, there exists a word automaton $\mathcal{A}^{k}_{\textsc{rob}}$ with $O(|\ConcProg|^{k+1})\cdot 2^{O(k^2\cdot(|\mm_1|^2+|\mm_2|^2)\ +\ k\log|\LabelDom|\ +\ |\ThreadDom|\ +\ |\LocationDom|)}$ states that accepts $w\in\Ops(k+1)^{*}$ iff $w$ is closed, $G^{w}\in\ExecutionGraphsOf{\ConcProg}{\mm_1}$, and $G^{w}\not\in\ExecutionGraphsOf{\ConcProg}{\mm_2}$.
\end{lemma}

\begin{proof}
    Let $\mathcal{A}^{k}_{\textsc{rob}}$ be the product of $\mathcal{A}^{k}_{\textsc{wf}}$, $\mathcal{A}^{k}_{\ConcProg}$, $\mathcal{A}^{k}_{\mm_1}$ and $\mathcal{A}^{k}_{\lnot\mm_2}$.
    Its number of states is the product of the bounds of \cref{LEM:AUT_WF,LEM:AUT_PROG,LEM:AUT_MM_YES}, where $2^{O(k)}$ and $2^{O(k\log k)}$ are absorbed by $2^{O(k^2\cdot(|\mm_1|^2+|\mm_2|^2))}$.
    By \cref{LEM:AUT_WF}, $\mathcal{A}^{k}_{\textsc{wf}}$ accepts $w$ iff $w$ is closed and $G^{w}$ is an execution graph.
    For such $w$, \cref{LEM:AUT_PROG} gives that $\mathcal{A}^{k}_{\ConcProg}$ accepts $w$ iff $\tuple{G^{w}.\Events,G^{w}.\seqbef}\in\GLang{\ConcProg}$, \cref{LEM:AUT_MM_YES} that $\mathcal{A}^{k}_{\mm_1}$ accepts $w$ iff $G^{w}\models\mm_1$, and that $\mathcal{A}^{k}_{\lnot\mm_2}$ accepts $w$ iff $G^{w}\not\models\mm_2$.
    Hence $\mathcal{A}^{k}_{\textsc{rob}}$ accepts $w$ iff $w$ is closed, $G^{w}\in\ExecutionGraphsOf{\ConcProg}{\mm_1}$, and $G^{w}\not\in\ExecutionGraphsOf{\ConcProg}{\mm_2}$, as the last holds for a graph of $\ConcProg$ iff $G^{w}\not\models\mm_2$.
\end{proof}

\Paragraph{Explicit programs.}
When the programs $\Prog_t$ are given explicitly, we can simply build $\mathcal{A}^{k}_{\textsc{rob}}$ and test it for emptiness.
This decides the existence of a witness of bounded pathwidth.

\begin{lemma}\label{LEM:EMP_MM_ROB}
    Given $k\in\Nats$, a concurrent program $\ConcProg$, and two memory models $\mm_1$ and $\mm_2$, whether there exists an execution graph $G$ such that $\pw(G)\leq k$, $G\in\ExecutionGraphsOf{\ConcProg}{\mm_1}$, and $G\not\in\ExecutionGraphsOf{\ConcProg}{\mm_2}$ is decidable in time $|\ConcProg|^{O(k)}\cdot 2^{O(k^2\cdot(|\mm_1|^2+|\mm_2|^2)\ +\ k\log|\LabelDom|\ +\ |\ThreadDom|\ +\ |\LocationDom|)}$.
\end{lemma}

\begin{proof}
    Let $\mathcal{A}^{k}_{\textsc{rob}}$ be the automaton of \cref{LEM:AUT_MM_ROB}.
    If $\mathcal{A}^{k}_{\textsc{rob}}$ accepts some $w$, then $G^{w}\in\ExecutionGraphsOf{\ConcProg}{\mm_1}\setminus\ExecutionGraphsOf{\ConcProg}{\mm_2}$, and $\pw(G^{w})\leq k$ by \cref{LEM:PATHWIDTH}.
    Conversely, if $G\in\ExecutionGraphsOf{\ConcProg}{\mm_1}\setminus\ExecutionGraphsOf{\ConcProg}{\mm_2}$ has $\pw(G)\leq k$, then $G=G^{w}$ for some closed $w\in\Ops(k+1)^{*}$ by \cref{LEM:PATHWIDTH}, and $\mathcal{A}^{k}_{\textsc{rob}}$ accepts $w$, as both sets of graphs are closed under the renaming of identifiers of \cref{SUBSEC:PRELIM_EXEC}.
    So it suffices to decide whether $\Lang{\mathcal{A}^{k}_{\textsc{rob}}}=\emptyset$.
    This takes time linear in the number of states and transitions of $\mathcal{A}^{k}_{\textsc{rob}}$.
    Let $n$ be the number of states given by \cref{LEM:AUT_MM_ROB}.
    The only nondeterministic component of $\mathcal{A}^{k}_{\textsc{rob}}$ is $\mathcal{A}^{k}_{\ConcProg}$, whose $\CFEIntro$ rule chooses one transition of $\ConcProg$, so each state has at most $|\Ops(k+1)|\cdot|\ConcProg|$ outgoing transitions.
    With $|\Ops(k+1)|=O(k\cdot|\LabelDom|+k^2)$ from \cref{SUBSEC:ALGEBRA_WORDS}, the time is $n\cdot|\Ops(k+1)|\cdot|\ConcProg|$, which is within the stated bound.
\end{proof}

By definition~(\cref{SUBSEC:PW_MM}), if $\mm$ admits $\ell$-$\pw$-bounded witnesses, then $\ConcProg$ is not robust against $\mm$ iff some $G\in\ExecutionGraphsOf{\ConcProg}{\mm}\setminus\ExecutionGraphsOf{\ConcProg}{\scmm}$ has $\pw(G)\leq\ell$.
In other words, deciding robustness is the special case of \cref{LEM:EMP_MM_ROB} with $\mm_1=\mm$, $\mm_2=\scmm$ and $k=\ell$.

\begin{theorem}\label{THE:ROB}
    Let $\mm$ be a memory model that admits $\ell$-$\pw$-bounded witnesses.
    Robustness of a concurrent program $\ConcProg$ against $\mm$ is decidable in time $|\ConcProg|^{O(\ell)}\cdot 2^{O(\ell^2\cdot|\mm|^2\ +\ \ell\log|\LabelDom|\ +\ |\ThreadDom|\ +\ |\LocationDom|)}$.
\end{theorem}

\begin{proof}
    If $\ConcProg$ is not robust against $\mm$, then, as $\mm$ admits $\ell$-$\pw$-bounded witnesses, some $G\in\ExecutionGraphsOf{\ConcProg}{\mm}\setminus\ExecutionGraphsOf{\ConcProg}{\scmm}$ has $\pw(G)\leq\ell$.
    Conversely, any such $G$ shows that $\ExecutionGraphsOf{\ConcProg}{\scmm}\neq\ExecutionGraphsOf{\ConcProg}{\mm}$.
    Hence $\ConcProg$ is robust against $\mm$ iff the instance of \cref{LEM:EMP_MM_ROB} with $k=\ell$, $\mm_1=\mm$ and $\mm_2=\scmm$ has a negative answer.
    As $|\scmm|$ is a constant, the bound of \cref{LEM:EMP_MM_ROB} becomes the one stated.
\end{proof}

Note that, for a fixed memory model and fixed domains, $\ell$ is a constant.
So, for explicit programs, robustness is decidable in time polynomial in $|\ConcProg|$.
Moreover, since prefix-closed memory models admit witnesses of pathwidth at most $\pwSC+1$, we have the following corollary.

\begin{corollary}\label{COR:ROB_PREFIX}
    Robustness of a concurrent program $\ConcProg$ against a prefix-closed memory model $\mm$ is decidable in time $|\ConcProg|^{O(|\ThreadDom|+|\LocationDom|)}\cdot 2^{O((|\ThreadDom|+|\LocationDom|)^2\cdot|\mm|^2\ +\ (|\ThreadDom|+|\LocationDom|)\log|\LabelDom|)}$.
\end{corollary}

\begin{proof}
    By \cref{LEM:PREFIX_BORDERLINE}, $\mm$ admits $\ell$-$\pw$-bounded witnesses for $\ell=\pwSC+1=|\ThreadDom|+|\LocationDom|+1$.
    Substituting this $\ell$ in \cref{THE:ROB}, the term $|\ThreadDom|+|\LocationDom|$ of the exponent is absorbed by $\ell\log|\LabelDom|$.
\end{proof}

\Paragraph{Polynomial space.}
We finally prove \cref{THE:ROB_PSPACE}.
Throughout, $N$ is the size of the input as in \cref{SUBSEC:AUT_ROB}, and $k$ is polynomial in $N$.
A state $\tuple{\pc,\phi}$ of $\Prog_t$ is encoded in $O(\log n_t+|\RegisterDom_t|\log(|\ValueDom|+1))$ bits, and a transition, as two states and a label, in $\mathrm{poly}(N)$ bits.

The procedure of \cref{THE:ROB_PSPACE} relies on the following properties of the components of $\mathcal{A}^{k}_{\textsc{rob}}$.

\begin{lemma}\label{LEM:SUCCINCT_STATES}
The following hold.
\begin{compactenum}
\item\label{ITEM:SUCCINCT_ENC} Every state of $\mathcal{A}^{k}_{\textsc{form}}$, $\mathcal{A}^{k}_{\seqbef}$, $\mathcal{A}^{k}_{\rf}$, $\mathcal{A}^{k}_{\mb}$, $\mathcal{A}^{k}_{\ConcProg}$ and $\mathcal{A}^{k}_{\mm}$ can be encoded using $\mathrm{poly}(N)$ bits, and membership in the respective accepting set is decidable in polynomial time.
\item\label{ITEM:SUCCINCT_ALPH} The alphabet $\Ops(k+1)$ has $(k+1)\cdot|\LabelDom| + (k+1) + 3\cdot (k+1)^2$ symbols, each encodable in $O(\log N)$ bits, where $|\LabelDom| = O(|\ThreadDom| \cdot |\LocationDom| \cdot |\ValueDom|^2 \cdot |\AnnDom|)$ is polynomial in $N$.
\item\label{ITEM:SUCCINCT_DET} The automata $\mathcal{A}^{k}_{\textsc{form}}$, $\mathcal{A}^{k}_{\seqbef}$, $\mathcal{A}^{k}_{\rf}$, $\mathcal{A}^{k}_{\mb}$ and $\mathcal{A}^{k}_{\mm}$ are deterministic: given a state $q$ and an operation $\mathsf{op} \in \Ops(k+1)$, the successor of $q$ under $\mathsf{op}$ is computable in polynomial time.
\item\label{ITEM:SUCCINCT_PROG} The automaton $\mathcal{A}^{k}_{\ConcProg}$ has out-degree exponential in $N$; nevertheless, given states $q, q'$ and an operation $\mathsf{op} \in \Ops(k+1)$, whether $q \xrightarrow{\mathsf{op}} q'$ is a transition of $\mathcal{A}^{k}_{\ConcProg}$ is decidable in $\PSPACE$.
\end{compactenum}
\end{lemma}

\begin{proof}
    For \cref{ITEM:SUCCINCT_ENC}, note that a state $\tuple{\ActSet,\EdgSet}$ of $\mathcal{A}^{k}_{\textsc{form}}$ consists of a subset of $\IntSet{k+1}$ and a subset of $\{\seqbef,\rf,\mb\}\times\IntSet{k+1}\times\IntSet{k+1}$, in $(k+1)+3(k+1)^2$ bits.
    A state $\tuple{\InSet,\OutSet,\FstSet,\LstSet,\RepMap,\ThrMap}$ of $\mathcal{A}^{k}_{\seqbef}$ takes $2(k+1)$ bits for $\InSet$ and $\OutSet$, $2|\ThreadDom|$ bits for $\FstSet$ and $\LstSet$, $O(k\log k)$ bits for $\RepMap$, and $O(k\log(|\ThreadDom|+1))$ bits for $\ThrMap$; $\mathcal{A}^{k}_{\mb}$ is the same with $\LocationDom$ and $\LocMap$ in place of $\ThreadDom$ and $\ThrMap$.
    A state of $\mathcal{A}^{k}_{\rf}$ stores a label and a bit per color, in $O(k\log|\LabelDom|)$ bits.
    A state $\tuple{\SrcSet,\TrMap}$ of $\mathcal{A}^{k}_{\ConcProg}$ stores a subset of $\IntSet{k+1}$ and a transition of some $\Prog_t$ per color, in $(k+1)\cdot\mathrm{poly}(N)$ bits.
    A state $\tuple{\SumPath,\SumCyc}$ of $\mathcal{A}^{k}_{\mm}$ is two subsets of $\IntSet{k+1}\times \IntSet{k+1}\times\States\times\States$, in $2(k+1)^2|\States|^2$ bits, where $|\States|$ is polynomial in $N$ by \cref{LEM:KAT_COMPILE}; the implicit sink takes one more bit.
    All of these are polynomial in $N$.
    Every accepting set is either the set of all states, as for $\mathcal{A}^{k}_{\seqbef}$, or described by emptiness of some components, as $\tuple{\emptyset,\emptyset}$ for $\mathcal{A}^{k}_{\textsc{form}}$, or, for the complement $\mathcal{A}^{k}_{\lnot\mm}$, by the negation of such a test, which is decidable in polynomial time.

    For \cref{ITEM:SUCCINCT_ALPH}, the number of symbols is the count of \cref{SUBSEC:ALGEBRA_WORDS}.
    A label is a kind, a thread, at most one location, at most two values and an annotation, so $|\LabelDom|\leq 4|\ThreadDom|\cdot|\LocationDom|\cdot|\ValueDom|^2\cdot|\AnnDom|$.
    A symbol names an operation, at most two colors and at most one label, in $O(\log k+\log|\LabelDom|)=O(\log N)$ bits.

    For \cref{ITEM:SUCCINCT_DET}, the automata $\mathcal{A}^{k}_{\textsc{form}}$, $\mathcal{A}^{k}_{\seqbef}$, $\mathcal{A}^{k}_{\rf}$ and $\mathcal{A}^{k}_{\mb}$ are deterministic by \cref{LEM:AUT_WF}, and $\mathcal{A}^{k}_{\mm}$ by \cref{LEM:AUT_MM_YES}.
    Their rules evaluate $\textsc{merge}$ and $\textsc{drop}$, which scan at most $k+1$ colors, compare labels, and, for $\mathcal{A}^{k}_{\mm}$, evaluate $\textsc{addTests}$, $\textsc{addEdges}$, $\textsc{tc}$, $\textsc{violationFree}$ and $\textsc{erase}$.
    The tests and transitions of $\mathcal{K}_{\mm}$ are computed in polynomial time, so $\textsc{addTests}$ and $\textsc{addEdges}$ take polynomial time.
    The sets of $\textsc{tc}$ are least fixed points over $\IntSet{k+1}\times \IntSet{k+1}\times\States\times\States$, computed by saturation in at most $2(k+1)^2|\States|^2$ rounds of polynomial time each.

    For \cref{ITEM:SUCCINCT_PROG}, note that a $\CFEIntro(\llab,c)$ from $\tuple{\SrcSet,\TrMap}$ may choose any transition $\delta$ of $\Prog_{\tid(\llab)}$ with $\llab(\delta)=\llab$, and there are exponentially many, one for each store; so the out-degree is exponential in $N$.
    Given $q=\tuple{\SrcSet,\TrMap}$, $q'=\tuple{\SrcSet',\TrMap'}$ and $\mathsf{op}$, we check the rules of \cref{SUBSEC:AUT_PROG}.
    Their premises are equalities between the components of $q$, $q'$ and $\mathsf{op}$, membership of $\delta$ in $\Trans_{\tid(\llab)}$, which takes polynomial time by the rules of the program semantics, and a premise $\epsilon\textsc{-reach}(u,v)$.
    The latter is reachability in the graph of silent transitions of $\Prog_t$, whose vertices are encoded in polynomially many bits and whose edges are decidable in polynomial time.
    It is decidable in nondeterministic polynomial space by guessing a path one vertex at a time, with a counter up to the number $2^{\mathrm{poly}(N)}$ of vertices, and hence in $\PSPACE$ by Savitch's theorem~\cite{Arora09}.
\end{proof}

We can now prove \cref{THE:ROB_PSPACE}.

\thmrobustnesspspaceexplicit*

\begin{proof}
    Let $\mathcal{A}^{\ell}_{\textsc{rob}}$ be the automaton of \cref{LEM:AUT_MM_ROB} with $\mm_1=\mm$, $\mm_2=\scmm$ and $k=\ell$, i.e., the product of $\mathcal{A}^{\ell}_{\textsc{form}}$, $\mathcal{A}^{\ell}_{\seqbef}$, $\mathcal{A}^{\ell}_{\rf}$, $\mathcal{A}^{\ell}_{\mb}$, $\mathcal{A}^{\ell}_{\ConcProg}$, $\mathcal{A}^{\ell}_{\mm}$ and $\mathcal{A}^{\ell}_{\lnot\scmm}$.
    As in the proofs of \cref{LEM:EMP_MM_ROB,THE:ROB}, $\ConcProg$ is not robust against $\mm$ iff $\Lang{\mathcal{A}^{\ell}_{\textsc{rob}}}\neq\emptyset$; these arguments do not depend on how $\ConcProg$ is given.

    Let $M$ be the number of states of $\mathcal{A}^{\ell}_{\textsc{rob}}$, which is $2^{\mathrm{poly}(N)}$ by \cref{ITEM:SUCCINCT_ENC} of \cref{LEM:SUCCINCT_STATES} with $k=\ell$.
    No component has an $\epsilon$-transition, so if $\Lang{\mathcal{A}^{\ell}_{\textsc{rob}}}\neq\emptyset$, then a shortest accepted word has an accepting run that visits no state twice, and thus has length less than $M$.
    The following nondeterministic procedure decides $\Lang{\mathcal{A}^{\ell}_{\textsc{rob}}}\neq\emptyset$.
    It stores the current state of each component, starting from the initial states, and a counter up to $M$.
    At each step, if every component is in an accepting state it accepts, and if the counter has reached $M$ it rejects.
    Otherwise, it guesses $\mathsf{op}\in\Ops(\ell+1)$, replaces the state of each deterministic component by its successor under $\mathsf{op}$, guesses a state $q'$ of $\mathcal{A}^{\ell}_{\ConcProg}$ and checks that $q\xrightarrow{\mathsf{op}}q'$ is a transition from its current state $q$, rejecting if not, and increments the counter.

    The sequence of states the procedure goes through on an accepting computation is an accepting run of $\mathcal{A}^{\ell}_{\textsc{rob}}$, and conversely every accepting run of length less than $M$ is followed by some computation, so the procedure is correct.
    By \cref{LEM:SUCCINCT_STATES}, the states, the operation and the counter take polynomial space, the deterministic successors and the acceptance test take polynomial time, and the check of the transition of $\mathcal{A}^{\ell}_{\ConcProg}$ takes polynomial space, which is reused across steps.
    Hence non-robustness is in $\NPSPACE$, which equals $\PSPACE$ by Savitch's theorem~\cite{Arora09}.
    As $\PSPACE$ is closed under complement, robustness is in $\PSPACE$.
\end{proof}

\section{Proofs of \cref{SEC:MON}}\label{SEC:APP_MON}

\subsection{Canonical Pathwidth}\label{SUBSEC:APP_MON_CPW}

This subsection proves the results on canonical pathwidth of \cref{SEC:MON}, starting with the canonical analogue of \cref{LEM:PATHWIDTH}.

\begin{lemma}\label{LEM:PATHWIDTH_CAN}
  Let $G$ be a free execution graph and $k \geq 0$.
  The following are equivalent.
  \begin{enumerate*}[label=(\roman*)]
    \item $G$ has a canonical path decomposition of width at most $k$.
    \item $G = G^{w}$ for some canonical closed $w \in \Ops(k+1)^{*}$.
  \end{enumerate*}
  Moreover, for a canonical closed $w$, the sequence $\dom{\chi^{w_0}}$, $\dom{\chi^{w_1}}$, \ldots, $\dom{\chi^{w_{|w|}}}$ is a canonical path decomposition of $G^{w}$ of width at most $k$.
\end{lemma}

\begin{proof}
    For the \emph{moreover} part, and hence (ii) $\Rightarrow$ (i), the sequence is a path decomposition of width at most $k$ by \cref{LEM:PATHWIDTH}.
    The first bag containing an event is the one following its $\CFEIntro$, so bags introduce events in the order of their $\CFEIntro$s in $w$, which follows $\seqbef$ as $w$ is canonical.

    For (i) $\Rightarrow$ (ii), take a canonical path decomposition $\Bag_1,\ldots,\Bag_n$ of width at most $k$, and write it as a word bag by bag.
    From $\Bag_i$ to $\Bag_{i+1}$, first forget the events of $\Bag_i\setminus\Bag_{i+1}$, then introduce those of $\Bag_{i+1}\setminus\Bag_i$ in an order extending $\seqbef$, then draw every edge of $G$ between events of $\Bag_{i+1}$ not drawn yet; finally forget the events of $\Bag_n$.
    Forgetting before introducing keeps at most $k+1$ events colored, so $k+1$ colors suffice, as in \cref{LEM:PATHWIDTH}.
    Every edge is drawn, as its endpoints share a bag.
    Events are introduced in the order of their first bags, with ties ordered by $\seqbef$, so the word is canonical, as the decomposition is.
\end{proof}

\lempwscreducedcan*

\begin{proof}
    The decomposition in the proof of \cref{LEM:PW_SC_REDUCED} has width at most $\pwSC$, and its first bag containing $\event_i$ is $\Bag_i$.
    It introduces events in the order $\sigma$, which extends $\po$, hence $\seqbef$.
\end{proof}

\lemprefixborderlinecan*

\begin{proof}
    Fix a prefix-closed $\mm$ and a program $\ConcProg$ that is not robust against $\mm$, and let $G$ be a witness with smallest $|G.\Events|$.
    As $G\not\models\scmm$, $G$ is non-empty, so by prefix-closedness some maximal event $\event$ of $G$ satisfies $G'\triangleq G\setminus\{\event\}\in\ExecutionGraphsOf{\ConcProg}{\mm}$.
    As $G'$ has fewer events than $G$, it is not a witness, so $G'\models\scmm$.

    As $\event$ is maximal, it has no $\seqbef$-successor and no $\rf$-successor in $G$.
    Hence the neighbours of $\event$ in $G$ are among its $\seqbef$-predecessor $\event_1$, its $\rf$-source $\event_2$, its $\mb$-predecessor $\event_3$ and its $\mb$-successor $\event_4$, whenever these exist.
    Every other edge of $G$ is an edge of $G'$.
    This is immediate for $\seqbef$ and $\rf$, and an $\mb$-edge of $G$ that avoids $\event$ has no event of $G'$ between its endpoints in $\mo$.

    We now build a decomposition of $G$.
Let $\Bag_1,\ldots,\Bag_n$ be the decomposition of $G'$ from the proof of \cref{LEM:PW_SC_REDUCED}.
    It is canonical of width at most $\pwSC$ by \cref{LEM:PW_SC_REDUCED_CAN}.
    The last event of a thread is retained by the first clause of that construction until $\Bag_n$.
    As $\event$ has no $\seqbef$-successor, $\event_1$ is the last event of $\tid(\event)$ in $G'$, and it lies in every bag from its first one to $\Bag_n$.
    For $j\in\{2,3,4\}$, add $\event_j$ to every bag after its last one, and append $\Bag_{n+1}\triangleq\{\event,\event_1,\event_2,\event_3,\event_4\}$, dropping the neighbours that do not exist.
    The result is a path decomposition of $G$.
    Every event lies in a contiguous range of bags, $\Bag_{n+1}$ covers the edges at $\event$, and the other edges are covered as in $G'$.
    It is canonical, as $\event$ is introduced last and its only $\seqbef$-neighbour $\event_1$ is introduced before.

    It remains to bound the width.
Each $\Bag_i$ with $i\leq n$ gains at most $|\{\event_2,\event_3,\event_4\}|$ events.
    A read has only $\event_2$, a write only $\event_3$ and $\event_4$, an RMW all three, and a fence none.
    So each $\Bag_i$ gains at most $3$ events, and at most $2$ under atomicity, where $\event_2=\event_3$ for an RMW as $G\models\mm$.
    Moreover $|\Bag_{n+1}|\leq 5$, and $|\Bag_{n+1}|\leq 4$ under atomicity, while $\pwSC\geq 2$ as $\ThreadDom$ and $\LocationDom$ are non-empty.
    Hence $\cpw(G)\leq\pwSC+3$, and $\cpw(G)\leq\pwSC+2$ under atomicity.
\end{proof}

\subsection{An Automaton That Executes the Program}\label{SUBSEC:APP_MON_PROG}

\Paragraph{Transition rules of $\mathcal{A}^{k}_S$.}
The first four rules execute instructions that emit no event.
The following five execute instructions that emit an event, and thus introduce it.
The last four handle $\CFEForget$ and the $\CFEIntroR{\rsym}$ operations.
As there, a rule applies only when the expressions it evaluates are not $\bot$.

\begin{mathparpagebreakable}
\inferrule{
    t \in \ThreadDom \quad S_t(\PC(t)) = \reg := \expr \\\\
    \PC' = \PC[t \mapsto \PC(t)+1] \\\\
    \Phi' = \Phi[\reg \mapsto \Phi(\expr)]
  }{
    \tuple{\PC, \Phi, \PndSet, \gamma} \xrightarrow{\epsilon} \tuple{\PC', \Phi', \PndSet, \gamma}
  }
\and
\inferrule{
    t \in \ThreadDom \quad S_t(\PC(t)) = \reg := \NDet(\ValueDom') \\\\
    v \in \ValueDom' \quad \PC' = \PC[t \mapsto \PC(t)+1] \\\\
    \Phi' = \Phi[\reg \mapsto v]
  }{
    \tuple{\PC, \Phi, \PndSet, \gamma} \xrightarrow{\epsilon} \tuple{\PC', \Phi', \PndSet, \gamma}
  }
\and
\inferrule{
     t \in \ThreadDom \quad S_t(\PC(t)) = \mathtt{if}\ \expr\ \mathtt{goto}\ m \\\\
    \Phi(\expr) \neq 0 \quad \PC' = \PC[t \mapsto m]
  }{
    \tuple{\PC, \Phi, \PndSet, \gamma} \xrightarrow{\epsilon} \tuple{\PC', \Phi, \PndSet, \gamma}
  }
\and
\inferrule{
     t \in \ThreadDom \quad S_t(\PC(t)) = \mathtt{if}\ \expr\ \mathtt{goto}\ m \\\\
    \Phi(\expr) = 0 \quad \PC' = \PC[t \mapsto \PC(t)+1]
  }{
    \tuple{\PC, \Phi, \PndSet, \gamma} \xrightarrow{\epsilon} \tuple{\PC', \Phi, \PndSet, \gamma}
  }
\and
\inferrule{
    \mathrm{op} = \CFEIntro(\llab,c) \quad \llab = \wt_a(t, x, \Phi(\expr)) \quad
    S_t(\PC(t)) = x :=_a \expr \\\\ \PC' = \PC[t \mapsto \PC(t)+1] \quad
    \PndSet' = \textsc{pend}(\PndSet,\gamma,t,c) \quad \gamma' = \gamma[t\mapsto c]
  }{
    \tuple{\PC, \Phi, \PndSet, \gamma} \xrightarrow{\mathrm{op}} \tuple{\PC', \Phi, \PndSet', \gamma'}
  }
\and
\inferrule{
    \mathrm{op} = \CFEIntro(\llab,c) \quad \llab = \rd_a(t, x, v) \quad
    S_t(\PC(t)) = \reg :=_a x \\\\ \PC' = \PC[t \mapsto \PC(t)+1] \quad
    \Phi' = \Phi[\reg \mapsto v] \quad
    \PndSet' = \textsc{pend}(\PndSet,\gamma,t,c) \quad \gamma' = \gamma[t\mapsto c]
  }{
    \tuple{\PC, \Phi, \PndSet, \gamma} \xrightarrow{\mathrm{op}} \tuple{\PC', \Phi', \PndSet', \gamma'}
  }
\and
\inferrule{
    \mathrm{op} = \CFEIntro(\llab,c) \quad \llab = \rmw_a(t, x, \Phi(\expr_R), \Phi(\expr_W)) \\\\
    S_t(\PC(t)) = \reg :=_a \mathsf{CAS}(x, \expr_R, \expr_W) \quad \PC' = \PC[t \mapsto \PC(t)+1] \\\\
    \Phi' = \Phi[\reg \mapsto \Phi(\expr_R)] \quad \PndSet' = \textsc{pend}(\PndSet,\gamma,t,c) \quad \gamma' = \gamma[t\mapsto c]
  }{
    \tuple{\PC, \Phi, \PndSet, \gamma} \xrightarrow{\mathrm{op}} \tuple{\PC', \Phi', \PndSet', \gamma'}
  }
\and
\inferrule{
    \mathrm{op} = \CFEIntro(\llab,c) \quad \llab = \rd_{\failann{a}}(t, x, v) \quad v \neq \Phi(\expr_R) \\\\
    S_t(\PC(t)) = \reg :=_a \mathsf{CAS}(x, \expr_R, \expr_W) \quad \PC' = \PC[t \mapsto \PC(t)+1] \\\\
    \Phi' = \Phi[\reg \mapsto v] \quad \PndSet' = \textsc{pend}(\PndSet,\gamma,t,c) \quad \gamma' = \gamma[t\mapsto c]
  }{
    \tuple{\PC, \Phi, \PndSet, \gamma} \xrightarrow{\mathrm{op}} \tuple{\PC', \Phi', \PndSet', \gamma'}
  }
\and
\inferrule{
    \mathrm{op} = \CFEIntro(\llab,c) \quad \llab = \fc_a(t) \\\\
    S_t(\PC(t)) = \mathsf{fence}_a \quad \PC' = \PC[t \mapsto \PC(t)+1] \\\\
    \PndSet' = \textsc{pend}(\PndSet,\gamma,t,c) \quad \gamma' = \gamma[t\mapsto c]
  }{
    \tuple{\PC, \Phi, \PndSet, \gamma} \xrightarrow{\mathrm{op}} \tuple{\PC', \Phi, \PndSet', \gamma'}
  }
\and
\inferrule{
  \mathrm{op} = \CFEForget(c) \\\\
  \nexists c'\in \IntSet{k+1}.\ \tuple{c, c'} \in \PndSet \lor \tuple{c',c} \in \PndSet \\\\
  \nexists t \in \ThreadDom.\ \gamma(t)=c
}{
    \tuple{\PC, \Phi, \PndSet, \gamma} \xrightarrow{\mathrm{op}} \tuple{\PC, \Phi, \PndSet, \gamma}
}
\and
\inferrule{
  \mathrm{op} = \CFEForget(c) \\\\
  \nexists c'\in \IntSet{k+1}.\ \tuple{c, c'} \in \PndSet \lor \tuple{c',c} \in \PndSet \\\\
  \gamma(t)=c \quad \gamma' = \gamma[t \mapsto \bot]
}{
    \tuple{\PC, \Phi, \PndSet, \gamma} \xrightarrow{\mathrm{op}} \tuple{\PC, \Phi, \PndSet, \gamma'}
}
\and
\inferrule{
  \mathrm{op} = \CFEIntrosb(c_1,c_2) \\\\
   \tuple{c_1,c_2}\in \PndSet \quad \PndSet' = \PndSet \setminus \{\tuple{c_1,c_2}\}
}{
    \tuple{\PC, \Phi, \PndSet, \gamma} \xrightarrow{\mathrm{op}} \tuple{\PC, \Phi, \PndSet', \gamma}
}
\and
\inferrule{
  \mathrm{op} = \CFEIntroR{\rsym}(c_1,c_2) \quad \rsym\in\{\rf,\mb\}
}{
    \tuple{\PC, \Phi, \PndSet, \gamma} \xrightarrow{\mathrm{op}} \tuple{\PC, \Phi, \PndSet, \gamma}
}
\end{mathparpagebreakable}
\lemautprogpl*

\begin{proof}
    Let $w$ be closed, with $G^{w}$ an execution graph, and fix $t\in\ThreadDom$.
    As $\seqbef$ restricted to $\Events^{t}$ is a successor relation, let $\event_1\LTo{\seqbef}\cdots\LTo{\seqbef}\event_m$ enumerate $\Events^{t}$.

    \SubParagraph{Soundness.}
    Suppose $\mathcal{A}^{k}_{S}$ has a run on $w$.
    A pair $\tuple{c_1,c_2}$ enters $\PndSet$ only at the $\CFEIntro$ of the event of $c_2$, with $c_1=\gamma(t)$ the color of an event of $t$ introduced before it, and $\CFEIntrosb(c_1,c_2)$ requires $\tuple{c_1,c_2}\in \PndSet$.
    Hence every $\seqbef$-edge of $G^{w}$ goes from an event to one introduced later, and $w$ is canonical.
    It follows that $\event_1,\ldots,\event_m$ are introduced in this order.
    The $\CFEIntro$ rules for $t$ and the silent rules for $t$ are the rules of \cref{SUBSEC:APP_PRELIM_PROG} applied to $\tuple{\PC(t),\Phi|_{\RegisterDom_t}}$, and the other rules leave this pair unchanged.
    So the steps of the run that concern $t$ form a run of $\Prog_t$ from $\InitState_t$ emitting $\llab(\event_1)\cdots\llab(\event_m)$, i.e., $\ProjectGraph{G^{w}}{\seqbef}{t}\in\Lang{\Prog_t}$.
    As $t$ is arbitrary, $\tuple{G^{w}.\Events,G^{w}.\seqbef}\in\GLang{\ConcProg}$.

    \SubParagraph{Completeness.}
    Suppose $w$ is canonical and $\tuple{G^{w}.\Events,G^{w}.\seqbef}\in\GLang{\ConcProg}$.
    By canonicity, $\event_1,\ldots,\event_m$ are introduced in this order, and some run $\rho_t$ of $\Prog_t$ from $\InitState_t$ emits $\llab(\event_1)\cdots\llab(\event_m)$.
    We build a run of $\mathcal{A}^{k}_{S}$ on $w$: before the $\CFEIntro$ of $\event_i$, apply the silent steps of $\rho_t$ that precede its $i$-th labeled transition, and at the $\CFEIntro$ apply the rule for the instruction at the current $\PC(t)$.
    That rule is the one executing the $i$-th labeled transition of $\rho_t$: the instruction is the same, and for a CAS the label, an $\rmw$ or a $\rd$ with a sibling annotation, selects the rule.
    It remains to check the premises of $\CFEIntrosb$ and $\CFEForget$.
    The edge $\event_i\LTo{\seqbef}\event_{i+1}$ is drawn when both events are colored, and $\event_{i+1}$ is introduced after $\event_i$, so $\event_i$ is still colored when $\event_{i+1}$ is introduced.
    At that point $\gamma(t)$ is the color $c_1$ of $\event_i$, so the rule adds $\tuple{c_1,c_2}$ to $\PndSet$, which $\CFEIntrosb(c_1,c_2)$ then finds.
    Conversely, each pending pair is removed by the unique $\CFEIntrosb$ drawing its edge, which precedes the $\CFEForget$ of either endpoint, so the premises of $\CFEForget$ hold.
    All states are accepting, so $w$ is accepted.

    Finally, we count the states.
    $\PC$ takes $\prod_t (n_t+1)\leq 2^{|\ThreadDom|}\cdot n^{|\ThreadDom|}$ values and $\Phi$ takes $(|\ValueDom|+1)^{|\RegisterDom|}$.
    Each color has at most one $\PndSet$-predecessor and at most one $\PndSet$-successor, both set at the $\CFEIntro$ of its event, so $\PndSet$ is a partial injection $\IntSet{k+1}\rightharpoonup \IntSet{k+1}$, with $2^{O(k\log k)}$ values, and $\gamma$ takes $(k+2)^{|\ThreadDom|}$ values.
\end{proof}

\section{Proofs of \cref{SEC:RELAXED}}\label{SEC:APP_RELAXED}

\subsection{Undecidability of Reachability}\label{SUBSEC:APP_RELAXED_PCP}

\therelaxedreach*

We prove \cref{THE:RELAXED_REACH} by a reduction from PCP~\cite{Post46}.

\begin{figure}
\centering
\newcommand{\pcpRow}[1]{\parbox[b][5.3cm][c]{\linewidth}{\centering #1}}
\begin{tabular*}{\textwidth}{@{}p{.55\textwidth}@{\extracolsep{\fill}}p{.4\textwidth}@{}}
\centering
$\vcenter{\hbox{%
\footnotesize
\pcpRow{%
\begin{tabular}{@{}l@{}}
\textbf{Thread} $t_1$\\ \hline
$j\gets\NDet(\IntSet{n})$\\
\textbf{loop}\\
\quad $\wt(\locid,\tuple{t_1,j})$\\
\quad $\rd(\locid,\tuple{t_2,j})$\\
\quad \textbf{for} $k\in\IntSet{|u_j|}$ \textbf{do}\\
\qquad $\wt(\loclt,\tuple{t_1,u_j[k]})$\\
\qquad $\rd(\loclt,\tuple{t_2,u_j[k]})$\\
\quad $j\gets\NDet(\IntSet{n}\cup\{0\})$\\
\quad \textbf{if} $j=0$ \textbf{then break}\\
\textbf{term}
\end{tabular}
\hspace{1.5em}
\begin{tabular}{@{}l@{}}
\textbf{Thread} $t_2$\\ \hline
$j\gets\NDet(\IntSet{n})$\\
\textbf{loop}\\
\quad $\rd(\locid,\tuple{t_1,j})$\\
\quad $\wt(\locid,\tuple{t_2,j})$\\
\quad \textbf{for} $k\in\IntSet{|v_j|}$ \textbf{do}\\
\qquad $\rd(\loclt,\tuple{t_1,v_j[k]})$\\
\qquad $\wt(\loclt,\tuple{t_2,v_j[k]})$\\
\quad $j\gets\NDet(\IntSet{n}\cup\{0\})$\\
\quad \textbf{if} $j=0$ \textbf{then break}\\
\textbf{term}
\end{tabular}%
}%
}}$ &
\centering
$\vcenter{\hbox{%
\begin{tikzpicture}[katpic, katlab/.append style={font=\scriptsize},
    evp/.style={inner sep=1.5pt, font=\scriptsize},
    evid/.style={evp, fill=colorLocId!15},
    evlt/.style={evp, fill=colorLocLt!15}]
  \node[font=\footnotesize] at (0,0.45)   {$t_1$};
  \node[font=\footnotesize] at (2.35,0.45) {$t_2$};
  \foreach \k/\l/\st in {1/{\wt(\locid,\tuple{t_1,1})}/evid, 2/{\rd(\locid,\tuple{t_2,1})}/evid, 3/{\wt(\loclt,\tuple{t_1,\mathtt{a}})}/evlt, 4/{\rd(\loclt,\tuple{t_2,\mathtt{a}})}/evlt, 5/{\wt(\loclt,\tuple{t_1,\mathtt{b}})}/evlt, 6/{\rd(\loclt,\tuple{t_2,\mathtt{b}})}/evlt, 7/{\wt(\locid,\tuple{t_1,2})}/evid, 8/{\rd(\locid,\tuple{t_2,2})}/evid, 9/{\wt(\loclt,\tuple{t_1,\mathtt{b}})}/evlt, 10/{\rd(\loclt,\tuple{t_2,\mathtt{b}})}/evlt}
    \node[\st] (a\k) at (0,-0.5*\k+0.5) {$\l$};
  \foreach \k/\l/\st in {1/{\rd(\locid,\tuple{t_1,1})}/evid, 2/{\wt(\locid,\tuple{t_2,1})}/evid, 3/{\rd(\loclt,\tuple{t_1,\mathtt{a}})}/evlt, 4/{\wt(\loclt,\tuple{t_2,\mathtt{a}})}/evlt, 5/{\rd(\locid,\tuple{t_1,2})}/evid, 6/{\wt(\locid,\tuple{t_2,2})}/evid, 7/{\rd(\loclt,\tuple{t_1,\mathtt{b}})}/evlt, 8/{\wt(\loclt,\tuple{t_2,\mathtt{b}})}/evlt, 9/{\rd(\loclt,\tuple{t_1,\mathtt{b}})}/evlt, 10/{\wt(\loclt,\tuple{t_2,\mathtt{b}})}/evlt}
    \node[\st] (b\k) at (2.35,-0.5*\k+0.5) {$\l$};
  \foreach \k/\m in {1/2,2/3,3/4,4/5,5/6,6/7,7/8,8/9,9/10} {
    \draw[evsb] (a\k) -- (a\m);
    \draw[evsb] (b\k) -- (b\m);
  }
  \foreach \s/\t in {a1.east/b1.west, b2.west/a2.east, a3.east/b3.west, b4.west/a4.east, a5.east/b7.west, b6.west/a8.east, a9.east/b9.west, b10.west/a10.east}
    \draw[evrf] (\s) -- (\t);
  \draw[evrf, line width=1.5pt, >={Stealth[length=5pt, width=5pt]}] (a7.east) -- (b5.west);
  \draw[evrf, line width=1.5pt, >={Stealth[length=5pt, width=5pt]}] (b8.west) -- (a6.east);
\end{tikzpicture}%
}}$
\tabularnewline[-9.4pt] %
\centering \subcaption{The program $\ConcProg_I$.}\label{SUBFIG:RELAXED_PCP_PROG} &
\centering \subcaption{An execution graph of $\ConcProg_I$.}\label{SUBFIG:RELAXED_PCP_GRAPH}
\tabularnewline
\end{tabular*}
\caption{The program $\ConcProg_I$ (\subref{SUBFIG:RELAXED_PCP_PROG}), and its execution graph for the instance $u=\tuple{\mathtt{ab},\mathtt{b}}$, $v=\tuple{\mathtt{a},\mathtt{bb}}$ and the solution $1,2$, with $\mb$-edges omitted, and the thick $\rf$-edges forming a $(\po\cup\rf)$-cycle (\subref{SUBFIG:RELAXED_PCP_GRAPH}).}
\label{FIG:RELAXED_PCP}
\end{figure}

\Paragraph{Post's Correspondence Problem.}
An instance $I$ of PCP consists of two tuples $\tuple{u_1,\dots,u_n}$ and $\tuple{v_1,\dots,v_n}$ of non-empty words over an alphabet $\Gamma$.
A \emph{solution} of $I$ is a non-empty sequence $j_1,\dots,j_m\in\IntSet{n}$ with $u_{j_1}\cdots u_{j_m}=v_{j_1}\cdots v_{j_m}$.
Deciding whether $I$ has a solution is undecidable~\cite{Post46}.
We write $u[k]$ for the $k$-th letter of a word $u$.

\Paragraph{The reduction.}
Given an instance $I$ of PCP, we construct the program $\ConcProg_I$ of \cref{SUBFIG:RELAXED_PCP_PROG}, with two threads $t_1,t_2$ and two locations $\locid,\loclt$.
Both threads guess a solution, one index at a time.
For each guessed index $j$, thread $t_1$ writes $j$ to $\locid$ and then the letters of $u_j$ to $\loclt$.
Similarly, thread $t_2$ writes $j$ to $\locid$ and then the letters of $v_j$ to $\loclt$.
Every value written is a pair $\tuple{t,v}$ of the writing thread $t$ and an index or a letter $v$.
Every read expects a value of the other thread, together with the index or letter that the reading thread has guessed at that point.
So, each read checks that the other thread has written the same index or letter.
Note that $t_1$ writes before it reads, while $t_2$ reads before it writes.
Finally, each thread ends by emitting $\mathbf{term}$.
Since the values range over a finite set, each thread of $\ConcProg_I$ is a finite LTS.
Note that the two threads advance on $\locid$ and on $\loclt$ at different rates, since $|u_j|\neq|v_j|$ in general.
This lag creates $(\po\cup\rf)$-cycles, as in \cref{SUBFIG:RELAXED_PCP_GRAPH}, but these are allowed by $\rlxmm$.

The correctness of the above construction is captured by the following lemma.
Intuitively, in each iteration, each thread reads and writes the same value on $\locid$ (e.g., $\wt(\locid,\tuple{t_1,j})$ and $\rd(\locid,\tuple{t_2,j})$ for $t_1$) or $\loclt$ (e.g., $\wt(\loclt,\tuple{t_1,u_j[k]})$ and $\rd(\loclt,\tuple{t_2,u_j[k]})$ for $t_1$).
The crux of the proof is in arguing that no thread can skip reading any of the writes of the other.
This implies that the sequences of indices and letters read by each thread are the same.
Therefore, whenever both threads reach $\mathbf{term}$, the sequence of indices written to $\locid$ constitutes a solution of $I$.

\begin{lemma}\label{LEM:RELAXED_PCP}
  Some $G\in\ExecutionGraphsOf{\ConcProg_I}{\rlxmm}$ reaches $\mathbf{term}$ iff $I$ has a solution.
\end{lemma}
\begin{proof}
    We first show that the two threads stay in sync.
    Let $G\in\ExecutionGraphsOf{\ConcProg_I}{\rlxmm}$ reach $\mathbf{term}$ and $z\in\{\locid,\loclt\}$.
    We claim that $t_1$ and $t_2$ write equally often on $z$, and that, for every $k$, the $k$-th read of each thread on $z$ reads from the $k$-th write of the other thread on $z$.
    As $G$ reaches $\mathbf{term}$, each thread completes every iteration it starts.
    Let $n_1$ and $n_2$ be the number of writes of $t_1$ and $t_2$ on $z$.
    We write $\wt^1_k,\rd^1_k$ for the $k$-th write and read of $t_1$ on $z$, and $\wt^2_k,\rd^2_k$ for those of $t_2$.
    By construction, $t_1$ accesses $z$ in the order $\wt^1_1,\rd^1_1,\wt^1_2,\rd^1_2,\dots$, and $t_2$ in the order $\rd^2_1,\wt^2_1,\rd^2_2,\wt^2_2,\dots$, each with as many reads as writes.
    The values force every $\rd^1_k$ to read from $t_2$ and every $\rd^2_k$ to read from $t_1$.

    As $G\models\rlxmm$, the relation $\poloc\cup\rf\cup\mo\cup\fr$ is acyclic on the accesses to $z$.
    Let $S$ be a total order on these accesses that extends it.
    Every read reads from the $S$-last write before it.
    Indeed, let $\wt\LTo{\rf}\rd$ and let $\wt'$ be a write with $\wt<_S\wt'<_S\rd$.
    As $S$ extends $\mo$, we have $\wt\LTo{\mo}\wt'$, and thus $\rd\LTo{\fr}\wt'$, contradicting $\wt'<_S\rd$.

    The write $\wt^1_k$ precedes $\rd^1_k$ in $S$, and the $S$-last write before $\rd^1_k$ belongs to $t_2$.
    Hence some write of $t_2$ lies between $\wt^1_k$ and $\rd^1_k$, and thus between $\wt^1_k$ and $\wt^1_{k+1}$.
    Symmetrically, some write of $t_1$ lies before $\rd^2_1$, and between $\wt^2_{k-1}$ and $\rd^2_k$, and thus between $\wt^2_{k-1}$ and $\wt^2_k$.
    So the writes on $z$ alternate in $S$, starting with $t_1$, as $\wt^1_1,\wt^2_1,\wt^1_2,\wt^2_2,\dots$.
    The read $\rd^1_{n_1}$ requires a write of $t_2$ after $\wt^1_{n_1}$, so $n_2\geq n_1$.
    The read $\rd^2_{n_2}$ requires a write of $t_1$ after $\wt^2_{n_2-1}$, so $n_1\geq n_2$.

    Finally, the read $\rd^1_k$ reads a write of $t_2$ between $\wt^1_k$ and $\wt^1_{k+1}$, and the only such write is $\wt^2_k$.
    The read $\rd^2_k$ reads a write of $t_1$ between $\wt^2_{k-1}$ and $\wt^2_k$, and the only such write is $\wt^1_k$.

    \SubParagraph{Soundness.}
    Let $G\in\ExecutionGraphsOf{\ConcProg_I}{\rlxmm}$ reach $\mathbf{term}$.
    Let $j_1,\dots,j_m$ and $j'_1,\dots,j'_{m'}$ be the indices guessed by $t_1$ and $t_2$, with $m,m'\geq 1$ by construction.
    By the claim on $\locid$, we have $m=m'$, and the $k$-th read of $t_1$ on $\locid$ reads $\tuple{t_2,j'_k}$ and expects $\tuple{t_2,j_k}$, so $j_k=j'_k$.
    Thread $t_1$ writes the letters of $u_{j_1}\cdots u_{j_m}$ on $\loclt$, and $t_2$ writes those of $v_{j_1}\cdots v_{j_m}$.
    By the claim on $\loclt$, the two words have equal length, and the $k$-th read of $t_1$ on $\loclt$ checks that their $k$-th letters coincide.
    Hence $j_1,\dots,j_m$ is a solution of $I$.

    \SubParagraph{Completeness.}
    Let $j_1,\dots,j_m$ be a solution of $I$.
    Let both threads guess $j_1,\dots,j_m$ and reach $\mathbf{term}$, and let $G$ be the execution graph with the following $\rf$ and $\mo$.
    On $z\in\{\locid,\loclt\}$, the $k$-th read of each thread reads from the $k$-th write of the other, and $\mo$ orders the writes as $\wt^1_1,\wt^2_1,\wt^1_2,\wt^2_2,\dots$, in the notation above.
    The values match, as both threads write $j_k$ as their $k$-th index and, the sequence being a solution, the same $k$-th letter.
    The order $\wt^1_1,\rd^2_1,\wt^2_1,\rd^1_1,\wt^1_2,\rd^2_2,\wt^2_2,\rd^1_2,\dots$ on the accesses to $z$ extends $\poloc$, $\rf$, $\mo$ and $\fr$ on $z$.
    So $\poloc\cup\rf\cup\mo\cup\fr$ is acyclic on each location, every realization of it stays on one location, and $G\models\rlxmm$.
\end{proof}

Since $\ConcProg_I$ has two threads and two locations, \cref{LEM:RELAXED_PCP} proves \cref{THE:RELAXED_REACH}.

\subsection{The Enclosure}\label{SUBSEC:APP_RELAXED_ENC}

\lemrelaxedenc*
\begin{proof}
    As $\ConcProg$ does not access $x$ and $y$, the only writes to them in $\ConcProg'$ are the final writes $\wt(y,1)$ of $t_1$ and $\wt(x,1)$ of $t_2$, which follow $\mathbf{term}$.

    Suppose that some $G\in\ExecutionGraphsOf{\ConcProg}{\rlxmm}$ reaches $\mathbf{term}$.
    By the definition of state reachability, we may take $G$ such that $\mathbf{term}$ is the last event of each thread.
    We build a witness $G'$ for $\ConcProg'$ on top of $G$, by adding $\rd(x,1)$ at the start and $\wt(y,1)$ at the end of $t_1$, $\rd(y,1)$ at the start and $\wt(x,1)$ at the end of $t_2$, and the $\rf$-edges from each write to the read of the other thread.
    Then $G'$ is a graph of $\ConcProg'$, and $G'\models\rlxmm$, as $x$ and $y$ each have one write and one read, in different threads, so no cycle on a single location arises.
    However, the two reads and the two writes form a $(\po\cup\rf)$-cycle, so $G'\not\models\scmm$, and $\ConcProg'$ is not robust.

    Conversely, let $G'$ be a witness for $\ConcProg'$.
    As the empty graph is $\scmm$-consistent, some thread, say $t_1$, has an event, which is $\rd(x,1)$.
    Its $\rf$-source is the $\wt(x,1)$ of $t_2$, which follows the $\mathbf{term}$ of $t_2$, and the $\rf$-source of the first event $\rd(y,1)$ of $t_2$ is in turn the $\wt(y,1)$ of $t_1$.
    Removing the two reads and the two writes yields a graph $G\in\ExecutionGraphsOf{\ConcProg}{\rlxmm}$ that reaches $\mathbf{term}$, as every cycle of $G$ is also one of $G'$.
\end{proof}

\end{document}